\documentclass[aps,superscriptaddress,twocolumn,dvips]{revtex4}
\usepackage{feynmp}
\usepackage{amsmath}
\usepackage{amssymb}
\usepackage{epsfig}
\usepackage{graphicx}
\usepackage{subfigure}
\usepackage{hyperref}
\usepackage{bbm}
\usepackage{multirow}
\usepackage{color}
\usepackage{longtable}
\newcommand{\tabincell}[2]{\begin{tabular}{@{}#1@{}}#2\end{tabular}}

\newcommand{\Rmnum}[1]{\expandafter\@slowromancap\romannumeral #1@}
\allowdisplaybreaks[3]

\begin{document}

\title{Excitonic pairing and insulating transition in two-dimensional semi-Dirac semimetals}

\author{Jing-Rong Wang}
\affiliation{High Magnetic Field Laboratory, Hefei Institutes of
Physical Science, Chinese Academy of Sciences, Hefei, Anhui 230031,
P. R. China} \affiliation{Department of Modern Physics, University
of Science and Technology of China, Hefei, Anhui 230026, P. R.
China}
\author{Guo-Zhu Liu}
\altaffiliation{Corresponding author: gzliu@ustc.edu.cn}
\affiliation{Department of Modern Physics, University of Science and
Technology of China, Hefei, Anhui 230026, P. R. China}
\author{Chang-Jin Zhang}
\altaffiliation{Corresponding author: zhangcj@hmfl.ac.cn}
\affiliation{High Magnetic Field Laboratory, Hefei Institutes of
Physical Science, Chinese Academy of Sciences, Hefei, Anhui 230031,
P. R. China} \affiliation{Collaborative Innovation Center of
Advanced Microstructures, Nanjing University, Nanjing 210093, P. R.
China}

\begin{abstract}
A sufficiently strong long-range Coulomb interaction can induce
excitonic pairing in gapless Dirac semimetals, which generates a
finite gap and drives semimetal-insulator quantum phase transition.
This phenomenon is in close analogy to dynamical chiral symmetry
breaking in high energy physics. In most realistic Dirac semimetals,
including suspended graphene, Coulomb interaction is too weak to
open an excitonic gap. The Coulomb interaction plays a more
important role at low energies in a two-dimensional semi-Dirac
semimetal, in which the fermion spectrum is linear in one component
of momenta and quadratic in the other, than a Dirac semimetal, and
indeed leads to breakdown of Fermi liquid theory. We study dynamical
excitonic gap generation in a two-dimensional semi-Dirac semimetal
by solving the Dyson-Schwinger equation, and show that a moderately
strong Coulomb interaction suffices to induce excitonic pairing.
Additional short-range four-fermion coupling tends to promote
excitonic pairing. Among the available semi-Dirac semimetals, we
find that TiO$_{2}$/VO$_{2}$ nanostructure provides a promising
candidate for the realization of excitonic insulator. We also apply
the renormalziation group method to analyze the strong coupling
between the massless semi-Dirac fermions and the quantum critical
fluctuation of excitonic order parameter at the semimetal-insulator
quantum critical point, and reveal non-Fermi liquid behaviors of
semi-Dirac fermions.
\end{abstract}

\maketitle


\section{Introduction\label{Sec:Introduction}}

The past decade has witnessed the appearance of a huge amount of
experimental and theoretical work on the physical properties of
various semimetals, in which the valence and conduction bands touch
at isolated points \cite{Vafek14, Wehling14, Weng16}. The low-energy
elementary excitations in these semimetals are various types of
massless fermions. The fermion density of state (DOS) vanishes at
band-touching points, so the Coulomb interaction remains
long-ranged. This is in sharp contrast to the ordinary metals
featuring a finite Fermi surface, where Coulomb interaction becomes
short-range due to static screening.

Graphene is a two-dimensional (2D) Dirac semimetal with massless
Dirac fermions being its low-energy excitations \cite{CastroNeto09,
Kotov12}. The surface state of three-dimensional (3D) topological
insulator (TI) is also a 2D Dirac semimetal \cite{Hasan10, Qi11}.
Apart from these two examples, there are also a number of other
systems that support low-energy bulk massless Dirac or Weyl
fermions. For instance, a 3D Dirac semimetal emerges at the quantum
critical point (QCP) between a trivial band insulator and a
topological insulator \cite{Xu11, Sato11, Brahlek12, Wu13}.
Moreover, stable 3D Dirac semimetal, protected by crystal symmetry,
has been found to exist in Na$_{3}$Bi \cite{Liu14} and
Cd$_{3}$As$_{2}$ \cite{Neupane14, Borisenko14, Jeon14}. 3D Weyl
semimetal, which hosts fermions with linear dispersion around pairs
of Weyl points with opposite chirality, was observed in TaAs
\cite{Xu15A, Lv15A, Lv15B, Yang15}, NbAs \cite{Xu15B}, TaP
\cite{Xu15C, XuN16}, and NbP \cite{XuFengGroup15, Souma16} by
angle-resolved photoemission spectroscopy (ARPES) experiments.
Several other types of semimetals, including 3D quadratic semimetal
\cite{Nakatsuji06, Machida10}, 3D anisotropic Weyl semimetal
\cite{Yang13}, 3D double Weyl semimetal \cite{Fang12, Huang16}, 3D
nodal-line semimetal \cite{Bian16, Wu16, Schoop16, Hu16}, and 2D
semi-Dirac semimetal \cite{Isobe16A, Cho16}, are also widely
investigated.

The effects of long-range Coulomb interaction have been studied in
various types of semimetals \cite{Kotov12, Sekine13, Sitte13,
Goswami11, Hosur12, Throckmorton15, Gonzalez14, Moon13, Herbut14,
Janssen15, Dumitrescu15, Janssen16A, Janssen16B, Yang14, Lai15,
Jian15, Huh16, Isobe16A, Cho16, Isobe16B}. The role of Coulomb
interaction depends crucially on the fermion dispersion and the
dimension. Extensive renormalization group (RG) analysis
\cite{Shankar94} have revealed that Coulomb interaction is
marginally irrelevant in 2D Dirac semimetal \cite{Kotov12, Sekine13,
Sitte13}, 3D Dirac/Weyl semimetal \cite{Goswami11, Hosur12,
Throckmorton15, Gonzalez14}, and also 3D double Weyl semimetal
\cite{Lai15, Jian15}. The Fermi liquid (FL) theory is valid in 2D
Dirac semimetal \cite{Kotov12}. However, the Coulomb interaction
cannot be simply neglected as it results in fermion velocity
renormalization and logarithmic-like correction to some of the
observable quantities \cite{Kotov12, Gonzalez94, Gonzalez99, Son07,
Hofmann14, Bauer15, Sharma16}. Indeed, the predicted velocity
renormalization has already been observed in ultra clean suspended
graphene \cite{Elias11}, quasi-freestanding graphene on silicon
carbide (SiC) \cite{Siegel11}, and graphene on boron nitride
substrate \cite{Yu13}. There is also experimental evidence for the
renormalization of fermion dispersion in TI-like system of Bi
bilayer grown on Bi$_{2}$Se$_{3}$ \cite{Miao13}, which seems to be
caused by Coulomb interaction. In a 3D semimetal with quadratic
dispersion, the Coulomb interaction is found to be relevant and
cause non-Fermi liquid (NFL) behaviors \cite{Moon13}. It is
interesting to notice that recent ARPES experiments have discovered
NFL behavior in a 3D quadratic semimetal material
Pr$_{2}$Ir$_{2}$O$_{7}$ \cite{Kondo15}. Moreover, in a 3D
anisotropic Weyl semimetal \cite{Yang14} and a 3D nodal-line
semimetal \cite{Huh16}, the FL description is robust because the
Coulomb interaction is irrelevant.

In an interacting 2D Dirac fermion system, an intriguing property is
that the strong Coulomb interaction might bind a gapless fermion and
a gapless hole to form an excitonic pair, which generates a finite
energy gap at the Dirac points and turns the Dirac semimetal into an
excitonic insulator \cite{Kotov12, Khveshchenko01,
CastroNetoPhysics09}. This picture is very similar to the
non-perturbative phenomenon of dynamical chiral symmetry breaking
(DCSB) that has been extensively investigated in high energy physics
\cite{Miransky} since the pioneering work of Nambu and Jona-Lasino
\cite{Nambu61}. In QCD, the current quarks are massless, but acquire
a dynamical mass due to the strong interaction mediated by gluons.
The dynamical quark mass breaks the chiral symmetry that is
preserved by massless quarks. In a 2D Dirac semimetal, the dynamical
fermion mass gap breaks the sublattice symmetry respected by
massless Dirac fermions.

According to theoretical analysis \cite{Kotov12, Khveshchenko01}, an
excitonic gap can be dynamically generated in zero external magnetic
field only when the effective strength of Coulomb interaction,
denoted by the parameter $\alpha$, is larger than certain critical
value. The critical value $\alpha_c$ defines a QCP that separates
the semimetallic and excitonic insulating phases. This issue has
been most extensively studied in the context of undoped graphene
\cite{Kotov12, Khveshchenko01, CastroNetoPhysics09}. Physically, the
excitonic insulating transition corresponds to the formation of a
charge density wave (CDW) order \cite{Kotov12, Khveshchenko01}. If
the ground state of suspended graphene is an excitonic insulator,
there would be more practical applications of graphene in the design
of electronic devies \cite{Kotov12, CastroNetoPhysics09}. From a
theoretic point of view, this provides an ideal laboratory to test
some important concepts of high energy physics \cite{Zhang11}, and
also gives us a nice platform to study the rich quantum critical
phenomena.

In graphene and other Dirac semimetals, the effective strength of
Coulomb interaction can be quantified by the ratio between the
Coulombic potential energy and kinetic energy of fermions, defined
as \cite{Kotov12} $\alpha = e^2/v\epsilon$, where $e$ is the
electric charge, $v$ fermion velocity, and $\epsilon$ dielectric
constant. Experiments have determined that the fermion velocity is
$v\approx c/300 \approx 10^{6}m/s$, where $c$ is the speed of light
in vacuum \cite{CastroNeto09, Kotov12}. For graphene on SiO$_{2}$
substrate, $\alpha \approx 0.78$. When graphene is suspended in
vacuum, $\alpha$ takes its largest value $\alpha \approx 2.2$
\cite{CastroNetoPhysics09}, which implies that excitonic pairing is
most possibly realized in suspended graphene. To specify whether
suspended graphene has a semimetallic or insulating ground state,
one needs to calculate the critical value $\alpha_c$ and then
compares it with $\alpha \approx 2.2$. The value of $\alpha_c$ has
been evaluated by means of several different methods, including
Dyson-Schwinger (DS) equation \cite{Khveshchenko01, Gorbar02,
Khveshchenko04, Liu09, Khveshchenko09, Gamayun10, Sabio10, Liu11,
WangLiu11A, WangLiu11B, WangLiu12, Popovici13, WangLiu14,
Gonzalez15, Carrington16}, Bethe-Salpeter (BS) equation
\cite{Gamayun09, WangJianhui11, Katanin16}, RG approach
\cite{Vafek08, Gonzalez10, Gonzalez12, Katanin16}, Monte Carlo
simulations \cite{Drut09A, Drut09B, Drut09C, Armour10, Armour11,
Buividovich12, Ulybyshev13, Smith14}, and some other methods
\cite{Juan12, Kotikov16}. Drut and L\"{a}hde claimed that
$\alpha_{c} \approx 1.1$ in a 2D Dirac fermion system \cite{Drut09A,
Drut09B}, which indicates that graphene placed on SiO$_{2}$
substrate is semimetallic but suspended graphene is insulating at
zero temperature. Extensive studies based on DS equation
\cite{Khveshchenko04, Liu09, Gamayun10, WangLiu11A, WangLiu11B}, BS
equation \cite{Gamayun09, WangJianhui11}, and RG approch
\cite{Vafek08, Gonzalez12} all showed that $\alpha_{c} < 2.2$ and
hence also predicted an insulating ground state of suspended
graphene. However, experiments clearly revealed that suspended
graphene remains a semimetal down to very low temperatures, without
any sign of insulating behavior \cite{Elias11, Mayorov12}, which is
apparently inconsistent with earlier theoretical results. A number
of improved studies \cite{WangLiu12, Ulybyshev13, Smith14,
Popovici13, Gonzalez15, Carrington16} have been accomplished to
reconcile this discrepancy. After taking into account the influence
of fermion velocity renormalization and fermion damping, it was
showed in a refined DS equation analysis \cite{WangLiu12} that the
critical value is $\alpha_c \approx 3.25$, which is much larger than
$\alpha = 2.2$. Thus it turns out that the Coulomb interaction in
suspended graphene is still not strong enough to generate a
dynamical gap. Subsequent DS equation studies \cite{Popovici13,
Gonzalez15, Carrington16} reached the same qualitative conclusion.
After considering the screening of Coulomb interaction due to the
$\sigma$-band electrons \cite{Wehling11}, recent Monte Carlo
simulations also found that $\alpha_{c}$ is greater than $2.2$
\cite{Ulybyshev13, Smith14}, which suggests a semimetallic ground
state of suspended graphene.

Since graphene cannot be an excitonic insulator, we turn to consider
other types of semimetal where the Coulomb interaction plays a
more important role. Among the existing semimetal materials, we find
that 2D semi-Dirac semimetal provides a better candidate for the
realization of excitonic insulating transition than graphene. The
dispersion for 2D semi-Dirac fermion is
\begin{eqnarray}
E=\pm\sqrt{ak_{x}^{4}+v^2k_{y}^{2}},
\end{eqnarray}
which is linear along one momentum component ($k_y$) but quadratic
along the other one ($k_x$). Such fermions can emerge at the QCP
between a 2D Dirac semimetal and a band insulator upon merging two
separate Dirac points into a single one. Generating 2D semi-Dirac
fermions by merging pairs of Dirac points was predicted to take
place in deformed graphene \cite{Hasegawa06, Dietl08, Goerbig08,
Montambaux09A, Montambaux09B}, pressured organic compound
$\alpha$-(BEDT-TTF)$_{2}$I$_{3}$ \cite{Goerbig08, Montambaux09B,
Kobayashi07, Kobayashi11}, few-layer black phosphorus subject to
pressure or perpendicular electric field \cite{Dolui15, Yuan16} or
doping \cite{Baik15}, and some sorts of artificial optical lattices
\cite{Wunsch08, Lim12}. Experimentally, the merging of distinct
Dirac points and the appearance of semi-Dirac fermions were recently
observed in ultracold Fermi gas of $^{\mathrm{40}}$K atoms in
honeycomb lattice \cite{Tarruell12}, and microwave cavities with
graphene-like structure \cite{Bellec13}. Kim \emph{et al.}
\cite{Kim15} realized semi-Dirac semimetals in few-layer black
phosphorus at critical surface doping with potassium. Robust
semi-Dirac semimetal state was also predicted to appear in
TiO$_{2}$/VO$_{2}$ nanostructure under suitable conditions
\cite{Pardo09, Pardo10, Banerjee09}. It was suggested by
first-principle calculations that semi-Dirac fermions may emerge in
strained puckered arsenene \cite{Kamal15, CanWang16}.

The influence of long-range Coulomb interaction in 2D semi-Dirac
fermion system was recently investigated by pertubative RG method to
one-loop order \cite{Isobe16A, Cho16}, which showed that Coulomb
interaction becomes anisotropic due to dynamical screening
\cite{Isobe16A, Cho16}. When the bare value of interaction strength
$\alpha$ is in the strong coupling regime, the Coulomb interaction
induces an anomalous dimension for fermions and produces NFL
behaviors over a wide range of intermediate energies
\cite{Isobe16A}. This property is qualitatively similar to graphene
\cite{Son07}. Regardless of the value of $\alpha$, the fermion
kinetic energy gets enhanced as the renormalized $a$ and $v$
increase with lowering energy. This then drives $\alpha$ to flow to
zero in the low-energy region very slowly, which means the Coulomb
interaction in 2D semi-Dirac semimetal is marginally irrelevant.
However, 2D semi-Dirac semimetal differs from graphene in one
important aspect. In graphene, the quasiparticle residue $Z_{f}$
flows to a finite value and thus the system is actually a normal FL
despite of the existence of strong velocity renormalization
\cite{Kotov12, Gonzalez99, Hofmann14}. In contrast, in a 2D
semi-Dirac semimetal, $Z_{f}$ flows to zero in the low-energy regime
quite slowly, and the system displays a marginal Fermi liquid like
behavior in the lowest energy limit \cite{Isobe16A}. This difference
indicates that the Coulomb interaction plays a more important role
in 2D semi-Dirac semimetal than graphene. It might be possible to
form excitonic insulator in some realistic 2D semi-Dirac semimetal.

Dynamical excitonic gap generation is a genuine non-perturbative
phenomenon, and cannot be obtained within the framework of ordinary
leading-order perturbative calculation \cite{Isobe16A, Cho16}. In
this paper, we study the possibility of dynamical gap generation in
a 2D semi-Dirac semimetal by solving the self-consistent DS integral
equation of the excitonic gap. The DS equation is formally very
complicated, so it is usually necessary to make some approximations.
After solving the DS equation by employing three frequently used
approximations, we show that a moderately strong Coulomb interaction
suffices to generate an excitonic gap, and that it is easier to
realize an excitonic insulating state in a 2D semi-Dirac semimetal
than in a 2D Dirac semimetal. Among the currently known semi-Dirac
semimetals, we find that the TiO$_{2}$/VO$_{2}$ nanostructure is a
particularly promising candidate to realize excitonic insulating
state. One reason is that the such nanostructure is an intrinsic
semi-Dirac fermion system, without necessity of fine tuning. The
other reason is that the physical value of $\alpha$ in this
nanostructure is either smaller or very close to the critical value
$\alpha_c$ obtained in our DS equation calculations. Apart from the
long-range Coulomb interaction, there may be some additional
short-range four-fermion couplings in real materials. Adding such
coupling to the system reduces the critical value $\alpha_c$, and
hence catalyzes dynamical excitonic gap generation.

In the insulating phase, the fermions are massive and the systems
exhibits different properties than the semimetal phase. We calculate
DOS and specific heat of the insulating phase, and compare them with
those of semimetal phase. Moreover, we perform a systematic RG
analysis of the Yukawa-type coupling between massless fermions and
the quantum fluctuation of excitonic order parameter at the QCP of
semimetal-insulator transition, and find NFL behaviors of massless
fermions. We also study the interplay of this Yukawa coupling with
long-range Coulomb interaction. In this case, the massless fermions
still exhibit NFL behaviors, but some model parameters behave quite
differently from those obtained in the case without Coulomb
interaction.

The rest of the paper will be organized as follows. We present the
Hamiltonians and the propagators in Sec.~\ref{Sec:Hamiltonian}. In
Sec.~\ref{Sec:DysonEquation}, we derive the self-consistent gap
equation, and numerically solve the gap equation in several
different approximations which were employed in the studies of the
excitonic gap equation in graphene. We compare our results for
semi-Dirac semimetal and previous results for graphene carefully in
this section. In Sec.~\ref{Sec:FourFermion}, we study the dynamical
gap generation including long-range Coulomb interaction and
additional four-fermion interaction. The impact of excitonic gap for
several observable quantities and the NFL behaviors of the fermions
at the QCP between semi-Dirac semimetal phase and excitonic
insulating phase are shown in Sec.~\ref{Sec:ObveQuan}. We summarize
the main results in Sec.~\ref{Sec:Discussion}.

\section{Model Hamiltonian\label{Sec:Hamiltonian}}

The Hamiltonian for free 2D semi-Dirac fermions is
\begin{eqnarray}
H_{\mathrm{f}} = \sum_{\sigma=1}^{N}\int d^2\mathbf{x}
\psi_{\sigma}^{\dag}(\mathbf{x})
\left[-a\nabla_{x}^{2}\tau_{1}-iv\nabla_{y}\tau_{2}\right]
\psi_{\sigma}(\mathbf{x}),
\end{eqnarray}
where $\psi_{\sigma}$ represents the two-component spinor field with
flavor index $\sigma = 1, 2, 3, ..., N$, and $\tau_{1, 2, 3}$ are
standard Pauli matrices. The spinor $\psi_{\sigma}$ can be written
as $\psi_{\sigma} = (\psi_{A\sigma}, \psi_{B\sigma})^{T}$, where $A$
and $B$ are two sublattice indices \cite{Dietl08, Montambaux09A,
Dora13}. Two model parameters $a$ and $v$ are introduced to
characterize the fermion energy spectrum. The fermions are subject
to a long-range Coulomb interaction, given by
\begin{eqnarray}
H_{\mathrm{C}} = \frac{1}{4\pi}\sum_{\sigma=1}^{N}\int d^2\mathbf{x} d^2
\mathbf{x}'\rho_{\sigma}(\mathbf{x}) \frac{e^2}{\epsilon\left|\mathbf{x} -
\mathbf{x}'\right|}\rho_{\sigma}^{\dag}(\mathbf{x}'),
\end{eqnarray}
where the fermion density operator is defined as
$$\rho_{\sigma}(\mathbf{x}) = \psi_{\sigma}^{\dag}(\mathbf{x})
\psi_{\sigma}(\mathbf{x}).$$ The model will be treated by making
perturbative expansion in powers of $1/N$.

The free fermion propagator reads
\begin{eqnarray}
G_{0}(\omega,\mathbf{k}) = \frac{1}{-i\omega+ak_{x}^{2}\tau_{1} +
vk_{y}\tau_{2}}.\label{Eq:FreeFermonPropagatorDef}
\end{eqnarray}
The bare Coulomb interaction is written in the momentum space as
\begin{eqnarray}
V_{0}(\mathbf{q}) = \frac{2\pi e^2}{\epsilon|\mathbf{q}|} =
\frac{2\pi\alpha v}{|\mathbf{q}|},
\end{eqnarray}
where $\alpha = e^2/\epsilon v$ represents the effective interaction
strength. After including the dynamical screening, the dressed
Coulomb interaction function can be written as
\begin{eqnarray}
V(\Omega,\mathbf{q}) = \frac{1}{V_{0}^{-1}(\mathbf{q}) +
\Pi(\Omega,\mathbf{q})},\label{Eq:DressedCoulomb}
\end{eqnarray}
in which the polarization function $\Pi(\Omega,\mathbf{q})$ is
given by
\begin{eqnarray}
\Pi(\Omega,\mathbf{q})& =& -N\int\frac{d\omega}{2\pi}\frac{d^2\mathbf{k}}{(2\pi)^{2}}
\mathrm{Tr}\left[G_{0}(\omega,\mathbf{k})\right.\nonumber
\\
&&\left.\times G_{0}(\omega +
\Omega,\mathbf{k} + \mathbf{q})\right]
\end{eqnarray}
to the leading order of $1/N$ expansion. It is technically quite
difficult to obtain a complete analytical expression of
$\Pi(\Omega,\mathbf{q})$. The recent work of Isobe \emph{et al.}
\cite{Isobe16A} found that $\Pi(\Omega,\mathbf{q})$ can be
approximated by the expression
\begin{eqnarray}
\Pi(\Omega,q_{x},q_{y}) &=&
\frac{N}{v}\frac{d_{x}a^{1/2}q_{x}^{2}} {\left(\Omega^2 +
c_{0}a^{2}q_{x}^{4}+v^2q_{y}^{2}\right)^{1/4}}\nonumber
\\
&+& \frac{N}{v}\frac{d_{y}a^{-1/2}v^2 q_{y}^{2}}{\left(\Omega^2 +
c_{0}a^2q_{x}^{4}+v^2q_{y}^{2}\right)^{3/4}}, \label{Eq:PolarizationApproExpression}
\end{eqnarray}
which produces the precise analytical expressions of
$\Pi(\Omega,\mathbf{q})$ in several different limits. In this
expression, $d_{x}$, $d_{y}$, and $c_{0}$ are three constants:
\begin{eqnarray}
&&d_{x} = \frac{1}{8\sqrt{\pi}}\frac{\Gamma(3/4)}{\Gamma(9/4)},
\quad d_{y} = \frac{1}{8\sqrt{\pi}}\frac{\Gamma(5/4)}{\Gamma(7/4)},
\nonumber \\
&&c_{0}=\left(\frac{2}{\sqrt{\pi}}
\frac{\Gamma(3/4)}{\Gamma(9/4)}\right)^4.
\end{eqnarray}

If the fermions dynamically acquire a finite mass $m$ due to Coulomb
interaction, a new term will be added to the total Hamiltonian
\begin{eqnarray}
H_{m} &=& m\sum_{\sigma=1}^{N}\int{\frac{d^2\mathbf{k}}{(2\pi)^{2}}}
\psi_{\sigma}^{\dag}\tau_{3}\psi_{\sigma}\nonumber \\
&=& m\sum_{\sigma=1}^{N}\int{\frac{d^2\mathbf{k}}{(2\pi)^{2}}}
\left(\psi_{A\sigma}^{\dag}\psi_{A\sigma} -
\psi_{B\sigma}^{\dag}\psi_{B\sigma}\right). \label{Eq:HamiltonianMassTerm}
\end{eqnarray}
It is easy to observe that the dynamically generated mass $m$ breaks
the exchanging symmetry between sublattices $A$ and $B$. Therefore,
the excitonic gap leads to the formation of CDW state.

\section{Dyson-Schwinger equation of excitnonic gap\label{Sec:DysonEquation}}

Since excitonic gap generation is a non-perturbative phenomenon, it
cannot be investigated by making ordinary perturbative calculations.
This issue will be studied by analyzing the DS equation, which is
non-perturbative in nature and provides an ideal tool of describing
various phase transitions. Since Nambu and Jona-Lasinio
\cite{Nambu61}, the DS equation approach has been widely applied to
study DCSB in QCD \cite{Roberts94, Roberts00} and QED$_{3}$
\cite{Appelquist88, Appelquist04, Herbut02, Liu03, Fischer04,
Feng06, WangLiuZhang15}. It also has been used to examine whether an
excitonic gap can be dynamically generated by the Coulomb
interaction in graphene \cite{Khveshchenko01, Gorbar02,
Khveshchenko04, Liu09, Khveshchenko09, Gamayun10, Sabio10, Zhang11,
Liu11, WangLiu11A, WangLiu11B, WangLiu12, Popovici13, WangLiu14,
Gonzalez15, Carrington16} and other closely related materials
\cite{Gonzalez14, Gonzalez15, Janssen16A}. The role played by the DS
equation in the studies of excitonic gap generation is similar to
that played by the gap equation in the studies of the formation of
superconductivity in BCS theory. In this section, we will compute
the critical interaction strength $\alpha_c$ for excitonic
insulating transition by solving the DS equation of fermion gap.

The full fermion propagator can be written as
\begin{eqnarray}
G_{F}(\omega,\mathbf{k}) = \frac{1}{-A_{0}i\omega +
A_{1}ak_{x}^{2}\tau_{1} + A_{2}vk_{y}\tau_{2}+m\tau_{3}}.
\label{Eq:FermionPropagatorFull}
\end{eqnarray}
Here, we introduce $A_{0,1,2}\equiv A_{0,1,2}(\omega,k_{x},k_{y})$
to represent the renormalized functions, and use $m \equiv
m(\omega,k_{x},k_{y})$ to represent the dynamically generated
fermion mass. The full and free fermion propagators are connected by
the following DS equation
\begin{eqnarray}
&&G_{F}^{-1}\left(\varepsilon,\mathbf{p}\right) =
G_{0}^{-1}\left(\varepsilon,\mathbf{p}\right) +
\Sigma(\varepsilon,\mathbf{p}),
\end{eqnarray}
where the fermion self-energy function is
\begin{eqnarray}
\Sigma(\varepsilon,\mathbf{p}) = \int\frac{d\omega}{2\pi}
\frac{d^2\mathbf{k}}{(2\pi)^{2}}G_{F}(\omega,\mathbf{k})
\Gamma_{\varepsilon,\mathbf{p},\omega,\mathbf{k}}
V\left(\Omega, \mathbf{q}\right)\label{Eq:GapEqOriginal}
\end{eqnarray}
with $\Omega = \varepsilon-\omega$ and $\mathbf{q} =
\mathbf{p}-\mathbf{k}$. The function
$\Gamma_{\varepsilon,\mathbf{p},\omega,\mathbf{k}}
\equiv\Gamma(\varepsilon,\mathbf{p};\omega,\mathbf{k})$ is the
vertex correction. In order to make the above equation tractable, it
is necessary to truncate the equation in a proper way. As the first
study in this field, here we employ the lowest order truncation.
Various higher order corrections will be systematically examined in
the subsequent works. Currently, we assume that $A_{0}\equiv 1$,
which is justified at large $N$ because the equation of $A_{0}$
contains a factor of $1/N$. We also assume $\Gamma \equiv 1$, which
naturally satisfies the Ward identity. To further simplify the
problem, we take $A_{1} = A_{2} \equiv 1$. Such truncation scheme
has previously been adopted to study dynamical gap generation in 2D
Dirac semimetal \cite{Khveshchenko01, Gorbar02, Khveshchenko09,
Gamayun10}, in QED$_{3}$ model \cite{Appelquist88, Herbut02}, and
also in 3D quadratical semimetal \cite{Janssen16A}. These studies
serve as a very useful starting point for further, improved
analysis.

After making these approximations, we obtain the following
non-linear integral equation of fermion mass
\begin{eqnarray}
m(\varepsilon,p_{x},p_{y}) &=& \int\frac{d\omega}{2\pi}
\int\frac{d^2\mathbf{k}}{(2\pi)^2} m(\omega,k_{x},k_{y}) \nonumber
\\
&&\times\frac{1} {\omega^2 + a^2k_{x}^{4} + v^2k_{y}^2 +
m^2(\omega,k_{x},k_{y})} \nonumber \\
&&\times V(\varepsilon - \omega,\mathbf{p} -
\mathbf{k}).\label{Eq:GapGeneral}
\end{eqnarray}
The integration ranges for $k_{x}$ and $k_{y}$ are chosen as
$k_{x}\in(-\Lambda_{x},\Lambda_{x})$ and
$k_{y}\in(-\Lambda_{y},\Lambda_{y})$, respectively. It is usually
sufficient to suppose that $\Lambda_{x}=\Lambda_{y}=\Lambda$, where
$\Lambda$ is the unit of momenta, and $v\Lambda$ is the unit of
energy. The solution is determined by three parameters: interaction
strength $\alpha$, fermion flavor $N$, and a tuning parameter $\beta
= a\Lambda/v$, where $\Lambda$ is a UV cutoff.

\begin{figure}[htbp]
\center
\includegraphics[width=2.6in]{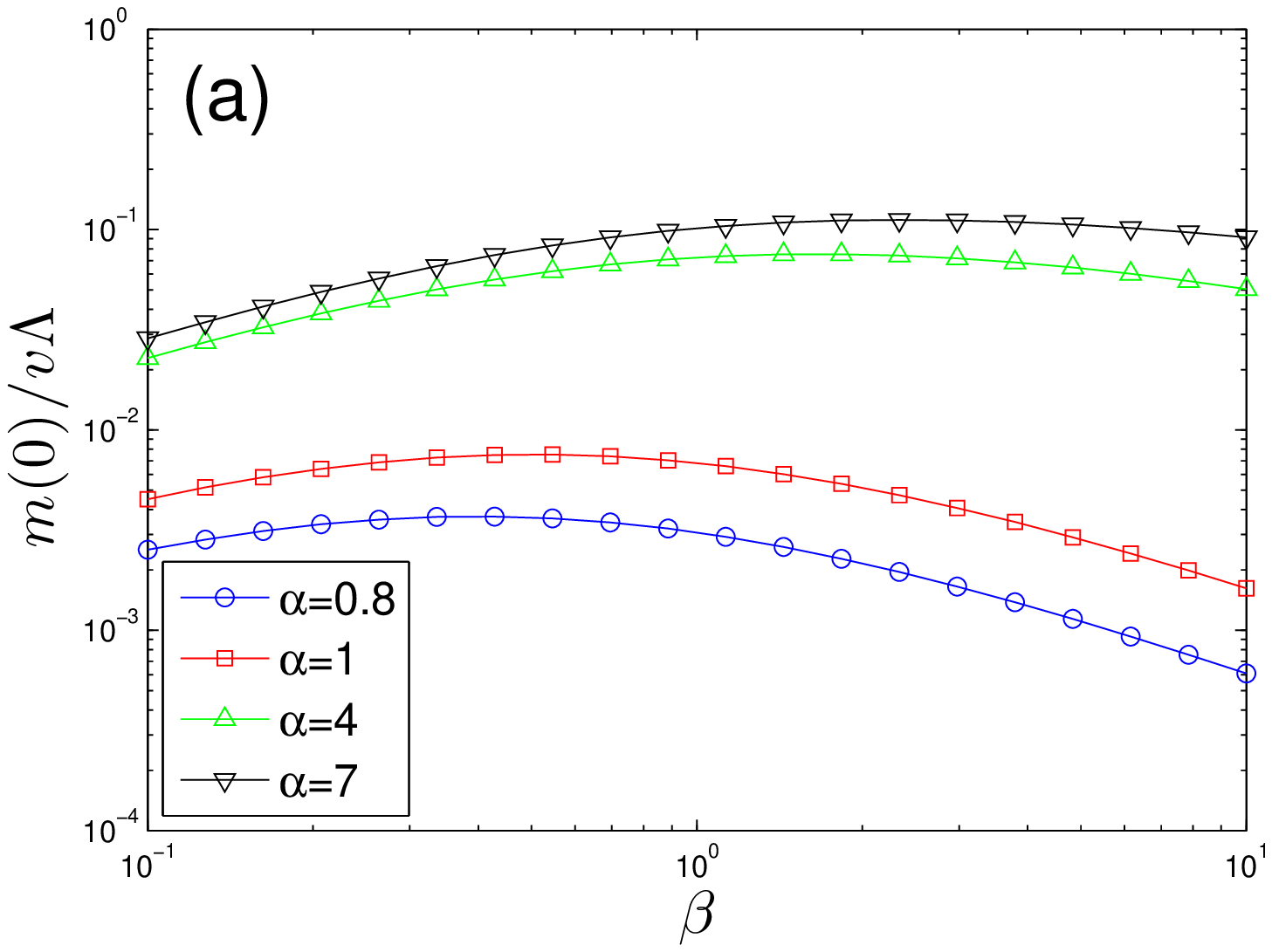}
\includegraphics[width=2.6in]{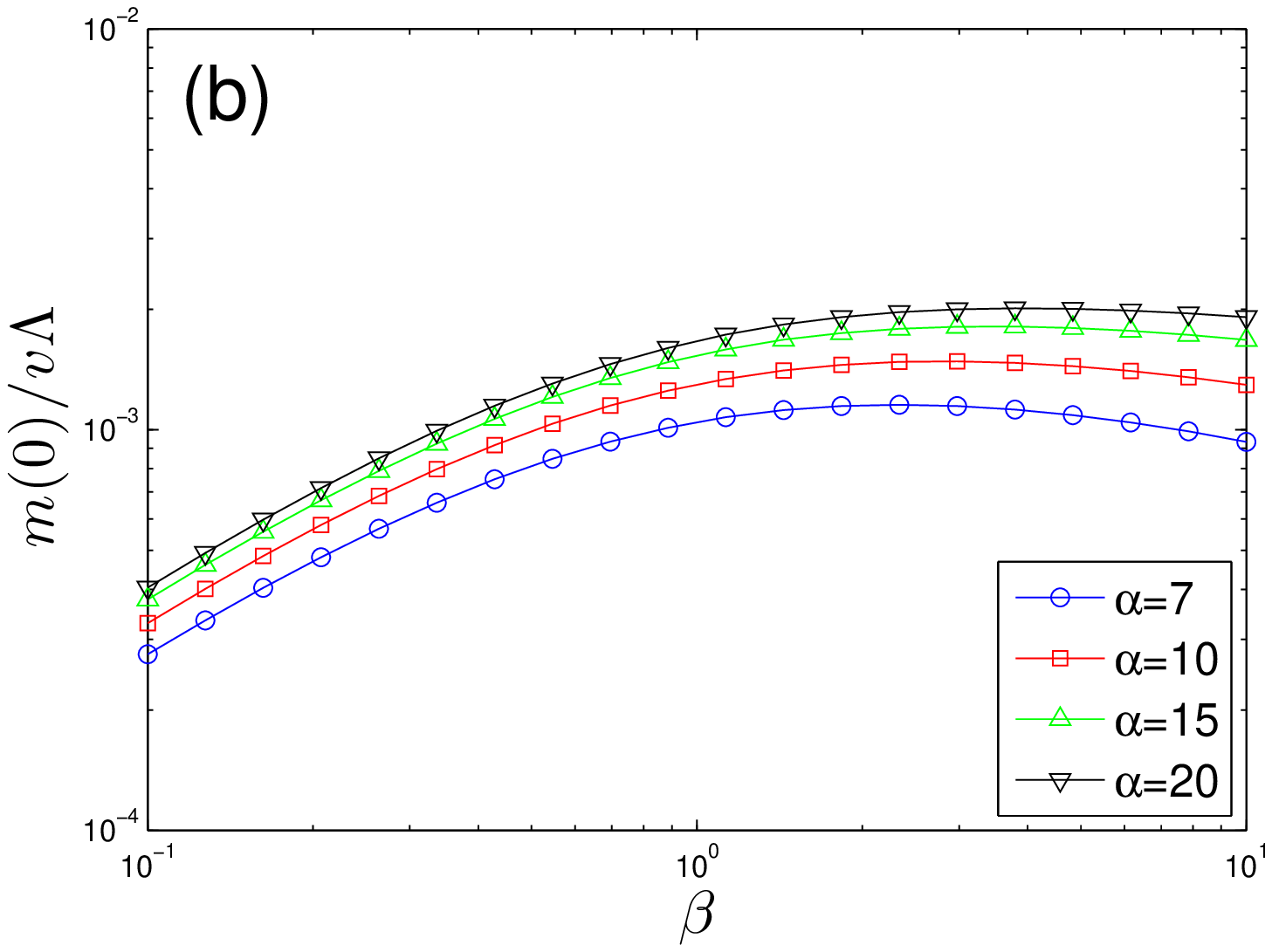}
\caption{Dependence of fermion mass $m(0)$ on $\beta$ obtained under
instantaneous approximation at different values of $\alpha$. (a)
$N=2$; (b) $N=4$. \label{Fig:Gap0Beta}}
\end{figure}

The above equation can be numerically solved by the iterative
method. Due to the explicit breaking of Lorentz invariance by the
Coulomb interaction and the anisotropic fermion dispersion,
the fermion mass gap $m(\varepsilon,p_{x},p_{y})$ depends on its
three free variables separately. Therefore, the above gap equation
is formally much more complicated than that in graphene, where the
gap equation contains only two independent variables, namely
$\varepsilon$ and $|p|$. To make sure that our numerical iterations
are under control, it is necessary to introduce further
approximations to the above gap equation. In the DS equation studies
of excitonic gap generation in graphene, there are three frequently
used approximations: instantaneous approximation
\cite{Khveshchenko01, Gorbar02, Khveshchenko04, Liu09}, Khveshchenko
approximation \cite{Khveshchenko09}, and Gamayun-Gorbar-Gusynin
(GGG) approximation \cite{Gamayun10}. We shall numerically solve the
DS equation (\ref{Eq:GapGeneral}) under these three approximations
separately, and then compare the results to those obtained in the
context of graphene \cite{Khveshchenko01, Gorbar02, Khveshchenko04,
Khveshchenko09, Gamayun10}.

\subsection{Instantaneous approximation}

\begin{figure}[htbp]
\center
\includegraphics[width=2.6in]{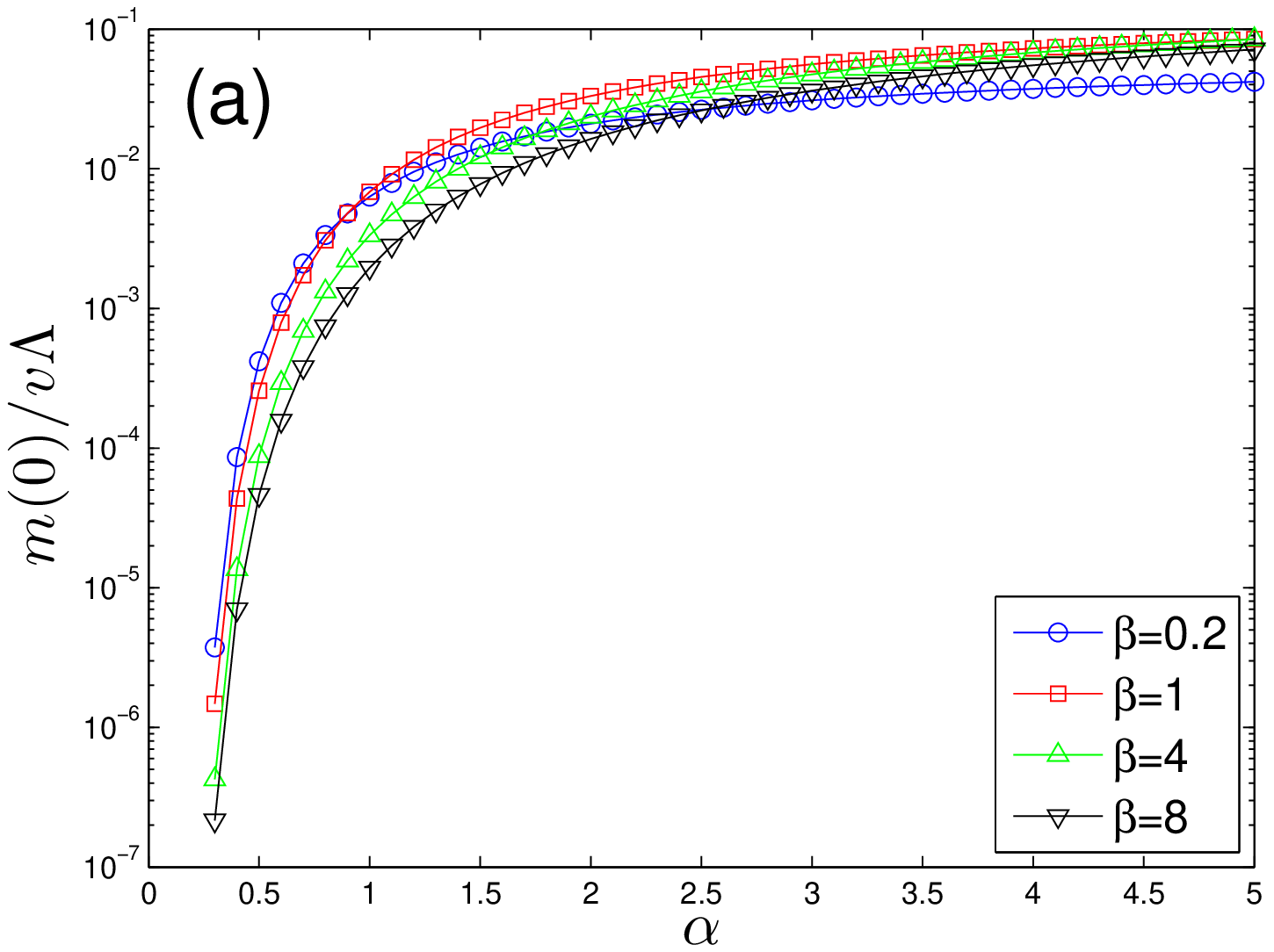}
\includegraphics[width=2.6in]{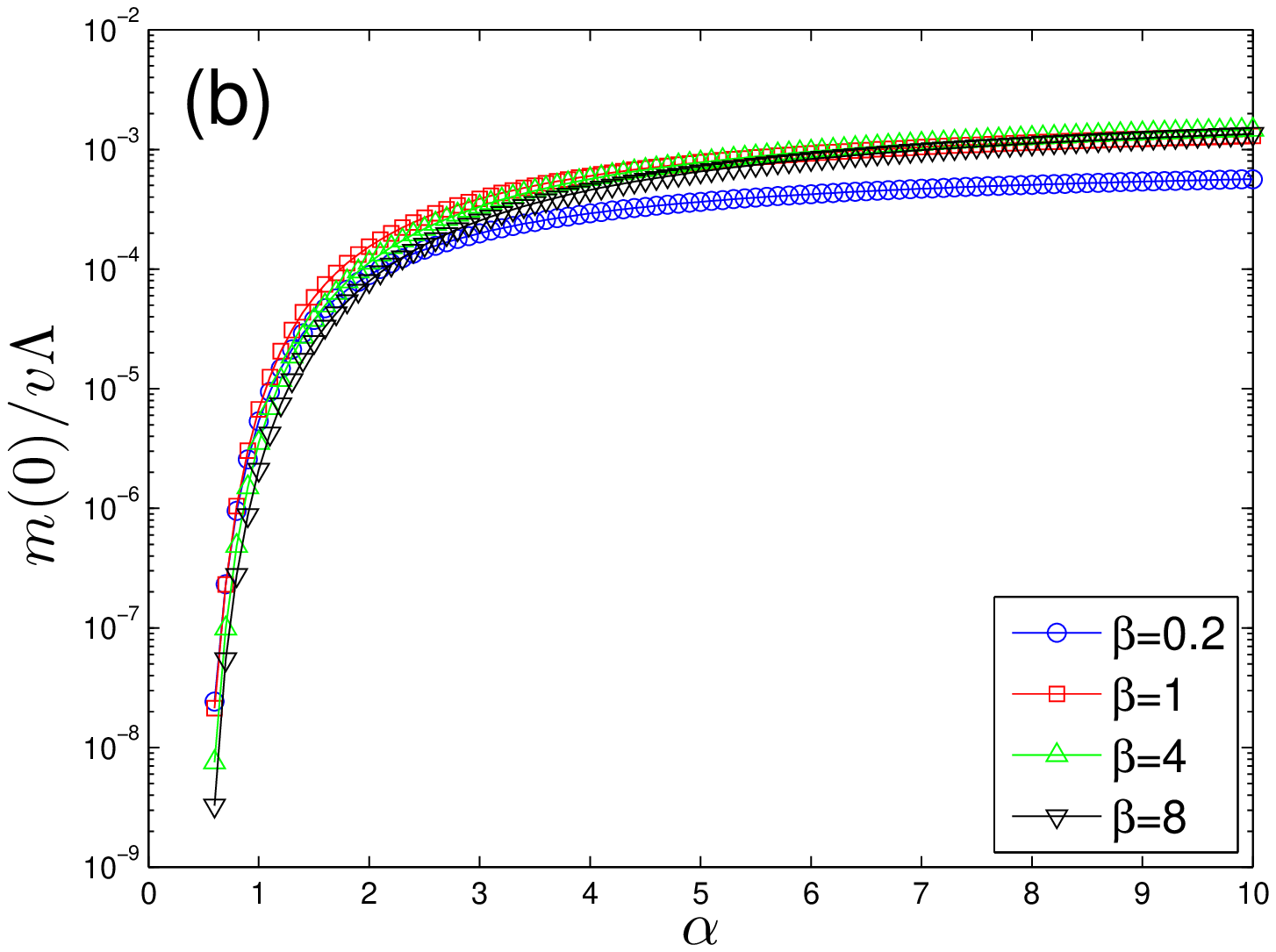}
\caption{Dependence of fermion mass $m(0)$ on $\alpha$ obtained
under instantaneous approximation at different values of $\beta$.
(a) $N=2$; (b) $N=4$.\label{Fig:Gap0Alpha}}
\end{figure}

The instantaneous approximation is widely employed to simplify the
self-consistent DS equation of dynamical fermion gap. It has been
used in such 2D Dirac semimetal as graphene \cite{Khveshchenko01,
Gorbar02, Khveshchenko04, Zhang11, Liu11, WangLiu11A, WangLiu14}, 3D
semimetal with quadratic touching points \cite{Janssen16A}, and
finite temperature QED$_{3}$ \cite{WangLiuZhang15}. A universal
feature shared by these systems is that the fermion mass gap depends
on energy and momentum separately due to the explicit breaking of
Lorentz invariance, which makes it very difficult to solve the
self-consistent gap equation numerically. In the instantaneous
approximation, the energy-dependence of fermion gap is dropped, but
the momentum-dependence is maintained. Under this approximation, the
interaction function becomes
\begin{eqnarray}
V(\Omega,\mathbf{q})&\rightarrow& V(\mathbf{q}) =
\frac{1}{V_{0}^{-1}(\mathbf{q})+\Pi(\mathbf{q})}\nonumber
\\
&=& \left[\frac{|\mathbf{q}|}{2\pi\alpha v}
+ \frac{N}{v}\frac{d_{x}a^{1/2}q_{x}^{2}}
{\left(c_{0}a^{2}q_{x}^{4} +
v^2q_{y}^{2}\right)^{1/4}}\right.\nonumber \\
&&\left.+\frac{N}{v}\frac{d_{y}a^{-1/2}v^2q_{y}^{2}}{
\left(c_{0}a^2q_{x}^{4}+v^2q_{y}^{2}\right)^{3/4}}\right]^{-1}.
\label{Eq:DressedCoulombInstan}
\end{eqnarray}
Accordingly, the gap equation (\ref{Eq:GapGeneral}) is simplified to
\begin{eqnarray}
m(p_{x},p_{y}) &=& \frac{1}{2}\int\frac{d^2\mathbf{k}}{(2\pi)^2}
\frac{m(k_{x},k_{y})}{\sqrt{a^2k_{x}^{4} +
v^2 k_{y}^2 + m^2(k_{x},k_{y})}} \nonumber \\
&&\times V(\mathbf{p}-\mathbf{k}).\label{Eq:GapEqInstanAppro}
\end{eqnarray}

\begin{figure}[htbp]
\center
\includegraphics[width=2.6in]{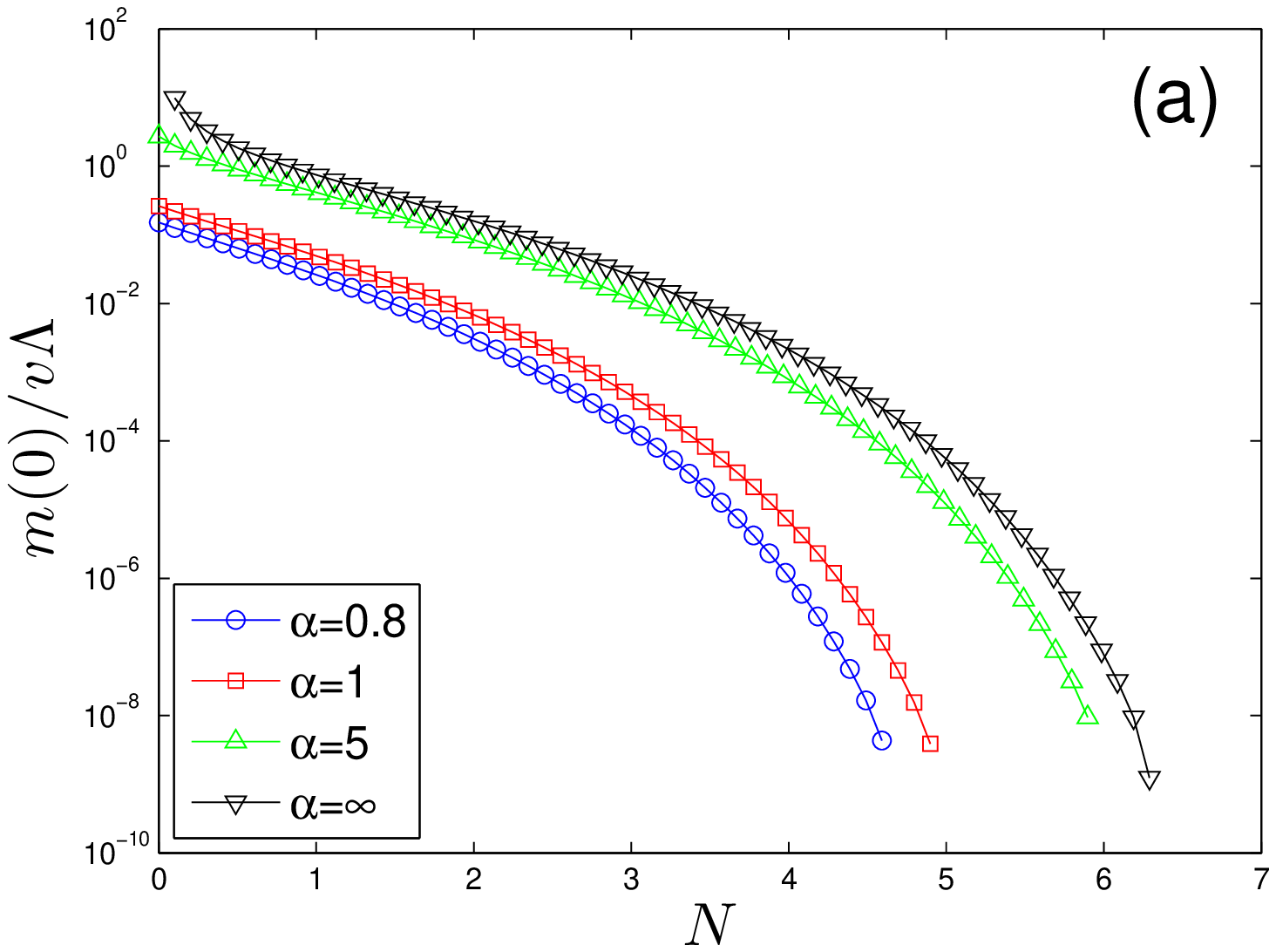}
\includegraphics[width=2.6in]{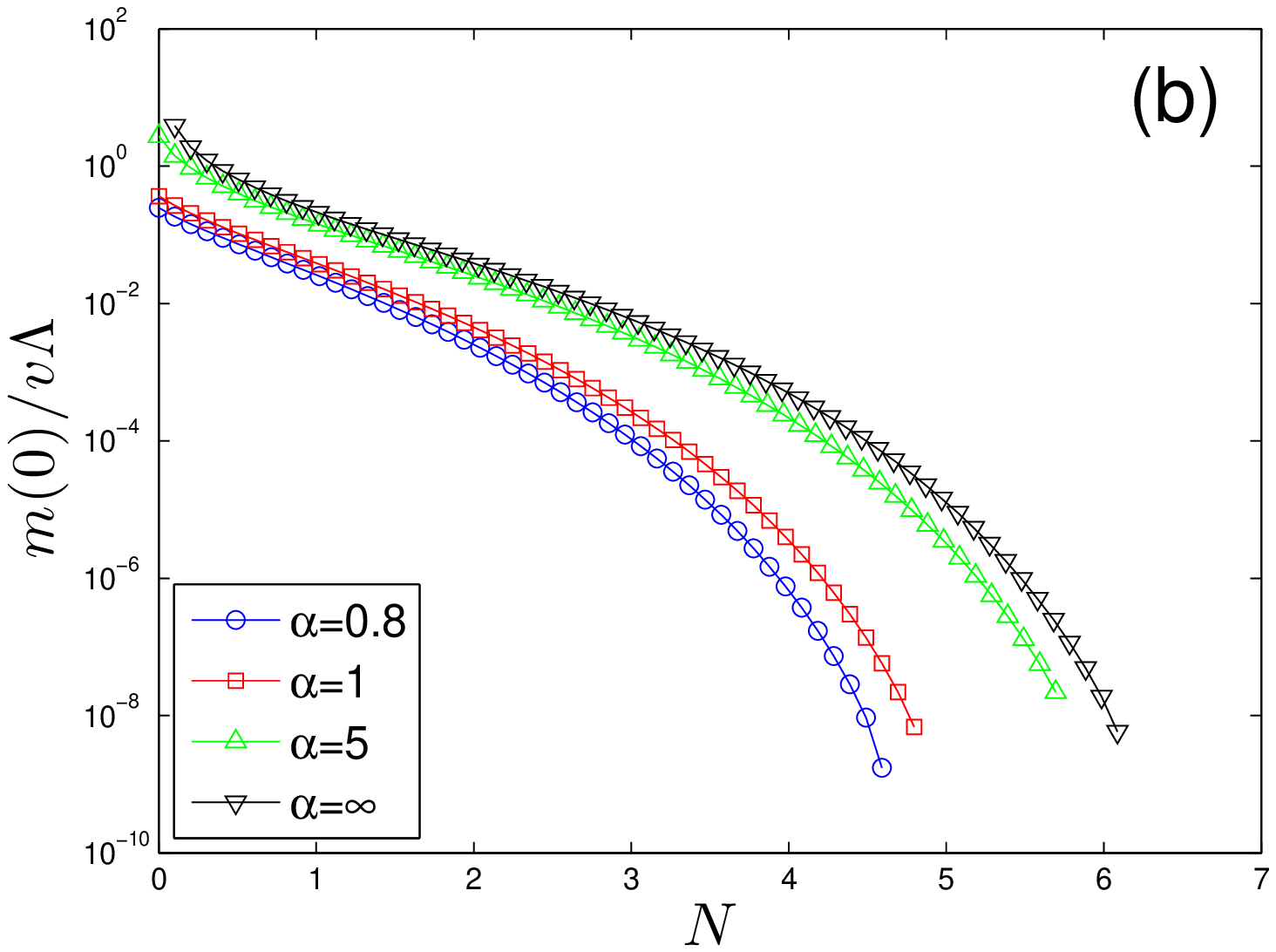}
\includegraphics[width=2.6in]{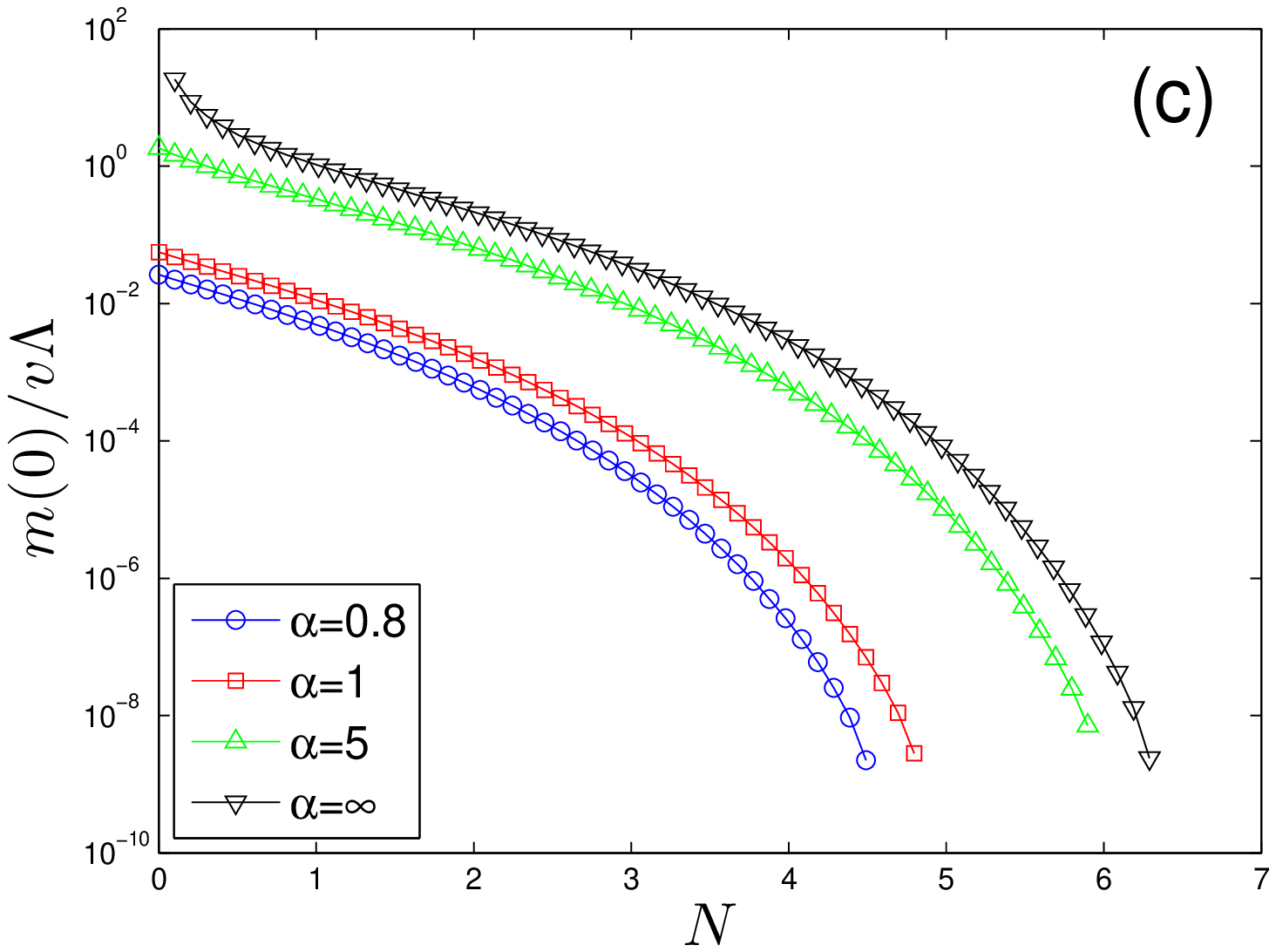}
\caption{Dependence of fermion mass $m(0)$ on $N$ obtained under
instantaneous approximation at different values of $\alpha$. (a)
$\beta=1$, (c) $\beta=0.1$, and (c) $\beta=10$. \label{Fig:Gap0N}}
\end{figure}

We show the dependence of $m(0)\equiv m(p_{x}=0, p_{y} = 0)$ on
$\beta$ obtained for $N = 2$ and $N = 4$ in
Fig.~\ref{Fig:Gap0Beta}(a) and Fig.~\ref{Fig:Gap0Beta}(b),
respectively. It is apparent that $m(0)$ exhibits a non-monotonic
dependence on $\beta$. As $\beta$ grows, $m(0)$ increases initially,
but begins to decrease once $\beta$ is greater than some critical
value. The dependence of $m(0)$ on parameter $\alpha$ in the cases
of $N = 2$ and $N=4$ is presented in Fig.~\ref{Fig:Gap0Alpha}. We
can clearly observe that $m(0)$ decreases as $\alpha$ is lowered,
and eventually vanishes once $\alpha$ is smaller than a critical
value $\alpha_{c}$. According to the results given in
Fig.~\ref{Fig:Gap0Alpha}, it turns out that the critical value
$\alpha_{c}$ is not very sensitive to the change in $\beta$. For a
series of different values of $\beta$, ranging from $0.2$ to $8$,
$\alpha_{c} \approx 0.6$ in the case of $N = 4$. In a 2D Dirac
semimetal, the critical interaction strength obtained under
instantaneous approximation is $\alpha_{c} \approx 2.33$ for $N = 4$
\cite{Khveshchenko01, Gorbar02}. An immediate indication is that the
excitonic gap can be much more easily generated in a 2D semi-Dirac
semimetal than a 2D Dirac semimetal.

\begin{figure}[htbp]
\center
\includegraphics[width=2.6in]{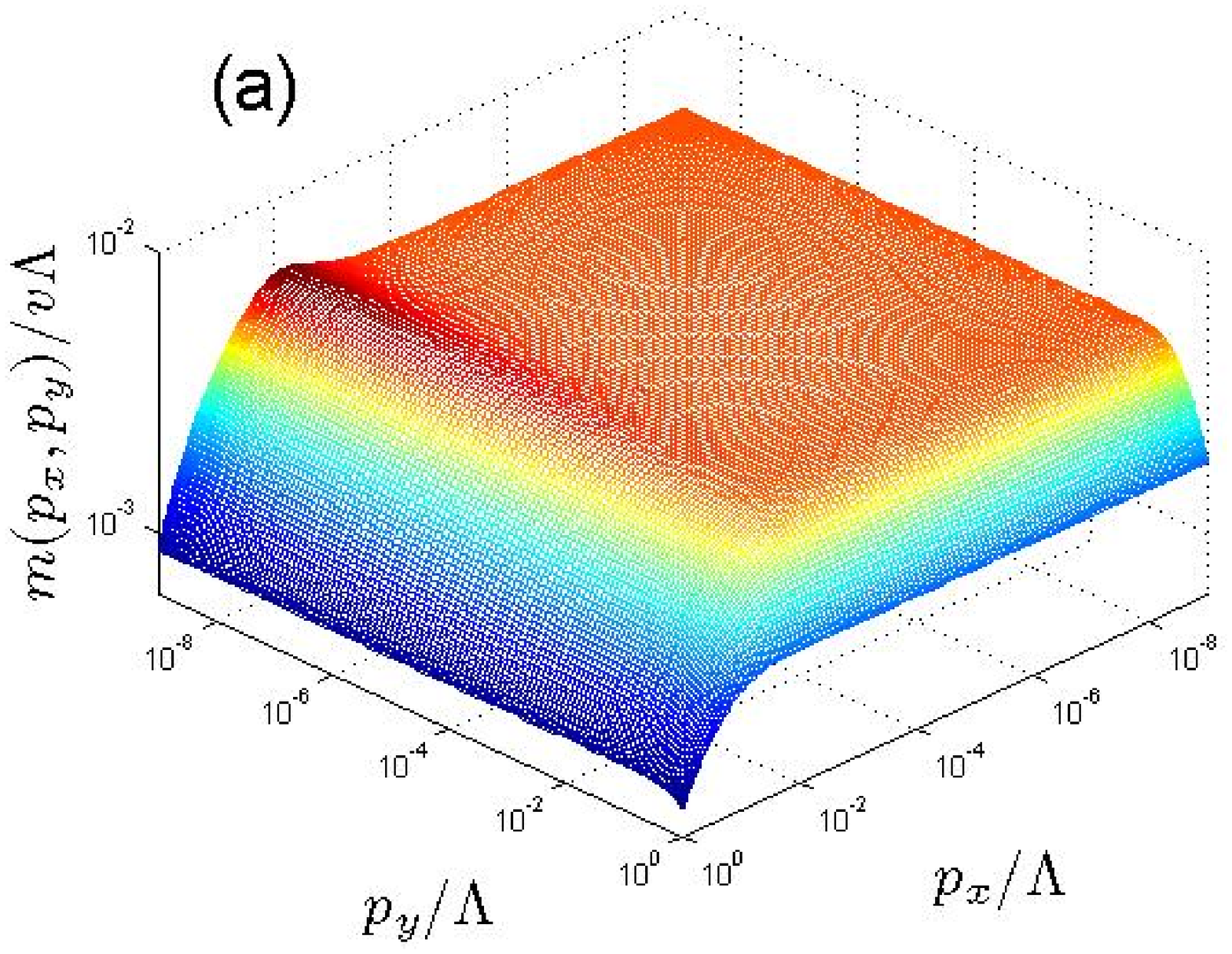}
\includegraphics[width=2.6in]{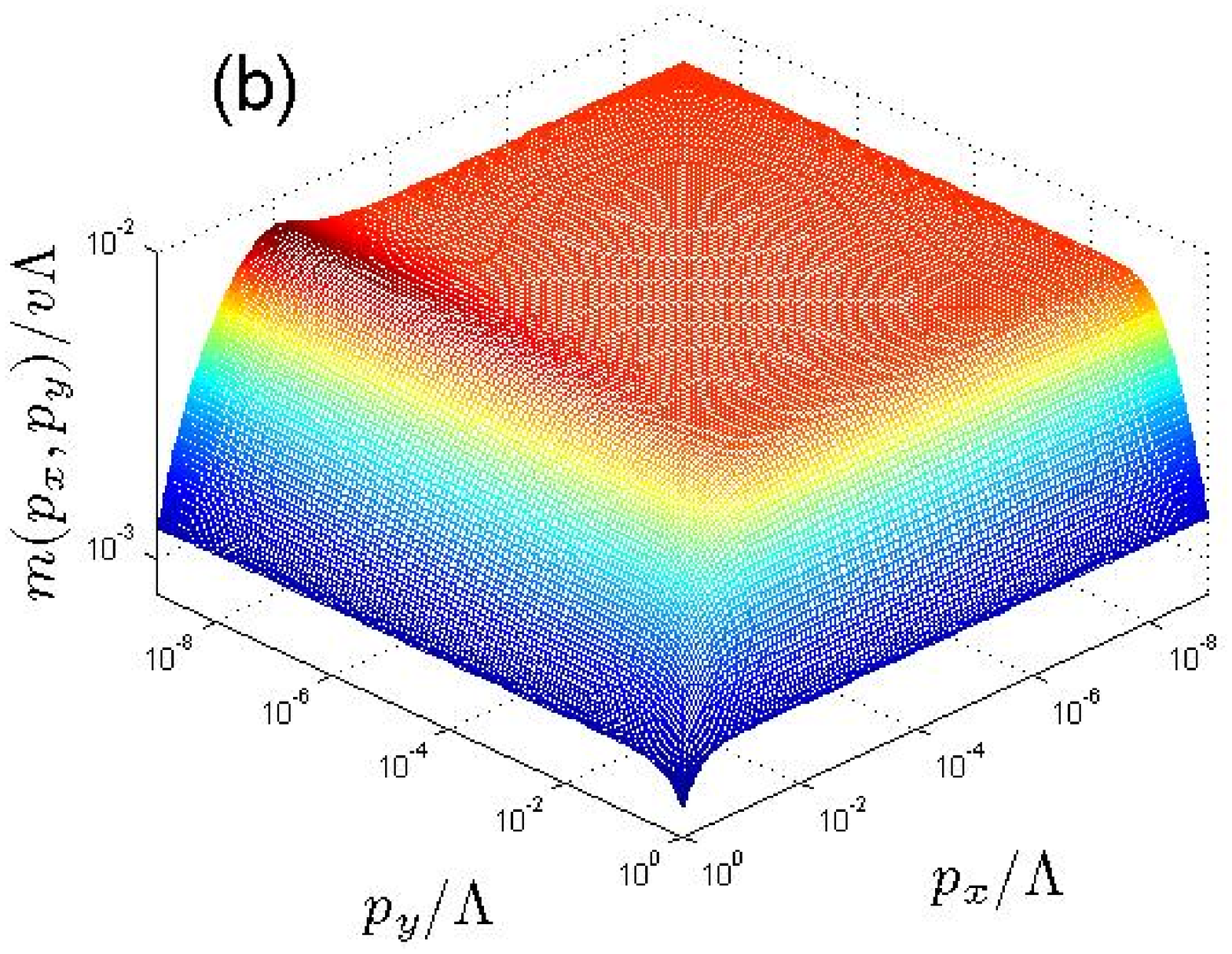}
\includegraphics[width=2.6in]{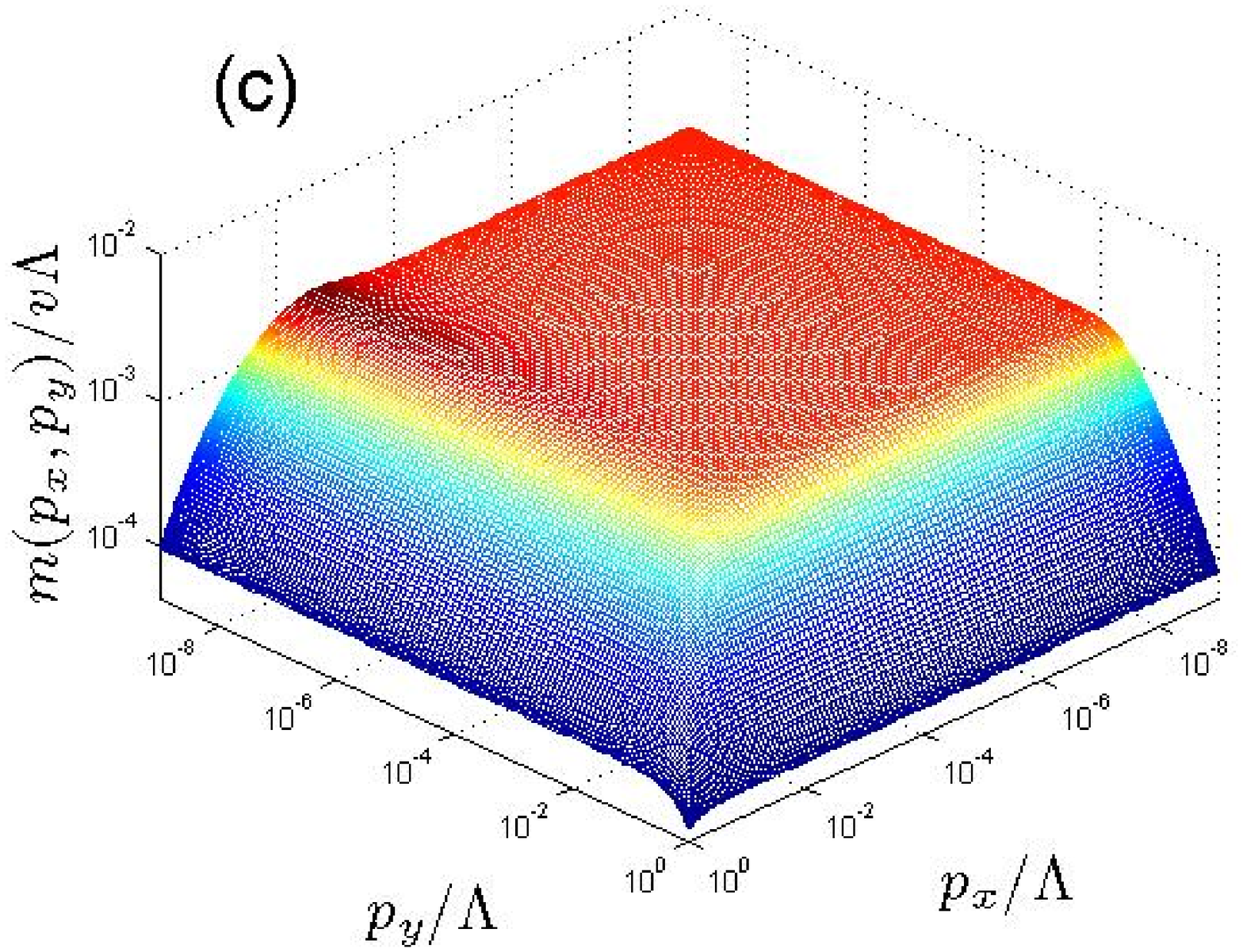}
\caption{Dependence of $m(p_{x},p_{y})$ on $p_{x}$ and $p_{y}$
obtained under instantaneous approximation at $\alpha = 1$ and $N =
2$. (a) $\beta = 0.1$; (b) $\beta = 1$; (c) $\beta = 10$.
\label{Fig:GapPInstan}}
\end{figure}

For fixed values of $\alpha$ and $\beta$, there exists a critical
fermion flavor $N_c$ that separates the semimetallic and excitonic
insulating phases. As shown in Fig.~\ref{Fig:Gap0N}, $m(0)$
decreases with growing $N$ and vanishes once $N$ exceeds $N_c$.
Specifically, $N_{c}$ takes a finite value even in the infinitely
strong coupling limit $\alpha \rightarrow \infty$. Thus, the flavor
$N$ must be sufficiently small for an excitonic gap to be opened.
Moreover, Fig.~\ref{Fig:Gap0N} tells us that, $N_{c}$ always falls
in the range of $(6,7)$ in the $\alpha \rightarrow \infty$ limit,
which is valid for three representative values of $\beta$. As a
comparison, we recall that $N_{c} = 8/\pi \approx 2.55$ in the limit
$\alpha \rightarrow \infty$ in a 2D Dirac semimetal, which was
obtained under the same approximation \cite{Khveshchenko01,
Gorbar02}. This result provides further evidence that it is much
easier to induce an excitonic gap in a 2D semi-Dirac semimetal than
a 2D Dirac semimetal.

To acquire a more quantitative knowledge of the fermion gap, we
present function $m(p_{x},p_{y})$ in Fig.~\ref{Fig:GapPInstan}. In
the limit $p_{x,y} \rightarrow 0$, $m(p_{x}, p_{y})$ approaches a
finite value $m(0,0)$. As $p_{x}$ or $p_{y}$ grows, $m(p_{x},p_{y})$
drops rapidly when $ap_{x}^{2}$ or $vp_{y}$ is larger than the
energy scale given by $m(0,0)$. It appears that $m(p_{x},p_{y})$
exhibits a weak non-monotonic dependence on $p_{x}$ in the
intermediate range of energies, but $m(p_{x},p_{y})$ decreases
monotonously with growing $p_{y}$. The strongly anisotropic behavior
of $m(p_{x},p_{y})$ clearly comes from the anisotropy in fermion
dispersion.

\subsection{Khveshchenko approximation}

The instantaneous approximation entirely neglects the
energy-dependence of the Coulomb interaction. In a 2D Dirac
semimetal, after including one-loop polarization, the dressed
Coulomb interaction becomes
\begin{eqnarray}
V_{\mathrm{Dirac}} = \frac{1}{\frac{|\mathbf{q}|}{2\pi \alpha
v}+\frac{N|\mathbf{q}|^2}{16\sqrt{\Omega^2+|\mathbf{q}|^{2}}}}.
\end{eqnarray}
Khveshchenko \cite{Khveshchenko09} proposed an improved
approximation
\begin{eqnarray}
V_{\mathrm{Dirac}} \rightarrow \frac{1}{\frac{|\mathbf{q}|}{2\pi
\alpha v}+\frac{N|\mathbf{q}|}{16\sqrt{2}}},
\end{eqnarray}
and then applied it to study dynamical gap generation in graphene
\cite{Khveshchenko09}. It was shown by Khveshchenko
\cite{Khveshchenko09} that $\alpha_{c} \approx 1.13$ for $N = 4$,
which is much smaller than $\alpha_{c}\approx 2.33$ obtained by
using the instantaneous approximation. It was also found
\cite{Khveshchenko09} that $N_{c} \approx 7.18$ in the strong
coupling limit $\alpha \rightarrow \infty$, which is much larger
than $N_{c}\approx 2.55$ obtained under instantaneous approximation.
It is therefore clear that the energy dependence of Coulomb
interaction plays an important role and needs to be seriously
incorporated in the DS equation.

\begin{figure}[htbp]
\center
\includegraphics[width=2.6in]{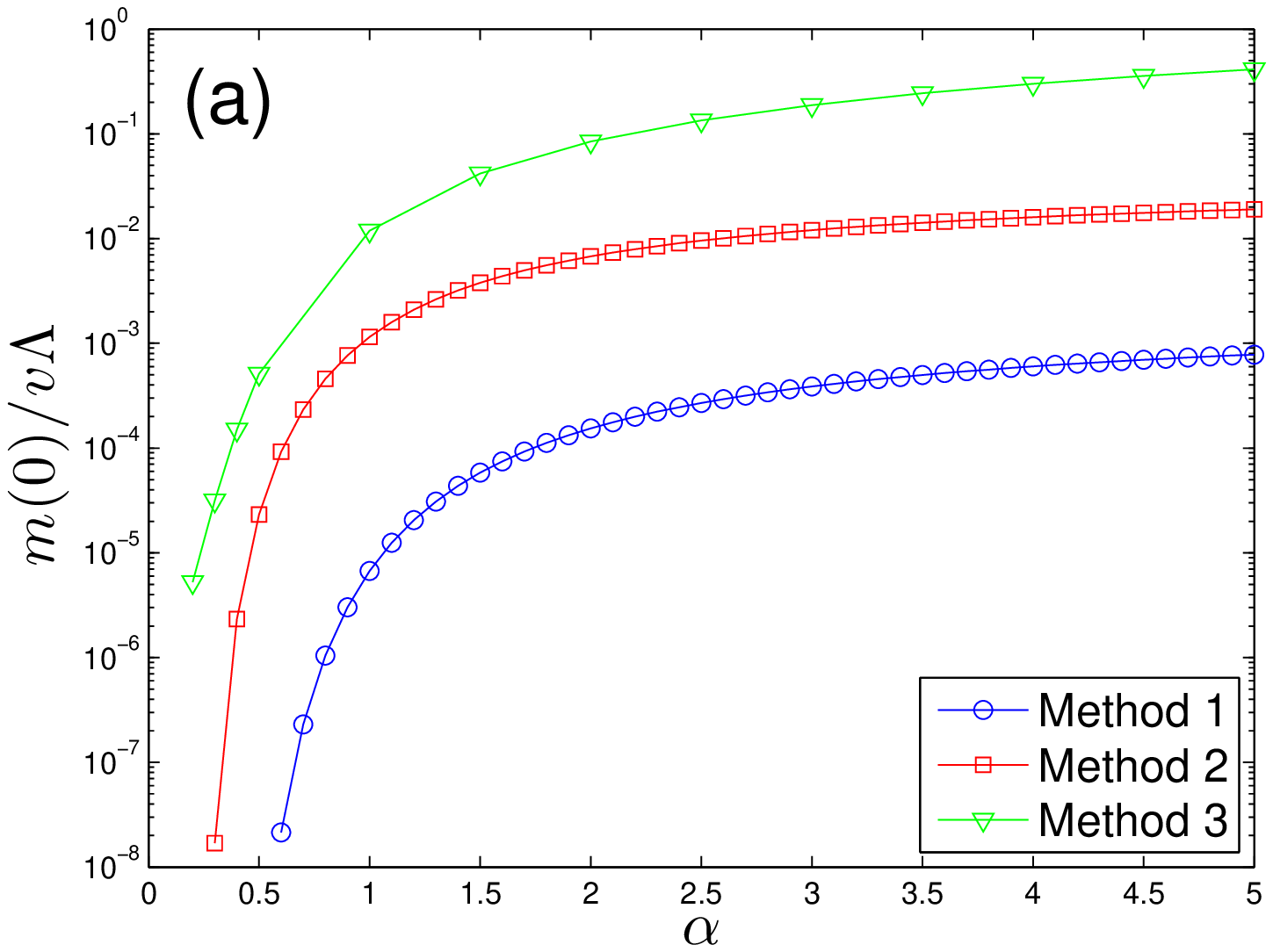}
\includegraphics[width=2.6in]{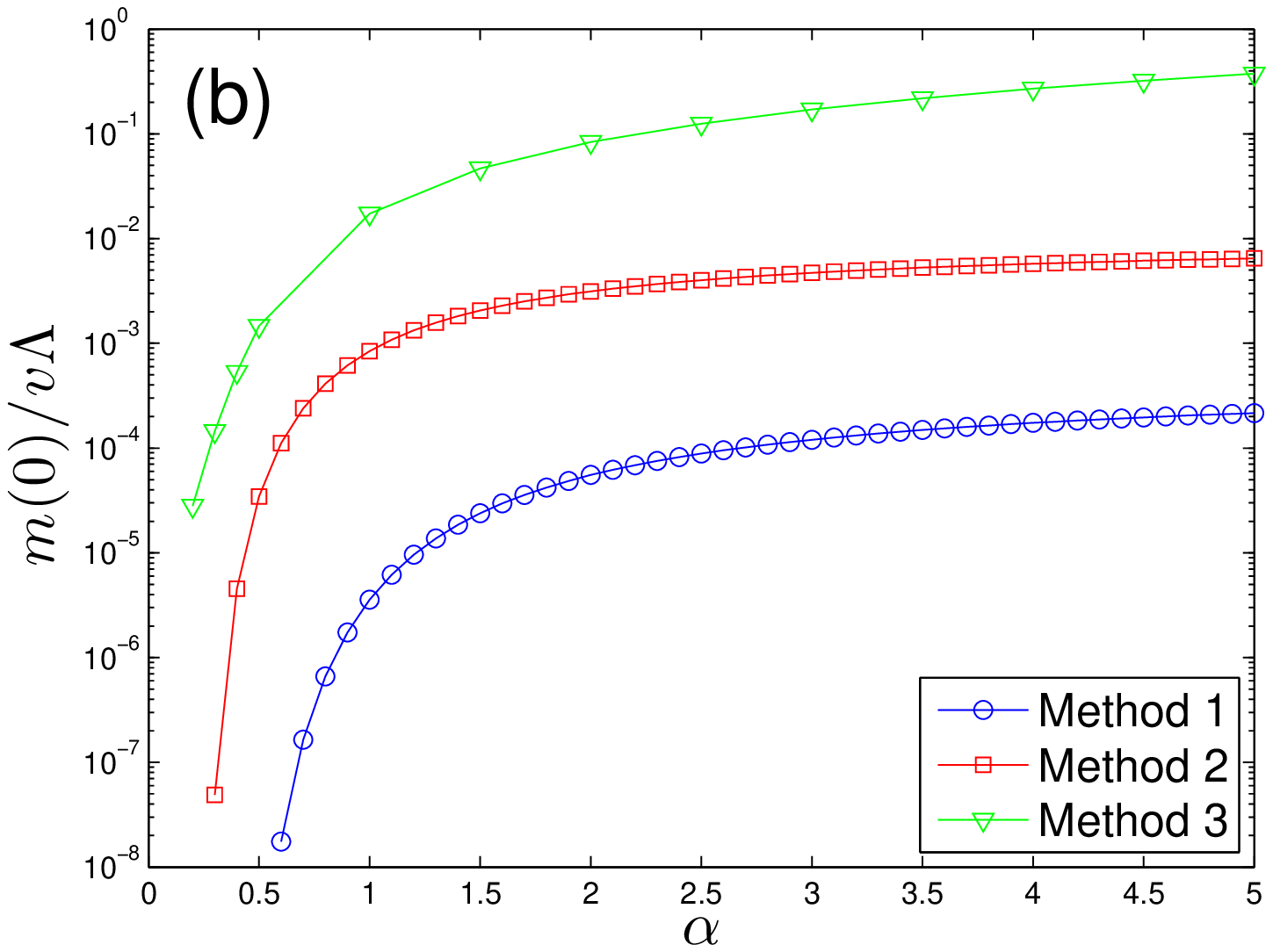}
\includegraphics[width=2.6in]{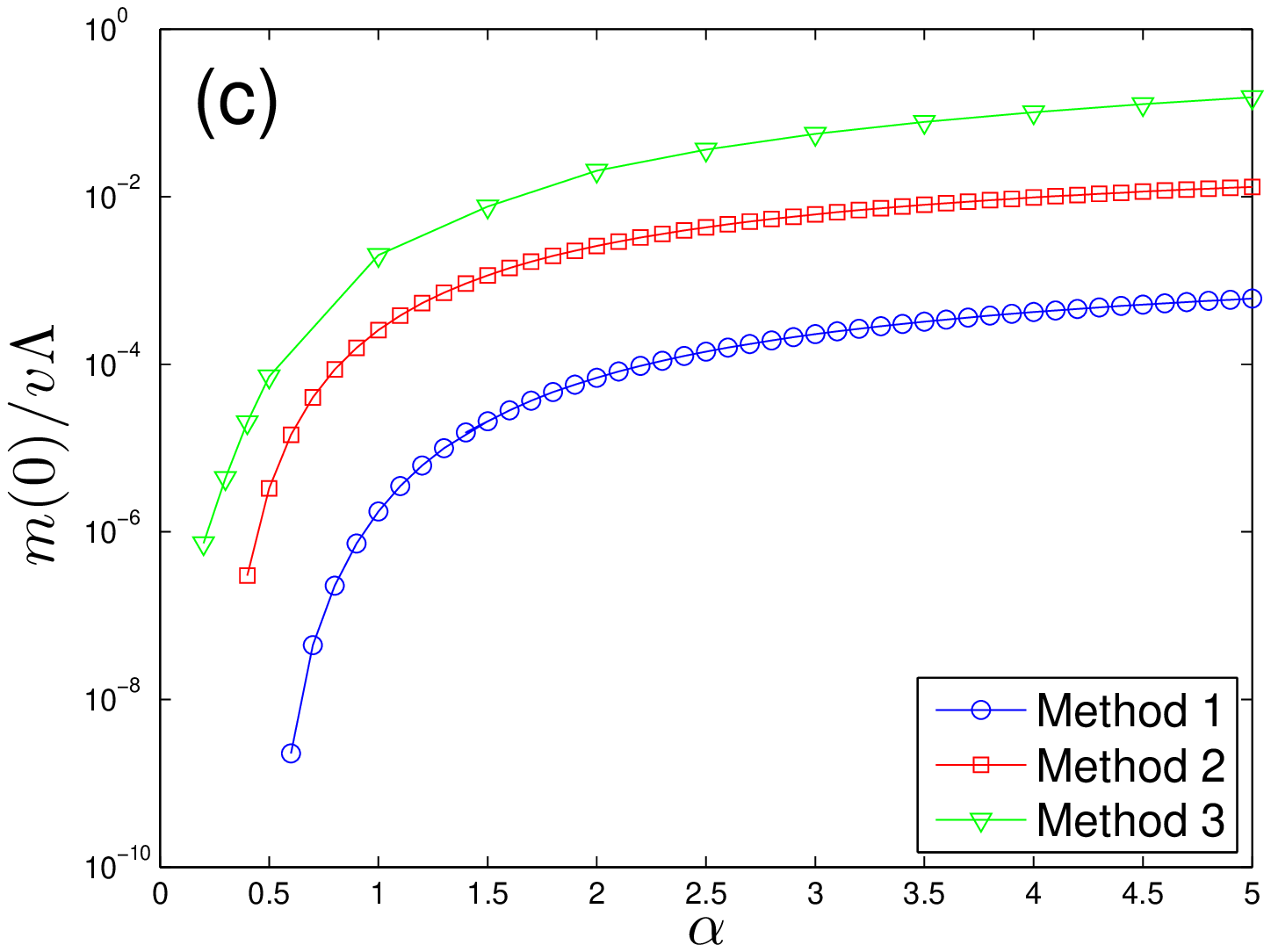}
\caption{Dependence of $m(0)$ on $\alpha$ with $\beta=1$ in (a),
$\beta=0.1$ in (b), and $\beta=10$ in (c). Method 1: Instantaneous
approximation; Method 2: Khveshchenko approximation; Method 3: GGG
approximation. This convention is also used in
Fig.~\ref{Fig:Gap0NEnergy}. Here, $N=4$.
\label{Fig:Gap0AlphaEnergy}}
\end{figure}

We now adopt the Khveshchenko approximation to study the DS gap
equation in 2D semi-Dirac semimetal. The dressed Coulomb interaction
function given by Eq.~(\ref{Eq:DressedCoulomb}) can be approximated
as
\begin{eqnarray}
V(\mathbf{q}) &\rightarrow& \left[\frac{|\mathbf{q}|}{2\pi\alpha v}
+ \frac{N}{v}\frac{d_{x}a^{1/2}q_{x}^{2}}
{\left(2\left(c_{0}a^{2}q_{x}^{4} +
v^2q_{y}^{2}\right)\right)^{1/4}}\right.\nonumber \\
&&\left.+\frac{N}{v}\frac{d_{y}a^{-1/2}v^2q_{y}^{2}}{\left(2
\left(c_{0}a^2q_{x}^{4}+v^2q_{y}^{2}\right)\right)^{3/4}}\right]^{-1}.
\label{Eq:DressedCoulombKhv}
\end{eqnarray}
Under the Khveshchenko approximation, the DS gap equation has the
same form as Eq.~(\ref{Eq:GapEqInstanAppro}) with $V(\mathbf{p} -
\mathbf{k})$ being given by Eq.~(\ref{Eq:DressedCoulombKhv}). We
then solve this DS equation numerically. At $N = 4$, the dependence
of $m(0,0)$ on $\alpha$ is shown in Fig.~\ref{Fig:Gap0AlphaEnergy}
for several values of $\beta$ in (a), (b), and (c) respectively,
represented by the red line with square mark. We can see that
$m(0,0)$ obtained by using Eq.~(\ref{Eq:DressedCoulombKhv}) is
larger than that obtained by using the instantaneous approximation,
which shows that including the energy dependence of Coulomb
interaction tends to favor the generation of excitonic gap in 2D
semi-Dirac semimetal. According to Fig.~\ref{Fig:Gap0AlphaEnergy},
within a wide range of values $\beta=1, 0.1, 10$, we find that
$\alpha_{c} \approx 0.3 \sim 0.4$, which is much smaller than
$\alpha_{c} \approx 1.13$ obtained in 2D Dirac semimetal under the
same approximation \cite{Khveshchenko09}. We thus see once again
that it is easier for Coulomb interaction to open an excitonic gap
in a 2D semi-Dirac semimetal. The relation between $m(0)$ and $N$ is
presented in Fig.~\ref{Fig:Gap0NEnergy}(a) and (b) with $\alpha =
0.1$ and $\alpha = \infty$ respectively. By using the Khveshchenko
approximation, it is found that $m(0)$ vanishes once $N$ is greater
than some critical value, even in the strong coupling limit $\alpha
\rightarrow \infty$. As shown in Fig.~\ref{Fig:Gap0NEnergy}(a),
$N_{c}\approx 9$ with $\beta = 1$ in the limit $\alpha \rightarrow
\infty$ under the Khveshchenko approximation, which is larger than
$N_{c} \approx 7.18$ obtained in the limit $\alpha \rightarrow
\infty$ under the same approximation in a 2D Dirac semimetal
\cite{Khveshchenko09}.

\subsection{Gamayun-Gorbar-Gusynin approximation}

In order to include the influence of energy dependence of Coulomb
interaction on dynamical gap generation in 2D Dirac semimetal,
Gamayun \emph{et al.} \cite{Gamayun10} introduced another
approximation, which assumes the mass gap to be energy independent,
namely
\begin{eqnarray}
m(\varepsilon,\mathbf{p})\rightarrow m(\mathbf{p}),
\end{eqnarray}
but retains the energy dependence of Coulomb interaction. Under the
GGG approximation, they solved the DS equation and found that
$\alpha_{c} \approx0.92$ for flavor $N = 4$, which is clearly
smaller than $\alpha_{c} \approx 2.33$ obtained under the
instantaneous approximation \cite{Khveshchenko01, Gorbar02}. This
result provides another signature that the energy dependence of
Coulomb interaction is in favor of dynamical excitonic gap
generation. They also found that, in the strong coupling limit
$\alpha \rightarrow \infty$, the critical fermion flavor $N_{c}
\rightarrow \infty$, which is quite different from the results
obtained under the instantaneous and Khveshchenko approximations.

In this subsection, we employ the GGG approximation to study the
dynamical excitonic gap in 2D semi-Dirac semimetal. By applying this
approximation, the DS gap equation (\ref{Eq:GapGeneral}) becomes
\begin{eqnarray}
m(p_{x},p_{y}) &=& \int\frac{d\omega}{2\pi}
\frac{d^2\mathbf{k}}{(2\pi)^{2}}m(k_{x},k_{y})\nonumber \\
&&\times\frac{1}{\omega^2 + a^2k_{x}^{4} +
v^2k_{y}^2+m^2(k_{x},k_{y})}\nonumber \\
&&\times V(\omega,\mathbf{p}-\mathbf{k}). \label{Eq:GapGorbarAppro}
\end{eqnarray}
In principle, the integration range of energy should be
$\omega\in(-\infty,\infty)$. In practical numerical computations, it
is necessary to introduce a cutoff. We choose to integrate over
energy within the range $(-\Lambda_{\omega}, \Lambda_{\omega})$,
where $\Lambda_{\omega}$ is taken to be sufficiently large so that
the magnitude of dynamical gap is nearly independent of varying
$\Lambda_{\omega}$. The relation between $m(0)$ and $\alpha$ with
$N=4$ is shown in Fig.~\ref{Fig:Gap0AlphaEnergy} by the green line
with the triangular mark. In Fig.~\ref{Fig:Gap0AlphaEnergy}(a), (b),
and (c), $\beta$ equal to $1$, $0.1$, and $10$ respectively. As
shown in Fig.~(\ref{Fig:Gap0AlphaEnergy}), $m(0)$ calculated through
Eq.~(\ref{Eq:GapGorbarAppro}) is clearly larger than the one
obtained under instantaneous and Khveshchenko approximations, which
indicates that energy dependence of Coulomb interaction enhances
dynamical gap generation. For $\beta=1, 0.1, 10$, we find that
$\alpha_{c} < 0.2$, which is also much smaller than $\alpha_{c}
\approx 0.92$ obtained in a 2D Dirac semimetal \cite{Gamayun10}. It
is also clear that an excitonic gap can be more easily opened by
Coulomb interaction in a 2D semi-Dirac semimetal.

\begin{figure}[htbp]
\center
\includegraphics[width=2.6in]{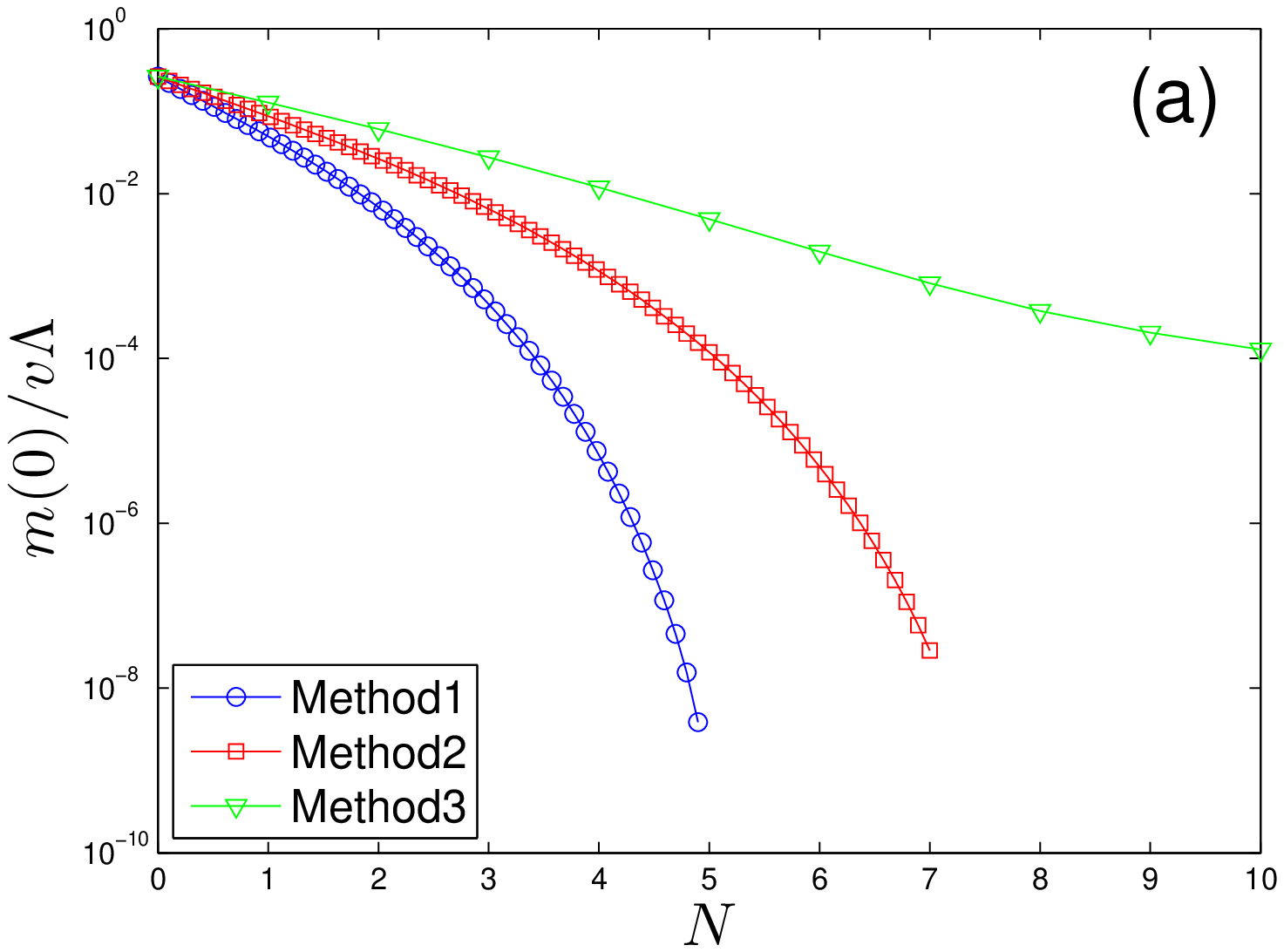}
\includegraphics[width=2.6in]{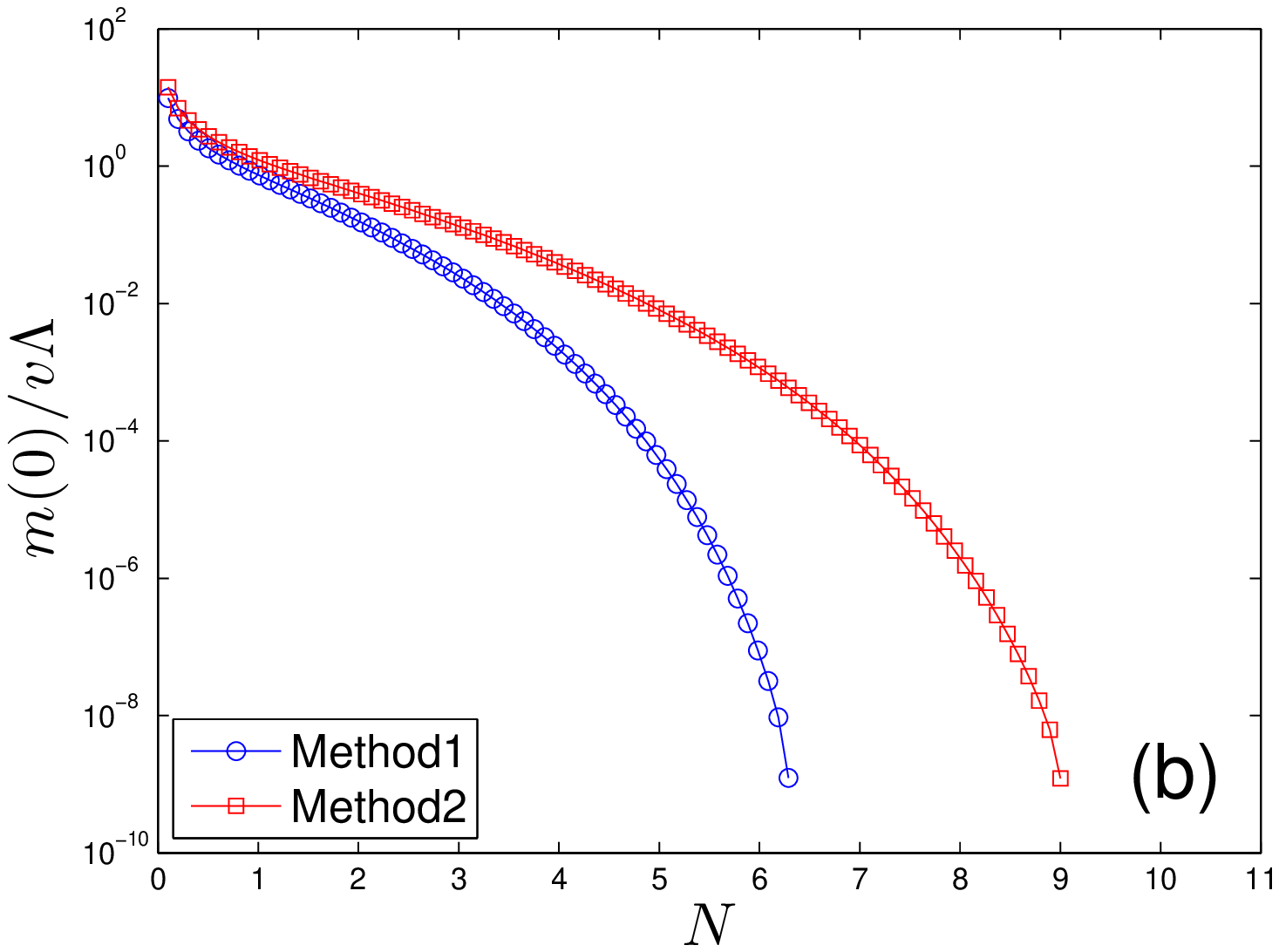}
\caption{Dependence of $m(0)$ as a function of flavor $N$. (a)
$\alpha=1$; (b) $\alpha = \infty$. Here, $\beta=1$.
\label{Fig:Gap0NEnergy}}
\end{figure}

The dependence of fermion gap on $N$ obtained at $\alpha = 1$ is
shown in Fig.~\ref{Fig:Gap0NEnergy}(a), represented by the green
line with triangular mark. The critical flavor $N_{c}$ is much
larger than that obtained under instantaneous and Khveshchenko
approximations. In the infinitely strong coupling limit $\alpha
\rightarrow \infty$, we find that $N_{c}$ goes to infinity. This
stems from an infrared divergence that is owing to the singular
contribution appearing in the regions of $aq_{x}^{2} \ll \Omega$ and
$vq_{y} \ll \Omega$. These results point towards the fact that the
energy dependence of Coulomb interaction is in favor of dynamical
gap generation.

As shown in Eq.~(\ref{Eq:DressedCoulomb}), the first term of the
denominator of dressed Coulomb interaction is the contribution from
bare Coulomb interaction, and depends only on momentum
$|\mathbf{q}|$. The second term arises from the dynamical screening
due to collective particle-hole excitations, and is a
function of energy $\Omega$ and momenta $q_{x,y}$. According to
Eq.~(\ref{Eq:PolarizationApproExpression}), it is easy to find that
$\Pi(\Omega,\mathbf{q})$ is smaller than $\Pi(0,\mathbf{q})$.
Therefore, the dressed Coulomb interaction becomes stronger after
including the energy dependence, which tends to promote dynamical
gap generation. Since $\Pi(\Omega,\mathbf{q})$ is proportional to
$N$, the difference between the fermion gaps obtained with and
without the energy dependence of Coulomb interaction is more
significant at larger $N$. This property can be seen from the
numerical results depicted in Fig.~\ref{Fig:Gap0NEnergy}(a).
Therefore, it is usually more important to incorporate the energy
dependence of Coulomb interaction for larger values of $N$.

It seems necessary to give a short summary here. We have numerically
solved the DS equation under three widely used approximations.
Although the precise value of $\alpha_c$ is approximation dependent,
we can infer from the extensive numerical solutions that $\alpha_c$
obtained in a 2D semi-Dirac semimetal is much smaller than that of
2D Dirac semimetal. We thus conclude that a moderately strong
Coulomb interaction suffices to open a finite excitonic gap in a 2D
semi-Dirac semimetal.

As mentioned in Sec.~\ref{Sec:Introduction}, there are indeed
several possible ways to obtain a realistic 2D semi-Dirac semimetal.
For instance, it can be created by merging two Dirac fermions at the
QCP between a 2D Dirac semimetal and a trivial band insulator.
Massless 2D semi-Dirac fermions also naturally emerge in a black
phosphorus placed in a perpendicular electric field. However, in
these cases the semi-Dirac semimetal is realized either by making a
delicate tuning of some special parameter or by introducing an
external field. It is therefore technically difficult to prepare a
robust and intrinsic semi-Dirac semimetal in these materials. In
contrast, an intrinsic 2D semi-Dirac semimetal can be more readily
achieved in a TiO$_{2}$/VO$_{2}$ nanostructure, which was proposed
by Pardo and his collaborators \cite{Pardo09, Pardo10, Banerjee09}.

We now estimate the actual value of the interaction parameter
$\alpha$. In VO$_{2}$, the fermion velocity is roughly $v_{F}
\approx1.5 \times 10^{5}\mathrm{m/s}$ \cite{Pardo09}, and the
fermion flavor is $N = 4$ \cite{Banerjee09}. The value of $\alpha$
relies crucially on the dielectric constant $\epsilon$. In VO$_{2}$,
the constant $\epsilon \approx 36$ at room temperature, and
increases with growing temperature \cite{Yang10}. It was predicted
\cite{Pardo09} that $\alpha \approx 0.41$ in VO$_{2}$ at room
temperature. At lower temperatures, the value of $\epsilon$ can be
made even smaller than $36$, which then gives rise to a value of
$\alpha$ that is larger than $\alpha \approx 0.41$. The value of
$\epsilon$ can also be further tuned by other non-thermal scenarios,
such as substrate. Our DS equation calculations have showed that the
critical parameter $\alpha_c \approx 0.6$, $\alpha_c \approx 0.3
\sim 0.4$, and $\alpha_c < 0.2$ under the instantaneous,
Khveshchenko, and GGG approximations, respectively. We can see that
the actual value of $\alpha$ at room temperature is already smaller
or close to the critical value $\alpha_c$. At nearly zero
temperatures, the physical value of $\alpha$ might become much
larger than $0.41$. We thus predict that the TiO$_{2}$/VO$_{2}$
nanostructure is an ideal candidate to realize excitonic insulator,
which can be probed by ARPES \cite{Hashimoto14} or other experiments
\cite{Elias11, Taillefer10}.

\begin{figure}[htbp]
\center
\includegraphics[width=2.6in]{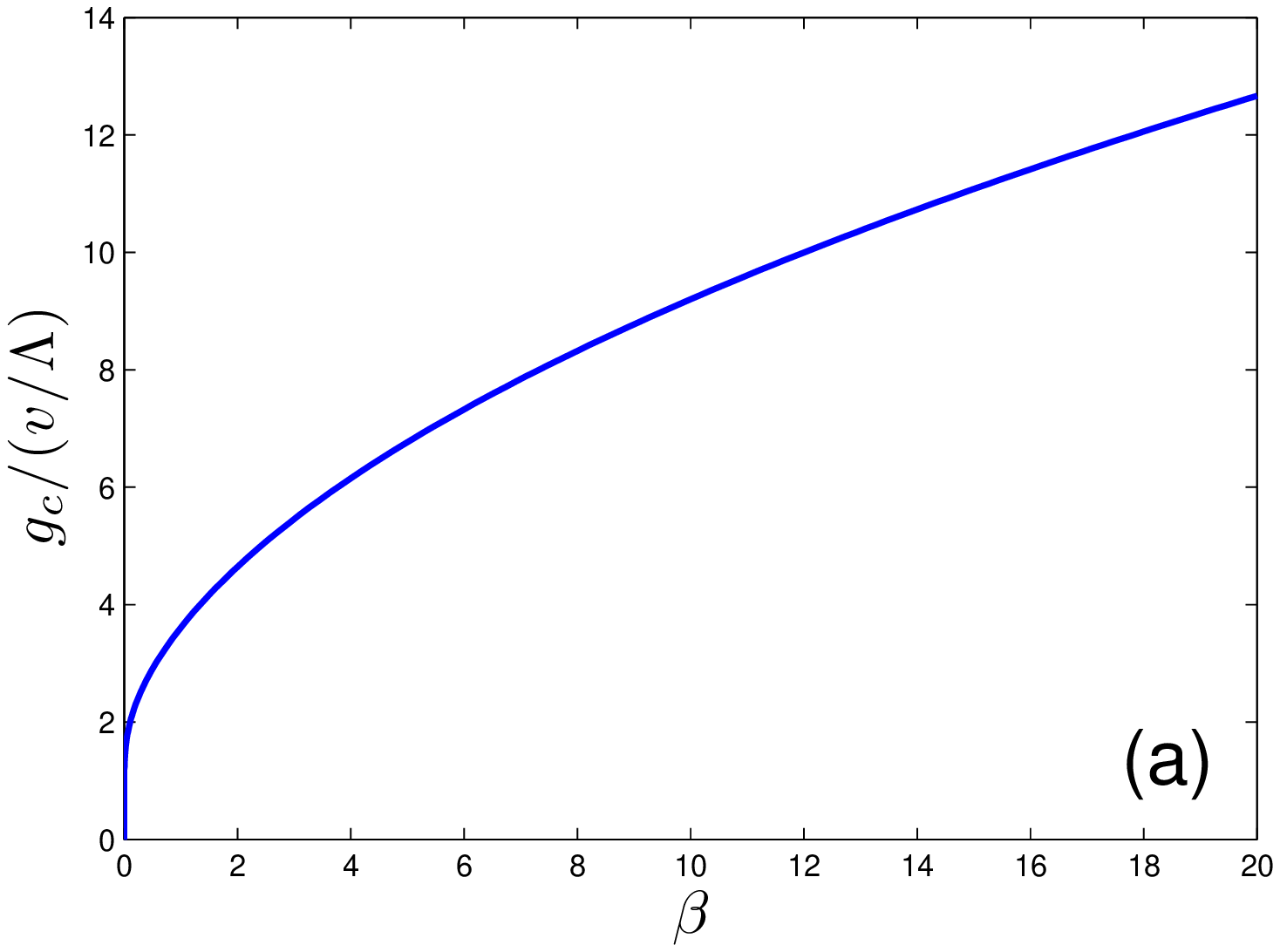}
\includegraphics[width=2.6in]{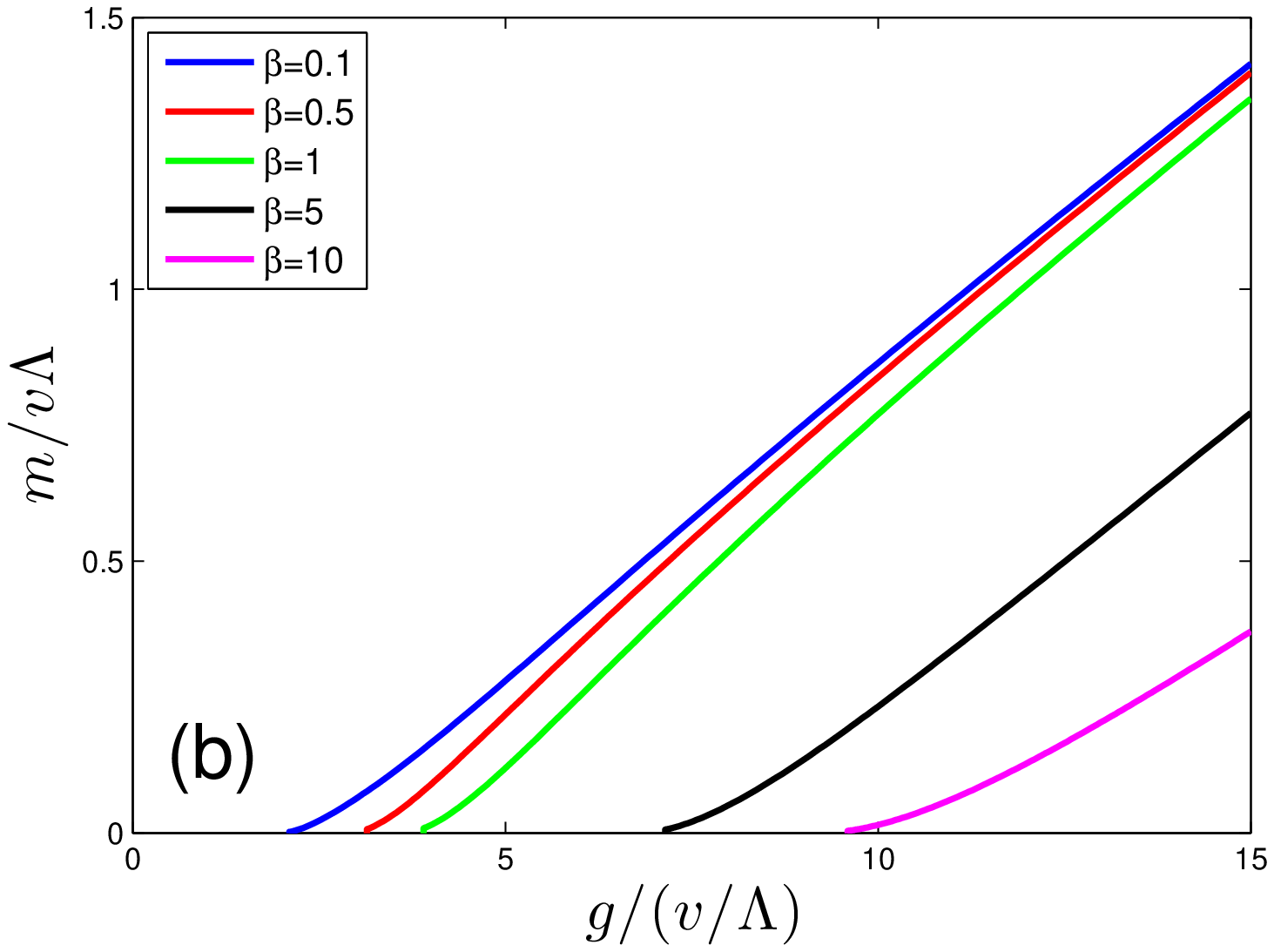}
\caption{(a) Dependence of $g_{c}$ on $\beta$. (b) Dependence of
$m(0)$ on $g$ at different values of $\beta$. Only the four-fermion
interaction is considered in this case.
\label{Fig:gcGapOnlyFourfrmion}}
\end{figure}

\section{Additional short-range four-fermion coupling\label{Sec:FourFermion}}

Besides the long-range Coulomb interaction, there may be additional
short-range four-fermion couplings \cite{Kotov12, Liu09, Gamayun10}.
In this section, we study the impact of such short-range couplings
on dynamical excitonic gap generation. In principle, there are a
number of possible four-fermion couplings. Their roles can be
classified by the symmetry and also the (ir)relevance of these
coupling terms. Here, we shall not consider all of the possible
coupling terms, but focus on the most simple one:
\begin{eqnarray}
H_{\mathrm{FF}} = \frac{g}{N}\sum_{\sigma=1}^{N}\int d^{2}\mathbf{x}
\left(\psi_{\sigma}^{\dag}(\mathbf{x})\tau_{3}
\psi_{\sigma}(\mathbf{x})\right)^{2}, \label{Eq:HamiltonianFF}
\end{eqnarray}
where $g$ is the quartic coupling constant. A simple power counting
shows that the four-fermion interaction is irrelevant in the
low-energy regime, which results from the vanishing of fermion DOS
at Fermi level. After performing RG calculations at the one-loop
level, we find that the four-fermion coupling is irrelevant if its
initial value is small. However, it becomes relevant when its
initial value is sufficiently large, which is usually interpreted as
the generation of an excitonic order parameter $\langle
\psi^{\dag}\tau_{3}\psi \rangle \neq 0$. Comparing to
Eq.~(\ref{Eq:HamiltonianMassTerm}), we see that the four-fermion
coupling shown in Eq.~(\ref{Eq:HamiltonianFF}) generates the same
order parameter as the one induced by Coulomb interaction. This is
the reason why we consider the additional four-fermion coupling
given by Eq.~(\ref{Eq:HamiltonianFF}). This strategy was previously
utilized in the studies of excitonic gap generation in graphene
\cite{Gamayun10}.

\begin{figure}[htbp]
\center
\includegraphics[width=2.6in]{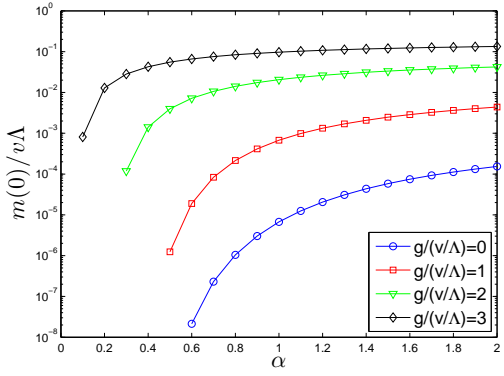}
\caption{Dependence of $m(0)$ on $\alpha$ at different values of
$g$. Both the Coulomb and four-fermion interactions are present. It
is clear that larger $g$ leads to larger gap. \label{Fig:GapBoth}}
\end{figure}

If we ignore the Coulomb interaction and retain the above
four-fermion coupling solely, the gap equation takes the form
\begin{eqnarray}
\frac{1}{2g}=\int\frac{d\omega}{(2\pi)}\frac{d^2\mathbf{k}}{(2\pi)^{2}}
\frac{1}{\omega^{2}+a^2k_{x}^{4}+v^2k_{y}^{2}+m^2},
\end{eqnarray}
where $m$ is supposed to be a constant. Integrating over $\omega$,
the gap equation can be further written as
\begin{eqnarray}
\frac{1}{g} = \int\frac{d^2\mathbf{k}}{(2\pi)^{2}}
\frac{1}{\sqrt{a^2k_{x}^{4}+v^2k_{y}^{2}+m^2}}.
\end{eqnarray}
By setting $m = 0$, we find a critical strength $g_{c}$ that is
determined by the equation
\begin{eqnarray}
\frac{1}{g_{c}}=\int\frac{d^2\mathbf{k}}{(2\pi)^{2}}
\frac{1}{\sqrt{a^2k_{x}^{4}+v^2k_{y}^{2}}}.
\end{eqnarray}
The fermions remain massless if $g < g_c$, but become massive once
$g$ exceeds $g_c$. The critical value $g_{c}$ is a function of the
parameter $\beta$, and the dependence of $g_{c}$ on $\beta$ is shown
in Fig.~\ref{Fig:gcGapOnlyFourfrmion}(a). It is clear that $g_{c}$
is an increasing function of $\beta$, with $g_{c} \rightarrow 0$
when $\beta \rightarrow 0$. As shown in
Fig.~\ref{Fig:gcGapOnlyFourfrmion}(b), the fermion mass $m$
increases as $g$ grows from a critical value $g_{c}$.

We then incorporate both long-range Coulomb interaction and
short-range four-fermion interaction, and obtain the following gap
equation:
\begin{eqnarray}
m(p_{x},p_{y})
&=&\frac{1}{2}\int\frac{d^2\mathbf{k}}{(2\pi)^2}\frac{m(k_{x},k_{y})}
{\sqrt{a^2k_{x}^{4}+v^2k_{y}^2+m^2(k_{x},k_{y})}}\nonumber
\\
&&\times V(\mathbf{p}-\mathbf{k})\nonumber
\\
&+&g\int\frac{d^2\mathbf{k}}{(2\pi)^{2}}
\frac{m\left(k_{x},k_{y}\right)}{\sqrt{a^2k_{x}^{4} +
v^2k_{y}^{2}+m^2\left(k_{x},k_{y}\right)}},\nonumber
\end{eqnarray}
where the instantaneous approximation has been adopted to simplify
numerical calculations. The momentum cutoff is chosen in the same
way as described below Eq.~(\ref{Eq:GapGeneral}). The dependence gap
including both of long-range Coulomb interaction and four-fermion
interaction is shown in Fig.~\ref{Fig:GapBoth}. We observe that, in
the presence of additional four-fermion coupling, the magnitude of
excitonic gap is enhanced and the critical value $\alpha_{c}$ is
lowered. If the system contains only four-fermion coupling, there is
a critical value $g_{c}(\beta)$. When both Coulomb and four-fermion
interactions are present, $\alpha_{c}$ takes a finite value smaller
than that obtained in the absence of four-fermion coupling, provided
that $g < g_{c}(\beta)$. In the special case of $g > g_{c}(\beta)$,
we find that $\alpha_{c} = 0$, so an arbitrary weak Coulomb
interaction makes an important contribution to excitonic pairing.

\begin{table}[htbp]
\begin{center}
\begin{tabular}{|c|c|c|c|c|}
\hline    $\frac{m(0)}{v\Lambda}$        & $\frac{g}{(v/\Lambda)}=0$
& $\frac{g}{(v/\Lambda)}=1$ & $\frac{g}{(v/\Lambda)}=2$ &
$\frac{g}{(v/\Lambda)}=3$
\\
\hline $\alpha=1$           & $1.19\times10^{-2}$   &
$4.54\times10^{-2}$  & $1.31\times10^{-1}$  & $2.55\times 10^{-1}$
\\
\hline $\alpha=2$         & $8.43\times 10^{-2}$   &
$1.84\times10^{-1}$   &   $3.2\times10^{-1}$  & $4.68\times 10^{-1}$
\\
\hline
\end{tabular}
\end{center}
\caption{Dynamical gap $m(0)$ evaluated by adopting the GGG
approximation in the presence of both Coulomb interaction and
four-fermion coupling. Here, $N = 4$ and $\beta = 1$.
\label{Table:CoulombFourFermion}}
\end{table}

In this section, we have utilized the instantaneous approximation to
solve the DS equation. The conclusion that dynamical gap generation
is enhanced by additional four-fermion coupling is still valid when
the energy dependence of Coulomb interaction is taken into account.
This conclusion is confirmed by the pertinent numerical results
presented in Table \ref{Table:CoulombFourFermion}.

\section{Quantum critical phenomena emergent at insulating transition\label{Sec:ObveQuan}}

Once an excitonic gap is opened at the Fermi level, the semi-Dirac
semimetal is converted to an insulator. In order to explicitly see
the difference between semimetallic and insulating phases, we shall
compute two important quantities, namely fermion DOS and specific
heat, in this section. At a given flavor $N$ and fixed $\beta$, the
insulating quantum phase transition happens at the QCP $\alpha_{c}$.
A bosonic order parameter $\phi$ develops a nonzero mean value
continuously as $\alpha$ becomes larger than $\alpha_c$. At such
QCP, the semi-Dirac fermions and the excitonic order parameter are
gapless, and can strongly affect each other. In this case, one
cannot simply integrate out the fermionic degrees of freedom
\cite{Hertz76, Millis93}. Instead, one should maintain gapless
fermions and gapless bosonic order parameter in an effective action,
and study the Yukawa coupling between them \cite{Huh08,
WangJingLiu11, Metlitski10, Garst10, Savary14, Jian15B}. In this
section, we perform a RG analysis of this coupling, and examine
whether the system exhibits NFL behaviors at the QCP.

\subsection{DOS and specific heat in insulating phase}

The dynamically generated fermion gap $m$ manifests itself in
several observable quantities, among which we are mainly interested
in fermion DOS and specific heat. In order to make analytic
computation simpler, we assume a constant gap $m$. After
including $m$, the retarded fermion propagator is written in the
form
\begin{eqnarray}
G^{R}(\omega,\mathbf{k}) = \frac{-1}{\omega - ak_{x}^{2}\tau_{1} -
vk_{y}\tau_{2} - m\tau_{3} + i\eta}.
\end{eqnarray}
The spectral function is given by
\begin{eqnarray}
A(\omega,\mathbf{k}) &=& \frac{1}{\pi}\mathrm{Tr}
\left[\mathrm{Im}\left[G^{R}(\omega,\mathbf{k})\right]\right]\nonumber
\\
&=& 2|\omega|\delta\left(\omega^2 - \left(a^{2}k_{x}^{4} +
v^{2}k_{y}^{2} + m^{2}\right)\right).
\end{eqnarray}
The DOS can be computed from spectral function:
\begin{eqnarray}
\nu(\omega) &=& N\int\frac{d^2\mathbf{k}}{(2\pi)^2}
A(\omega,\mathbf{k}) \nonumber \\
&=&\frac{K\left(\frac{1}{2}\right)N}{\sqrt{2}\pi^2 \sqrt{a}v}
\frac{|\omega|}{\left(\omega^{2} -
m^{2}\right)^{\frac{1}{4}}}\theta\left(|\omega| - m\right),
\label{Eq:DosInsulting}
\end{eqnarray}
where $K(x)$ is complete elliptic integrals of the first kind. It is
easy to see that DOS is significantly suppressed by the finite gap
in the low-energy regime.

To compute specific heat, we find it convenient to work in the
Matsubara Green function formalism. The Matsubara propagator of
massive semi-Dirac fermions is
\begin{eqnarray}
G(\omega_{n},\mathbf{k}) &=& \frac{1}{-i\omega_{n} +
ak_{x}^{2}\tau_{1} + vk_{y}\tau_{2} + m\tau_{3}}\nonumber
\\
&=& \frac{i\omega_{n} + ak_{x}^{2}\tau_{1} + vk_{y}\tau_{2} +
m\tau_{3}}{\left(\omega_{n}^{2} + a^{2}k_{x}^{4} + v^{2}k_{y}^{2} +
m^{2}\right)},
\end{eqnarray}
where $\omega_{n}=(2n+1)\pi T$ with $n$ being integers. Accordingly,
the free energy takes the form
\begin{eqnarray}
F(T) = 2NT\sum_{\omega_{n}}\int\frac{d^2\mathbf{k}}{(2\pi)^2}
\ln\left[\left(\omega_{n}^2+E_{k}^{2}+m^2\right)^{\frac{1}{2}}\right]
\end{eqnarray}
with $E_{k}=\sqrt{a^{2}k_{x}^{4}+v^{2}k_{y}^{2}}$. After carrying
out the summation over imaginary frequency $\omega_n$, we obtain
\begin{eqnarray}
F(T) = -4NT\int\frac{d^2\mathbf{k}}{(2\pi)^2} \ln\left[1 +
e^{-\frac{\sqrt{E_{k}^2+m^2}}{T}}\right],
\end{eqnarray}
where a temperature independent term has been dropped. The specific
heat is connected to the free energy through the relation:
\begin{eqnarray}
C_{V}(T) = -T\frac{\partial^{2}}{\partial T^2}F(T).
\end{eqnarray}
In the gapless semimetallic phase with $m = 0$, $C_{V}(T)$ depends
on temperature as
\begin{eqnarray}
C_{V}(T) = \frac{15\sqrt{2}\left(4 - \sqrt{2}\right)
K\left(\frac{1}{2}\right)\zeta\left(\frac{5}{2}\right)}{16\pi^{\frac{3}{2}}}
\frac{N}{\sqrt{a}v}T^{\frac{3}{2}},
\end{eqnarray}
where $\zeta(x)$ is the Riemann zeta function. In the insulating
phase with $m \neq 0$, $C_V(T)$ is strongly suppressed by fermion
gap comparing to the semimetallic phase. In particular, in the limit
$m \gg T$, $C_{V}(T)$ is given by
\begin{eqnarray}
C_{V}(T) &\approx& \frac{4\sqrt{2}K\left(\frac{1}{2}\right)}{3\pi^2}
\frac{N}{\sqrt{a}v} \frac{m^{\frac{7}{2}}}{T^{2}}e^{-\frac{m}{T}},
\label{Eq:CvInsulating}
\end{eqnarray}
which decreases rapidly with lowering $T$.

From Eq.~(\ref{Eq:DosInsulting}) and Eq.~(\ref{Eq:CvInsulating}), we
observe that both DOS and specific heat are significantly suppressed
in the low-energy region in the insulating phase.

\subsection{Non-Fermi liquid behaviors at semimetal-insulator QCP\label{Sec:SubSecNFLSoleQuantumFL}}

In this subsection, we analyze the interaction between the quantum
fluctuation of bosonic order parameter $\phi$ and the gapless
semi-Dirac fermions, which is described by a Yukawa-type coupling:
\begin{eqnarray}
S_{fb} = \lambda_{0}\sum_{\sigma=1}^{N}\int
d\tau d^2\mathbf{x} \phi\psi_{\sigma}^{\dag}\tau_{3}\psi_{\sigma}.
\end{eqnarray}
where $\lambda_{0}$ is the coupling coefficient. The free action of
$\phi$ takes the standard form
\begin{eqnarray}
S_{\phi} = \int d\tau d^2\mathbf{x}
\left[\frac{1}{2}\left(\partial_{\tau}\phi\right)^{2} +
\frac{\kappa}{2}\left(\mathbf{\nabla}\phi\right)^{2} +
\frac{r}{2}\phi^{2}+\frac{u_{0}}{24}\phi^{4}\right],
\end{eqnarray}
where the varying parameter $r$ tunes the excitonic insulating
transition with $r=0$ being QCP. The free propagator of $\phi$ is
given by
\begin{eqnarray}
D_{0}(\omega,\mathbf{q}) = \frac{1}{\omega^2 + \kappa\mathbf{q}^{2}
+ r}.
\end{eqnarray}
Following the treatment of Ref.~\cite{Huh08}, we now make the
replacements: $\phi \rightarrow \phi/\lambda_{0}$ and $r \rightarrow
r\lambda_{0}^2$. Performing such a re-scaling manipulation leads to
\begin{eqnarray}
S_{fb} = \sum_{\sigma=1}^{N}\int d\tau d^2\mathbf{x}
\phi\psi_{\sigma}^{\dag}\tau_{3}\psi_{\sigma}.
\end{eqnarray}
It is important to remember that both $\phi$ and $\psi$ particles
are gapless at QCP. The quantum critical behaviors cannot be studied
within the Hertz-Millis theory \cite{Hertz76, Millis93}.
Alternatively, we need to treat $\phi$ and $\psi$ on an equal
footing and carefully study their coupling \cite{Huh08,
WangJingLiu11, Metlitski10, Garst10, Savary14, Jian15B}. We now
perform a detailed RG analysis of this coupling by employing a $1/N$
expansion, and examine whether the fermions exhibit NFL behaviors.

At the QCP with $r=0$, including the polarization function
contributed from fermions leads to the following propagator for
$\phi$ field:
\begin{eqnarray}
D(\Omega,\mathbf{q}) = \frac{1}{\omega^2 +
\mathbf{\kappa}\mathbf{q}^{2} + \Pi_{33}(\Omega,\mathbf{q})},
\end{eqnarray}
where the polarization $\Pi_{33}$ is given by
\begin{eqnarray}
\Pi_{33}(\Omega,\mathbf{q}) &=& N\int\frac{d\omega}{2\pi}
\int\frac{d^2\mathbf{k}}{(2\pi)^{2}}
\mathrm{Tr}\left[\tau_{3}G_{0}(\omega,\mathbf{k})\tau_{3}\right.
\nonumber \\
&&\left.\times G_{0}\left(\omega +
\Omega,\mathbf{k}+\mathbf{q}\right)\right]
\label{Eq:Pi33DefMainText}
\end{eqnarray}
to the leading order of $1/N$ expansion. After straightforward
calculations, which are presented in Appendix
\ref{App:Polarization33}, we find that $\Pi_{33}$ can be
approximated by
\begin{eqnarray}
\Pi_{33}(\Omega,\mathbf{q}) = \frac{N}{\sqrt{a}v}\left[b_{1}
\left(\Omega^2 + v^{2}q_{y}^{2}\right) +
b_{2}a^{2}q_{x}^{4}\right]^{\frac{1}{4}},
\end{eqnarray}
where $b_{1}$ and $b_{2}$ are two constants. In the low-energy
regime, $\Pi_{33}$ dominates over the free term of $\phi$. We then
drop the free term, and write the propagator as
\begin{eqnarray}
D(\Omega,\mathbf{q}) = \frac{1}{\Pi_{33}(\Omega,\mathbf{q})}.
\end{eqnarray}
The leading-order fermion self-energy induced by the Yukawa coupling
is
\begin{eqnarray}
\Sigma_{fb}(\omega,\mathbf{k}) &=& \int'\frac{d\Omega}{2\pi}
\frac{d^2\mathbf{q}}{(2\pi)^{2}}\tau_{3}
G_{0}\left(\omega+\Omega,\mathbf{k}+\mathbf{q}\right)\nonumber \\
&&\times\tau_{3}D(\Omega,\mathbf{q}).\label{Eq:FermionSelfEnergyDefMainText}
\end{eqnarray}
Here, the integration $\int'\frac{d\Omega}{2\pi}
\frac{d^2\mathbf{q}}{(2\pi)^{2}}$ is made by choosing a suitable
momentum shell for some related parameter. To be specific, we now
employ the following RG scheme:
\begin{eqnarray}
-\infty < \Omega < \infty, \qquad b\Lambda < E_{q} < \Lambda,
\label{Eq:RGSchemeAMainText}
\end{eqnarray}
where $E_{q} = \sqrt{a^2q_{x}^{4}+v^2q_{y}^{2}}$, and $b = e^{-l}$
with $l$ being a running length scale. According to Appendix
\ref{App:FermionSelfEnergy}, the fermion self-energy takes the
approximate form
\begin{eqnarray}
\Sigma_{fb}(\omega,\mathbf{k}) &\approx&\left[-i\omega
C_{1}+ak_{x}^{2}C_{2}\tau_{1}+vk_{y}C_{3}\tau_{2}\right]\ln(b^{-1}).
\nonumber \\
\label{Eq:FermionSelfEnergyResult}
\end{eqnarray}
The expressions of constants $C_{1,2,3}$ are presented in
Eqs.~(\ref{Eq:C1RGSchemeAAppen})-(\ref{Eq:C3RGSchemeAAppen}).
Numerical calculation leads us to
\begin{eqnarray}
C_{1} &\approx& -\frac{0.0434929}{N}, \label{Eq:C1Num}
\\
C_{2} &\approx& -\frac{0.00149482}{N},\label{Eq:C2Num}
\\
C_{3} &\approx& -\frac{0.0434929}{N}.\label{Eq:C3Num}
\end{eqnarray}
Here notice that $C_{1} = C_{3}$.

We now proceed to derive the RG equations. The action of free
semi-Dirac fermions is
\begin{eqnarray}
S_{\psi} &=& \int\frac{d\omega}{2\pi}\frac{dk_{x}}{(2\pi)}
\frac{dk_{y}}{(2\pi)}\psi^{\dag}(\omega,\mathbf{k})
\left(-i\omega+ak_{x}^{2}\tau_{1}\right.\nonumber \\
&&\left.+vk_{y}\tau_{2}\right)\psi(\omega,\mathbf{k}).
\end{eqnarray}
After including the self-energy corrections, this action becomes
\begin{eqnarray}
S_{\psi} &=& \int\frac{d\omega}{2\pi}\frac{dk_{x}}{(2\pi)}
\frac{dk_{y}}{(2\pi)}\psi^{\dag}(\omega,\mathbf{k})
\left(-i\omega+ak_{x}^{2}\tau_{1}\right.\nonumber
\\
&&\left.+vk_{y}\tau_{2}-\Sigma_{fb}(i\omega,\mathbf{k})\right)
\psi(\omega,\mathbf{k})
\\
&\approx&\int\frac{d\omega}{2\pi}\frac{dk_{x}}{(2\pi)}
\frac{dk_{y}}{(2\pi)}\psi^{\dag}(\omega,\mathbf{k})\left(-i\omega
e^{-C_{1}l}\right.\nonumber \\
&&\left.+ak_{x}^{2}\tau_{1}e^{-C_{2}l} + vk_{y}\tau_{2}
e^{-C_{3}l}\right)\psi(\omega,\mathbf{k}).
\end{eqnarray}
We then make the following scaling transformations:
\begin{eqnarray}
k_{x}&=&k_{x}'e^{-\frac{l}{2}},\label{Eq:kxscaling}
\\
k_{y}&=&k_{y}'e^{-l}, \label{Eq:kyscaling}
\\
\omega&=&\omega'e^{-l}, \label{Eq:omegascaling}
\\
\psi&=&\psi' e^{\left(\frac{7}{4}+\frac{C_{1}}{2}\right)l},\label{Eq:Psiscaling}
\\
a&=&a'e^{(C_{2}-C_{1})l},\label{Eq:Ascaling}
\\
v&=&v'e^{(C_{3}-C_{1})l}.\label{Eq:vscaling}
\end{eqnarray}
Now the action is recast in the form
\begin{eqnarray}
S_{\psi'} &=& \int\frac{d\omega'}{2\pi}\frac{dk_{x}'}{(2\pi)}
\frac{dk_{y}'}{(2\pi)}\psi'^{\dag}(\omega',\mathbf{k'})
\left(-i\omega'+a'k_{x}'^{2}\tau_{1}\right.\nonumber \\
&&\left.+v'k_{y}'\tau_{2}\right)\psi'(\omega',\mathbf{k}'),
\end{eqnarray}
which is formally the same as the action of free fermions. According
to Eq.~(\ref{Eq:Psiscaling}), the RG equation for wave function
renormalization $Z_{f}$ satisfies
\begin{eqnarray}
\frac{dZ_{f}}{dl} = C_{1}Z_{f}.
\label{Eq:RGEqZf}
\end{eqnarray}
Based on Eqs.~(\ref{Eq:Ascaling}) and (\ref{Eq:vscaling}), we derive
the RG equations for $a$ and $v$:
\begin{eqnarray}
\frac{da}{dl} &=& (C_{1}-C_{2})a,\label{Eq:RGEqA} \\
\frac{dv}{dl} &=& (C_{1}-C_{3})v.\label{Eq:RGEqv}
\end{eqnarray}
Solving the above three RG equations, we obtain
\begin{eqnarray}
Z_{f}&=&Z_{f0}e^{C_{1}l},\label{Eq:SolutionZf}
\\
a&=&a_{0}e^{(C_{1}-C_{2})l},\label{Eq:SolutionA}
\\
v&=&v_{0}, \label{Eq:Solutionv}
\end{eqnarray}
where $Z_{f0}=1$. In the long wavelength limit $l \rightarrow
\infty$, it is easy to find that $Z_{f}$ and $a$ both flow to zero,
whereas $v$ doest not flow and remains a constant. The behavior
$\lim_{l \rightarrow \infty}Z_{f} \rightarrow 0$ clearly indicates
the breakdown of FL theory. The wave function renormalization
$Z_{f}$ can also be obtained from the identity:
\begin{eqnarray}
Z_{f}(\omega) = \frac{1}{\left|1 -
\frac{\partial}{\partial\omega}\mathrm{Re}
\Sigma^{R}(\omega)\right|}, \label{Eq:ZfDef2}
\end{eqnarray}
where $\Sigma^{R}(\omega)$ is the retarded fermion self-energy
function. Combining Eqs.~(\ref{Eq:SolutionZf}) and
(\ref{Eq:ZfDef2}), and then using the scaling relationship $\omega =
\omega_{0}e^{-l}$, one can find that
\begin{eqnarray}
\mathrm{Re}\Sigma^{R}(\omega) \sim \omega^{1-\eta_{f}},
\end{eqnarray}
where $\eta_{f} = -C_{1}$ is a positive quantity. By virtue of
Kramers-Kronig (KK) relation, it is easy to obtain
\begin{eqnarray}
\mathrm{Im}\Sigma^{R}(\omega) \sim \omega^{1-\eta_{f}},
\end{eqnarray}
which is apparently typical NFL behavior.

\begin{figure}[htbp]
\center
\includegraphics[width=2.8in]{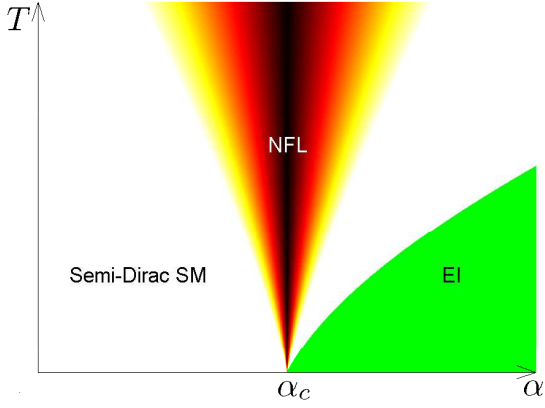}
\caption{Schematic phase diagram of semi-Dirac fermion system on
plane $(\alpha,T)$ with fixed values of $N$ and $\beta$. Here, SM,
EI, and NFL stand for semimetal, excitonic insulating, and non-Fermi
liquid phases, respectively. The zero-temperature QCP is broadened
by thermal fluctuation to a finite quantum critical region on the
phase diagram at finite temperatures. \label{Fig:PhaseSemiDirac}}
\end{figure}

We present a schematic phase diagram in
Fig.~\ref{Fig:PhaseSemiDirac} for 2D interacting semi-Dirac fermion
system on the plane spanned by $\alpha$ and $T$, with $\alpha_c$
defining the semimetal-insulator QCP. The system stays in the
gapless semimetal phase for $\alpha < \alpha_c$, and becomes an
excitonic insulator for $\alpha > \alpha_c$. The fermions exhibit
NFL behaviors at zero-temperature QCP $\alpha_c$, which becomes a
finite quantum critical region at finite temperature.

RG method has recently been applied to study the influence of
Coulomb interaction in several semimetals which are distinguished
mainly by the fermion dispersions \cite{Yang14, Lai15, Jian15,
Huh16, Isobe16A, Cho16}. In a 3D anisotropic Weyl semimetal where
the spectrum displays quadratic dependence on momentum in one
direction but linear dependence on other components of momenta, Yang
\emph{et al.} supposed a shell of $(b\Lambda, \Lambda)$ for the
integration over the quadratic component of momenta \cite{Yang14}.
Lai \cite{Lai15} used the similar RG scheme to examine the impact of
Coulomb interaction in a 3D double Weyl semimetal, in which the
dispersion is quadratic in two components of momenta but linear in
the third component. Different from Lai \cite{Lai15}, Jian and Yao
\cite{Jian15} considered the same model but made use of a different
RG scheme which defines the momentum shell by $b \Lambda < E_{k} <
\Lambda$ with $E_{k} = \sqrt{\frac{1}{4m^2}(k_{x}^{2} +
k_{y}^{2})^{2} + v^2k_{z}^{2}}$. There are some differences between
the results obtained by Lai \cite{Lai15} and Jian and Yao
\cite{Jian15}. When studying the role of Coulomb interaction in a 2D
semi-Dirac semimetal, Isobe \emph{et al.} \cite{Isobe16A} introduced
a shell for energy $b \Lambda < \Omega < \Lambda$, whereas Cho and
Moon \cite{Cho16} defined a shell for the linearly dependent
momentum component.

\begin{figure*}[htbp]
\center
\includegraphics[width=2.8in]{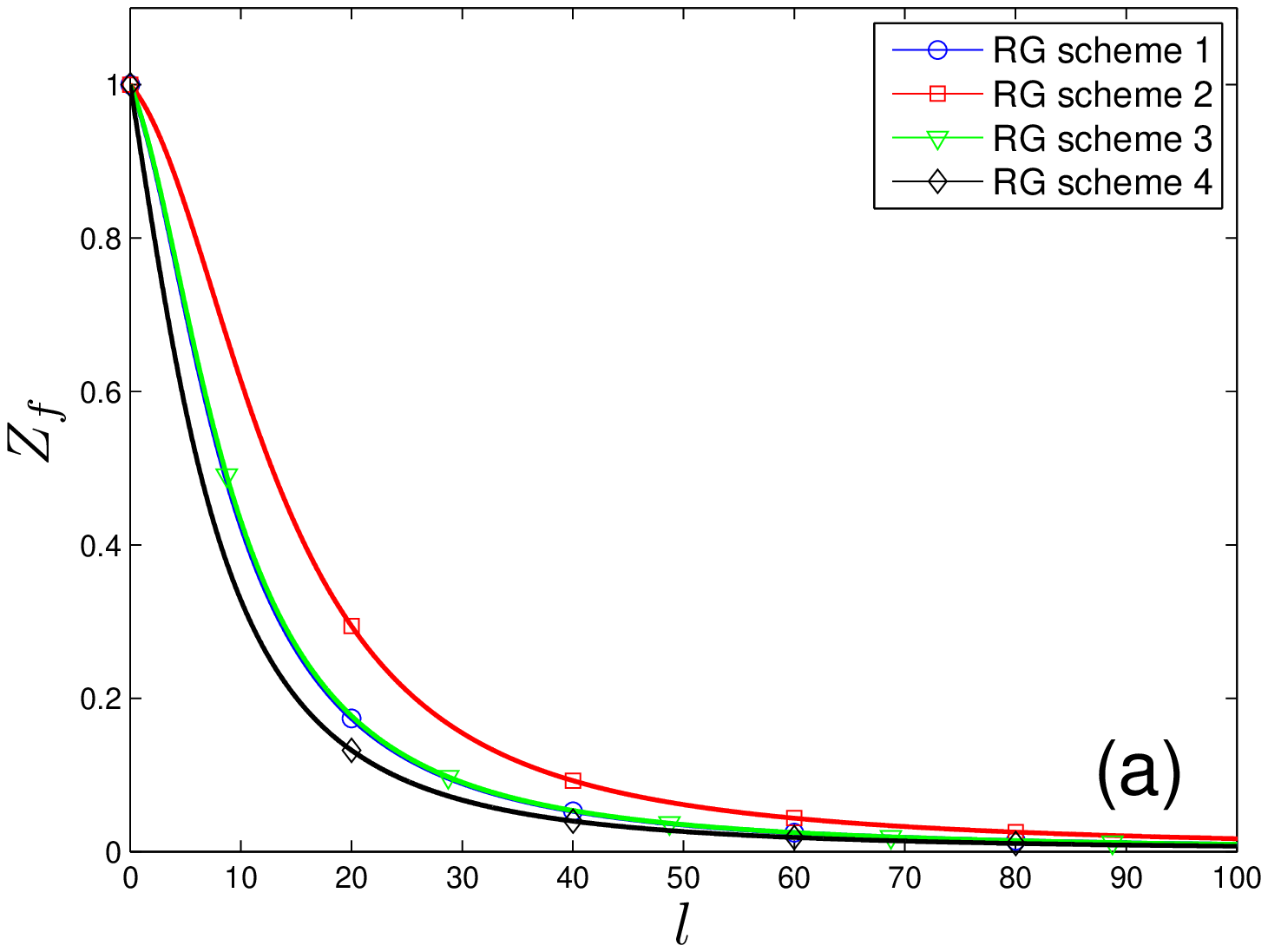}
\includegraphics[width=2.8in]{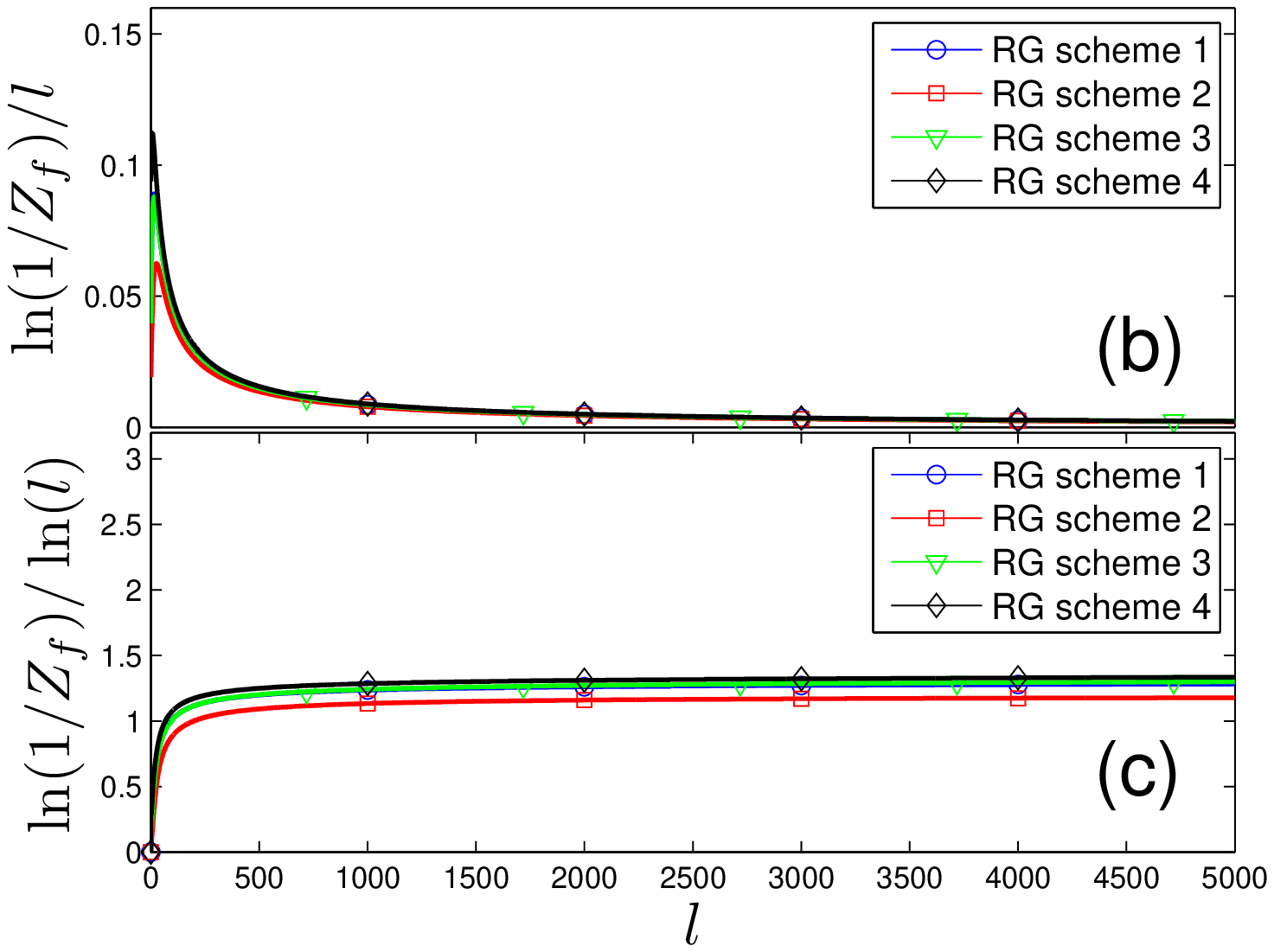}
\includegraphics[width=2.8in]{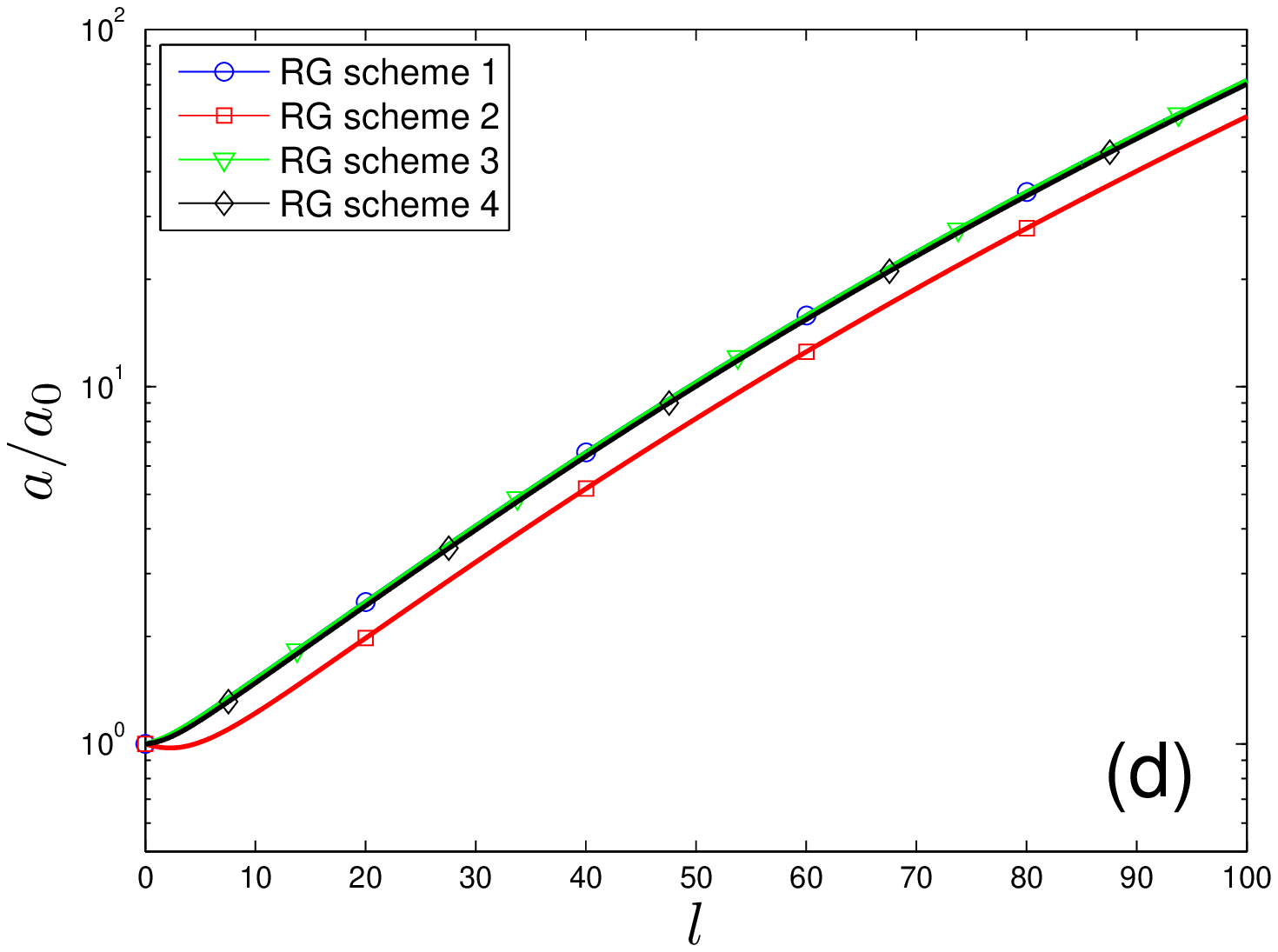}
\includegraphics[width=2.8in]{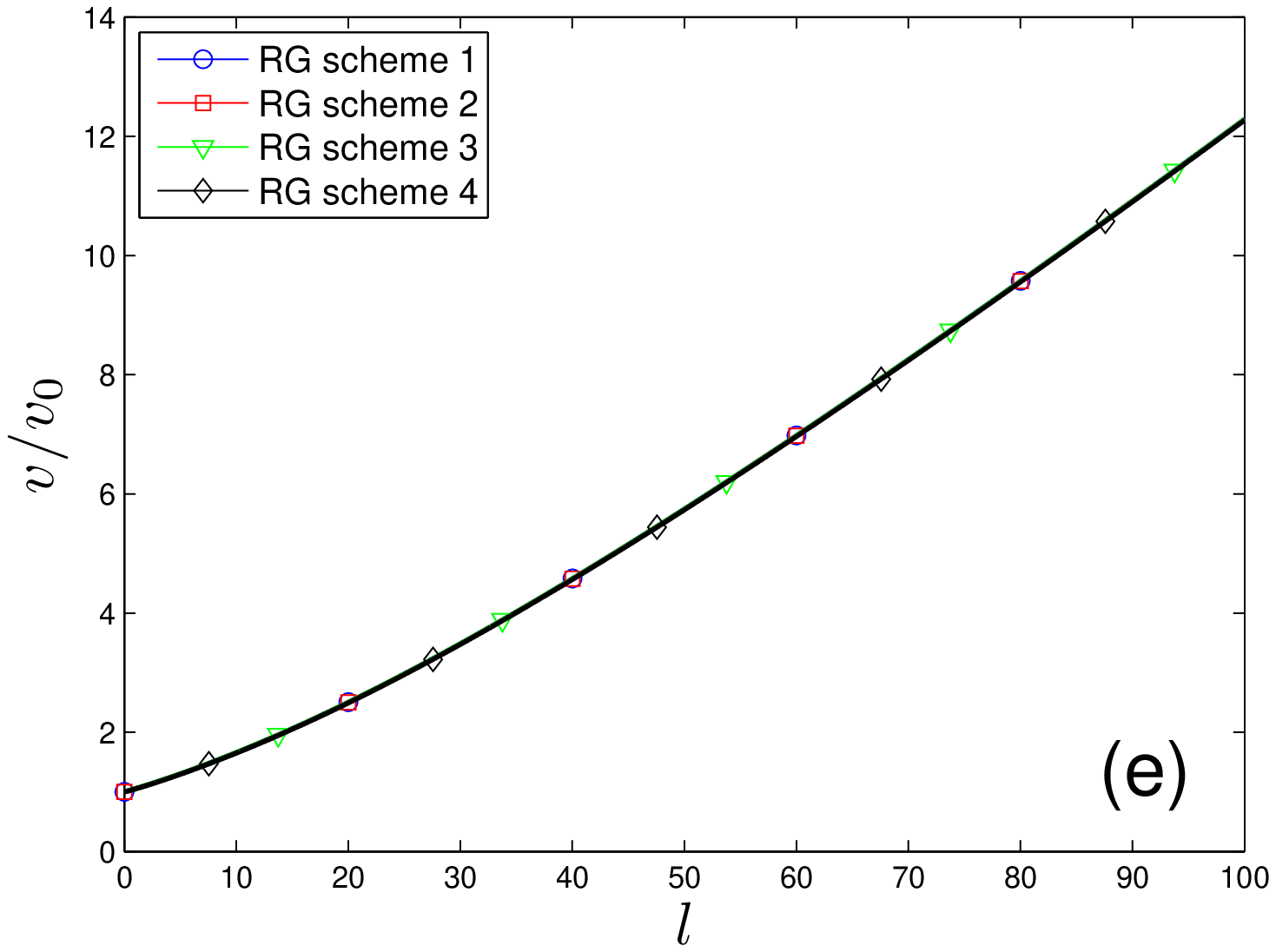}
\caption{Flows of $Z_{f}$$, a$, and $v$ considering long-range
Coulomb interaction in semimetal phase in different RG schemes. The
initial condition $\alpha_{0}=1$ and $\beta_{0}=1$ are taken.
\label{Fig:CoulombVRGCompare}}
\end{figure*}

Since recent studies showed that different RG schemes may result in
quantitative differences \cite{Lai15, Jian15}, it is important to
carry out RG calculations by employing several possible RG schemes
and testify the reliability of our RG results. In the present
problem, we also consider the following three RG schemes:
\begin{eqnarray}
\int_{\omega,\mathbf{q}}' =\int_{-\infty}^{+\infty} d\Omega
\left(\int_{-\Lambda}^{-\sqrt{b}\Lambda}+\int_{\sqrt{b}\Lambda}^{\Lambda}\right)
dq_{x} \int_{-\infty}^{+\infty}dq_{y},\label{Eq:RGSchemeBMainText}
\\
\int_{\omega,\mathbf{q}}'=\int_{-\infty}^{+\infty}d\Omega
\int_{-\infty}^{+\infty}dq_{x} \left(\int_{-\Lambda}^{-b\Lambda} +
\int_{b\Lambda}^{\Lambda}\right)dq_{y},\label{Eq:RGSchemeCMainText}
\\
\int_{\omega,\mathbf{q}}'=\left(\int_{-\Lambda}^{-b\Lambda} +
\int_{b\Lambda}^{\Lambda}\right)d\Omega
\int_{-\infty}^{+\infty}dq_{x}\int_{-\infty}^{+\infty}dq_{y},
\label{Eq:RGSchemeDMainText}
\end{eqnarray}
where $\int_{\omega,\mathbf{q}}'\equiv\int'd\omega d^2\mathbf{q}$.
For convenience, we hereinafter use RG scheme $1$, $2$, $3$, and $4$
to represent Eqs.~(\ref{Eq:RGSchemeAMainText}),
(\ref{Eq:RGSchemeBMainText}), (\ref{Eq:RGSchemeCMainText}), and
(\ref{Eq:RGSchemeDMainText}), respectively. In the RG schemes 2, 3,
and 4, the fermion self-energy can still be written as
Eq.~(\ref{Eq:FermionSelfEnergyResult}), with the expressions of
$C_{1, 2, 3}$ being given in Appendix \ref{App:C1C2C3Expressions}.
Numerical calculations show that the values of $C_{1,2,3}$ computed
using RG schemes $2$, $3$, and $4$ are precisely the same as those
given by Eqs.~(\ref{Eq:C1Num})-(\ref{Eq:C3Num}). Thus, these four
different RG schemes lead to exactly the same results.

\subsection{Interplay of Yukawa coupling and Coulomb interaction}

At semimetal-insulator QCP, semi-Dirac fermions not only couple to
the quantum fluctuation of excitonic order parameter $\phi$, but
interact with each other through the Coulomb interaction. For
completeness, it is necessary to consider both the Yukawa coupling
and the Coulomb interaction and treat them equally. From the above
analysis, we know that Yukawa coupling tends to drive parameter $a$
to decrease upon lowering the energy scale. In contrast, Coulomb
interaction can increase $a$ in the low-energy regime. When both
Yukawa coupling and Coulomb interaction are present, these two
opposite tendencies might give rise to interesting low-energy
behaviors of semi-Dirac fermions. This issue will be addressed in
this subsection.

\begin{figure*}[htbp]
\center
\includegraphics[width=2.8in]{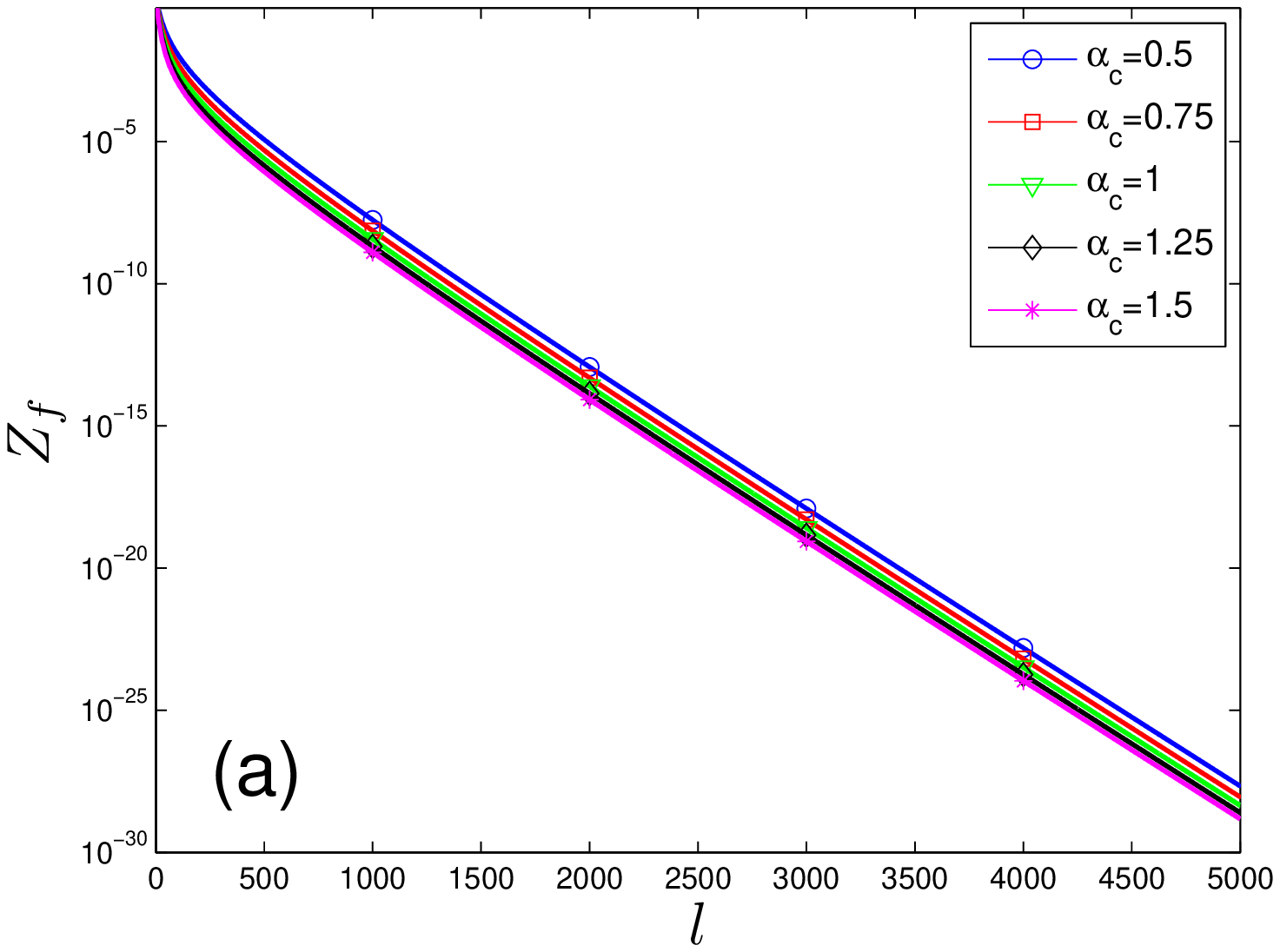}
\includegraphics[width=2.8in]{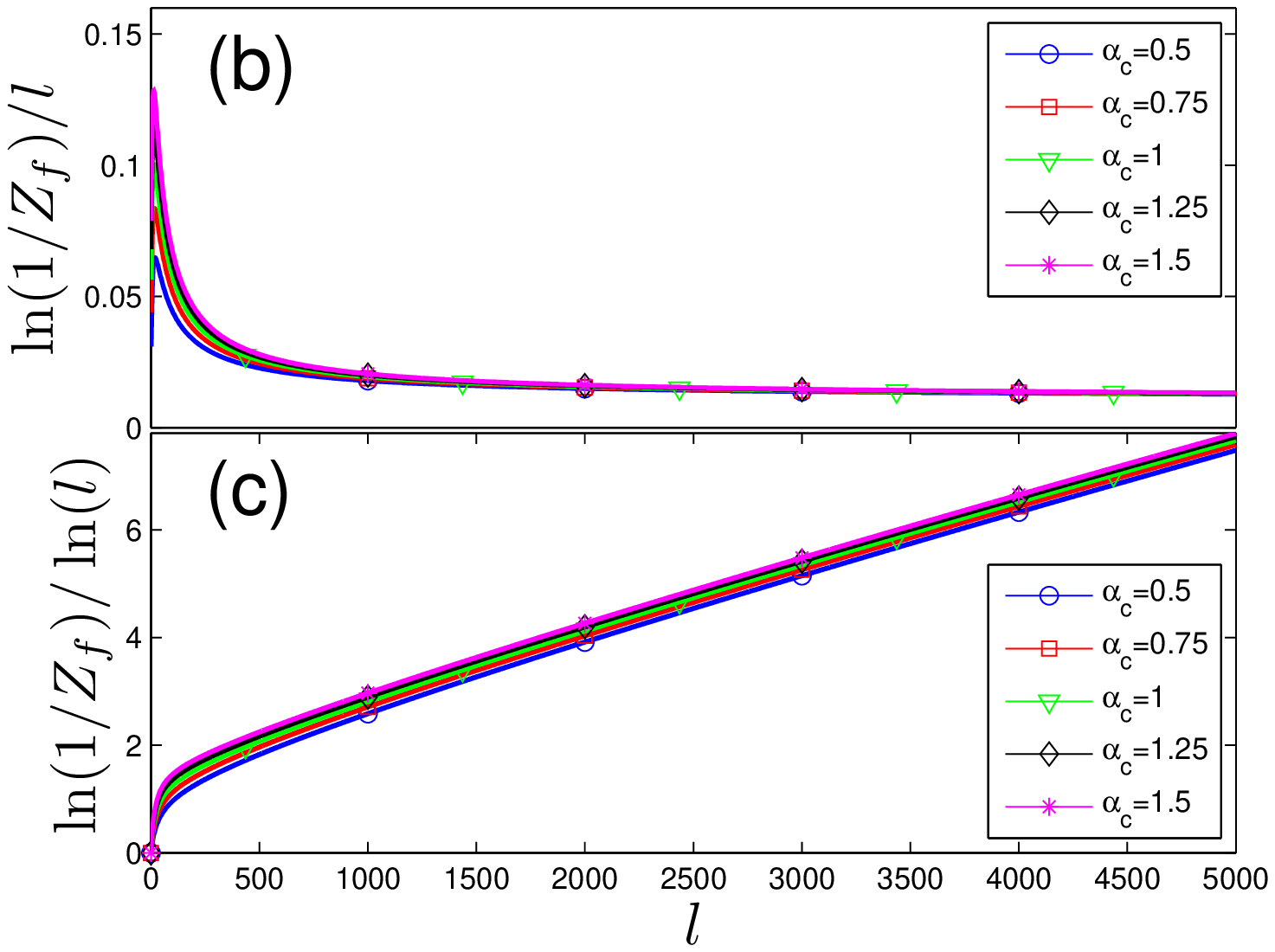}
\includegraphics[width=2.8in]{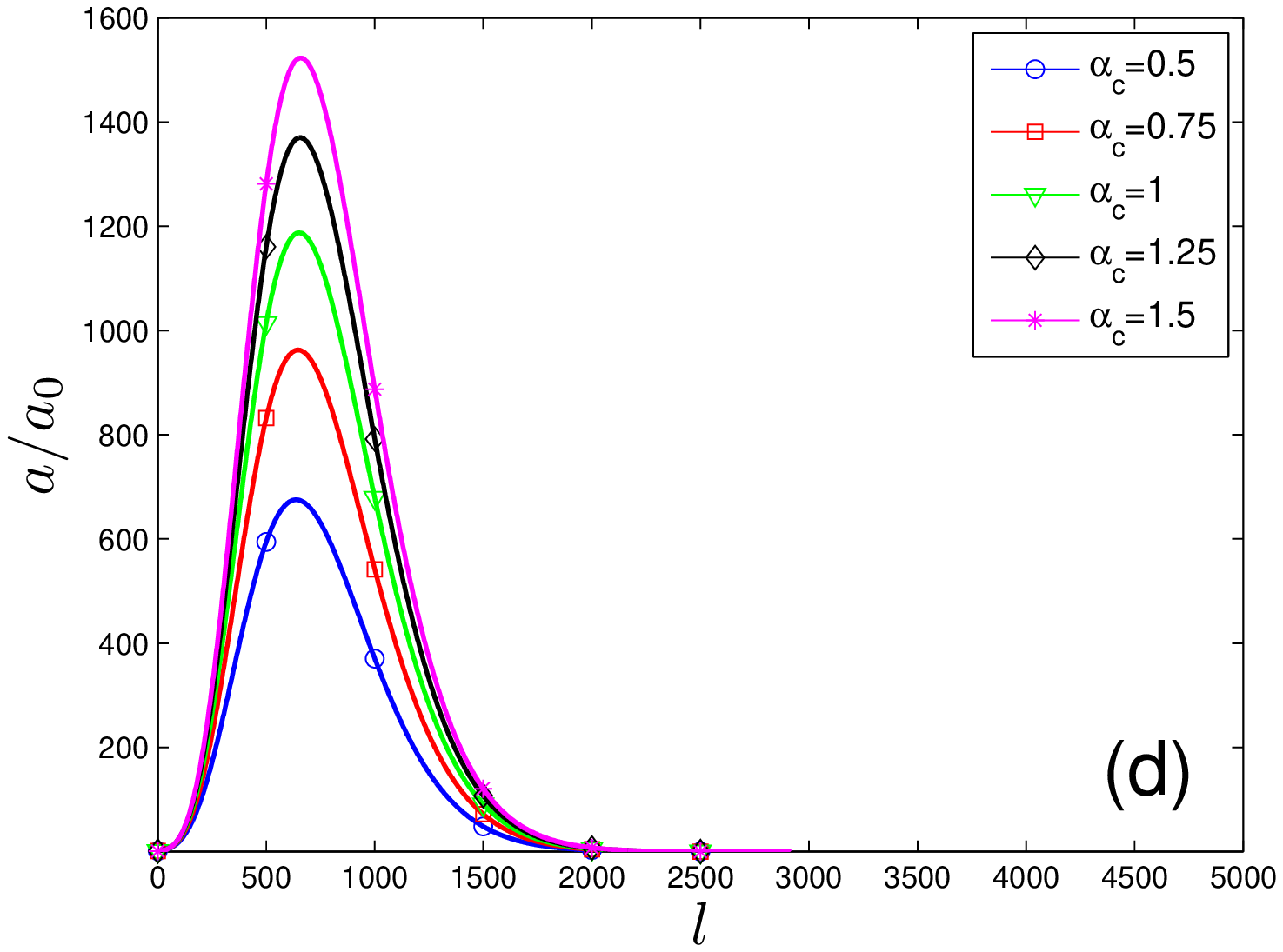}
\includegraphics[width=2.8in]{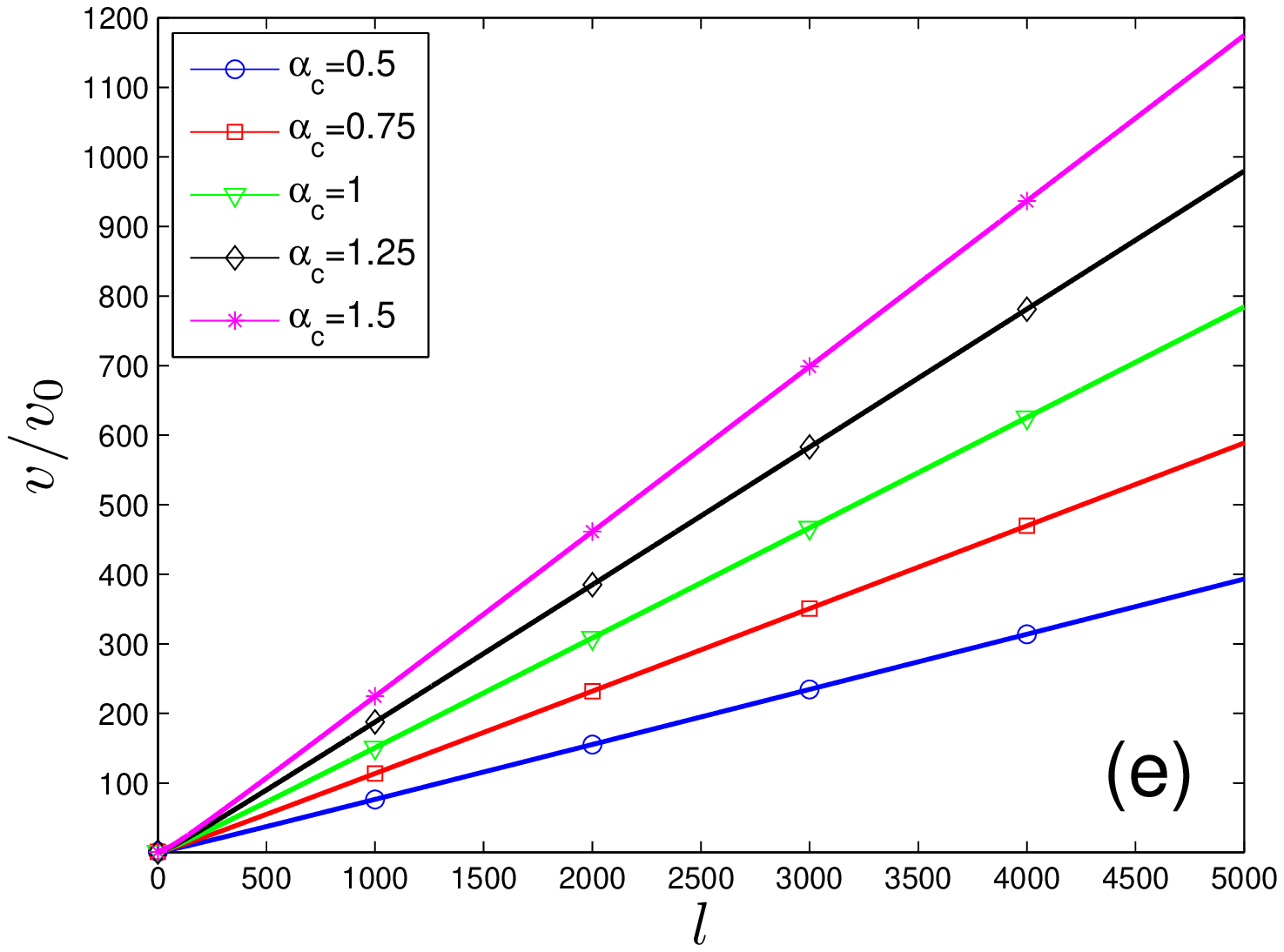}
\caption{Flows of $Z_{f}$$, a$, and $v$ considering both of quantum
fluctuation of insulating phase and long-range Coulomb interaction
at QCP in RG scheme 1. The initial condition $\beta_{0}=1$ is taken.
\label{Fig:VRGBoth}}
\end{figure*}

In a previous work, Isobe \emph{et al.} studied the influence of
Coulomb interaction on semi-Dirac fermions by utilizing the RG
scheme Eq.~(\ref{Eq:RGSchemeDMainText}). Because different RG
schemes might give rise to unidentical results \cite{Lai15, Jian15},
we feel it helpful to revisit the effect of Coulomb interaction by
employing the four RG schemes defined by
Eq.~(\ref{Eq:RGSchemeAMainText}), and (\ref{Eq:RGSchemeBMainText}),
(\ref{Eq:RGSchemeCMainText}), and (\ref{Eq:RGSchemeDMainText})
respectively. Detailed analytic calculations lead us to
\begin{eqnarray}
\frac{dZ_{f}}{dl} &=& C_{1}'Z_{f},
\\
\frac{da}{dl} &=& \left(C_{1}'-C_{2}'\right)a,
\\
\frac{dv}{dl} &=& \left(C_{1}'-C_{3}'\right)v,
\\
\frac{d\alpha}{dl} &=& \left(C_{3}'-C_{1}'\right)\alpha.
\end{eqnarray}
For RG schemes $1$ and $4$, $\beta$ is defined as $\beta =
\frac{a\Lambda}{v^2}$, and the RG equation of $\beta$ can be written
as
\begin{eqnarray}
\frac{d\beta}{dl} = \left(2C_{3}' - C_{1}' - C_{2}'-1\right)\beta.
\end{eqnarray}
For RG scheme $2$, $\beta$ is defined as $\beta =
\frac{a\Lambda}{v}$, whose RG equation is
\begin{eqnarray}
\frac{d\beta}{dl} = \left(C_{3}' - C_{2}'-\frac{1}{2}\right)\beta.
\end{eqnarray}
For RG scheme $3$, $\beta$ is defined as $\beta =
\frac{a\Lambda}{v}$, and the corresponding RG equation is
\begin{eqnarray}
\frac{d\beta}{dl} = \left(C_{3}' - C_{2}'-1\right)\beta.
\end{eqnarray}
The expressions of $C_{1}'$, $C_{2}'$, and $C_{3}'$ obtained by
employing different RG schemes are given in
Appendix~\ref{App:C1C2C3PrimeExpressions}. The RG flows of $Z_{f}$,
$a$, $v$ are shown in Fig.~\ref{Fig:CoulombVRGCompare}. According to
Fig.~\ref{Fig:CoulombVRGCompare}(a), $Z_{f}$ obtained by employing
four RG schemes all flow to zero at large $l$, thus normal FL
description becomes invalid. The $l$-dependence of $Z_{f}$ for RG
scheme $1$ is close to that for RG scheme $3$ over a wide range of
energies, but different from those obtained by means of RG schemes
$2$ and $4$. Based on Fig.~\ref{Fig:CoulombVRGCompare}(b) and (c),
it is easy to verify that
\begin{eqnarray}
\lim_{l \rightarrow \infty} \frac{\ln(1/Z_{f})}{l}\rightarrow 0,
\label{Eq:ZfLimitCoulombA}
\end{eqnarray}
but that
\begin{eqnarray}
\lim_{l \rightarrow \infty} \frac{\ln(1/Z_{f})}{\ln(l)}\rightarrow
\mathrm{constant.} \label{Eq:ZfLimitCoulombB}
\end{eqnarray}
We see from the above two expressions that Coulomb interaction gives
rise to MFL like behaviors in the lowest energy limit. This result is
qualitatively the same within four different RG schemes, and agrees
with the conclusion reached in Ref.~\cite{Isobe16A}. As can be seen
from Fig.~\ref{Fig:CoulombVRGCompare}(d) and (e), both $a$ and $v$
increase with growing length scale $l$. In particular, the numerical
results for $a$ are almost identical for RG schemes $1$, $3$, and
$4$, but quantitatively larger than the result obtained by RG scheme
$2$. The numerical results for $v$ in for different RG schemes are
all very close to each other.

We then incorporate both the Yukawa coupling and Coulomb
interaction, and study their interplay by means of RG method
combined with $1/N$ expansion. We will not give the calculational
detail here, but only list the final RG equations:
\begin{eqnarray}
\frac{dZ_{f}}{dl} &=& \left(C_{1}+C_{1}'\right)Z_{f},
\\
\frac{da}{dl} &=& \left(C_{1}+C_{1}'-C_{2}-C_{2}'\right)a,
\\
\frac{dv}{dl} &=& \left(C_{1}'-C_{3}'\right)v,
\\
\frac{d\alpha}{dl} &=& \left(C_{3}'-C_{1}'\right)\alpha.
\end{eqnarray}
The RG equation of $\beta$ are found to be
\begin{eqnarray}
\frac{d\beta}{dl} = \left(C_{3} - C_{2} + 2C_{3}' - C_{1}' -
C_{2}'-1\right)\beta
\end{eqnarray}
for RG schemes $1$ and $4$,
\begin{eqnarray}
\frac{d\beta}{dl} = \left(C_{3}+C_{3}'-C_{2}-C_{2}'-\frac{1}{2}\right)\beta
\end{eqnarray}
for RG scheme $2$, and
\begin{eqnarray}
\frac{d\beta}{dl} = \left(C_{3}+C_{3}'-C_{2}-C_{2}'-1\right)\beta
\end{eqnarray}
for RG scheme $3$.

We present the low-energy behaviors of parameters $Z_{f}$, $a$, and
$v$ in Fig.~\ref{Fig:VRGBoth} by adopting RG scheme $1$ as an
example. As shown by Fig.~\ref{Fig:VRGBoth}(a), $Z_{f}$ flows
monotonously to zero in the lowest energy limit. Using the results
displayed in Fig.~\ref{Fig:VRGBoth}(b) and (c), we find that $Z_{f}$
manifests the following asymptotic behavior:
\begin{eqnarray}
\lim_{l\rightarrow \infty} \frac{\ln(1/Z_{f})}{l} \rightarrow
\mathrm{constant},
\end{eqnarray}
and
\begin{eqnarray}
\lim_{l\rightarrow \infty} \frac{\ln(1/Z_{f})}{\ln(l)}\rightarrow
\infty.
\end{eqnarray}
Obviously, the semi-Dirac fermions exhibit NFL behaviors at the
semimetal-insulator QCP, so the schematic phase diagram depicted in
Fig.~\ref{Fig:VRGBoth} is still applicable. We observe from
Fig.~\ref{Fig:VRGBoth}(d) that the parameter $a$ displays a
non-monotonic dependence on the length scale $l$: as $l$ grows, $a$
first increases and then decreases. However,
Fig.~\ref{Fig:VRGBoth}(e) shows that parameter $v$ increases
monotonously with growing $l$. Since $v$ does not flow at all in
case there is only Yukawa coupling, the low-energy behavior of $v$
is indeed determined by Coulomb interaction. The numerical results
obtained by applying the RG schemes $2$, $3$, and $4$, which are not
presented here, are qualitatively the same as those obtained based
on RG scheme $1$.

\begin{table*}[htbp]
\begin{center}
\begin{tabular}{|c|c|c|c|c|c|c|}
\hline  \multirow{2}{*}{} & \multicolumn{3}{|c|}{Lowest order
truncation} & \multicolumn{3}{|c|}{\multirow{2}{*}{Higher order
corrections}}
\\ \cline {2-4} & Instan. & Khv.  & GGG & \multicolumn{3}{|c|}{}
\\
\hline Dirac SM     & \tabincell{c}{$\alpha_{c}\approx 2.33$ \\
\cite{Khveshchenko01, Gorbar02}} & \tabincell{c}{$\alpha_{c} \approx
1.13$
\\ \cite{Khveshchenko09}} &  \tabincell{c}{$\alpha_{c} \approx 0.92$ \\
\cite{Gamayun10} } & \tabincell{c}{$3.2<\alpha_{c}<3.3$ \\
\cite{WangLiu12}}  & \tabincell{c}{$\alpha_{c}\approx3.1$ \\
\cite{Gonzalez15}} &
\tabincell{c}{$\alpha_{c,\mathrm{min}}\approx2.889$,
$\alpha_{c,\mathrm{max}}\approx3.19$ \\ \cite{Carrington16}}
\\
\hline semi-Dirac SM         & $\alpha_{c}\approx0.6$   &
$\alpha_{c}\approx 0.3\sim0.4$   & $\alpha_{c}<0.2$   &
\multicolumn{3}{|c|} {in progress}
\\
\hline
\end{tabular}
\end{center}
\caption{The critical value $\alpha_{c}$ obtained by adopting
various approximations in the cases of 2D Dirac semimetal and 2D
semi-Dirac semimetal, where $N = 4$. For semi-Dirac semimetal, we
choose $\beta = 0.1, 1, 10$. We use Instan., Khv., and GGG to stand
for instantaneous, Khveshchenko, and GGG approximations.
\label{Table:AlphaC}}
\end{table*}

\section{Summary and Discussions\label{Sec:Discussion}}

In summary, we have studied dynamical excitonic gap generation
induced by long-range Coulomb interaction in a 2D semi-Dirac
semimetal. The critical Coulomb interaction strength $\alpha_c$ has
been calculated by solving the self-consistent DS equation of
dynamical excitonic gap. By adopting three frequently used
approximations, we have showed that a moderately strong Coulomb
interaction suffices to generate a finite excitonic gap. It is
therefore much easier for Coulomb interaction to trigger excitonic
pairing in a 2D semi-Dirac semimetal than 2D Dirac semimetal. We
also have found that additional short-range interaction reduces the
critical value $\alpha_c$ and hence catalyzes excitonic gap
generation.

Among all the currently known 2D semi-Dirac semimetals, we find that
the TiO$_2$/VO$_2$ nanostructure is the most promising candidate for
the realization of the anticipated excitonic insulating state. There
are two reasons. Firstly, 2D semi-Dirac fermions naturally emerge in
such a nanostructure, and it is not necessary to elaborately adjust
some model parameters. Secondly, the physical value of $\alpha$ in
this material is smaller or at least very close to the critical
value $\alpha_c$ obtained in our DS equation analysis. It is
certainly also possible to open an excitonic gap in other 2D
semi-Dirac semimetal materials, which deserves a systematic
investigation. We hope that experiments, including but not
restricted to ARPES, would be performed in the future to search the
predicted excitonic insulating state in various 2D semi-Dirac
materials, with TiO$_2$/VO$_2$ nanostructure being the most probable
candidate.

We also have showed that 2D semi-Dirac fermions exhibit distinct
behaviors in the massless semimetal phase, excitonic insulating
phase, and at the semimetal-insulator QCP. Specifically, the
massless semi-Dirac fermions couple strongly to the quantum
fluctuation of excitonic order parameter by a Yukawa-type coupling
at the QCP. We have examined the impact of this coupling on the
low-energy properties of fermions by carrying out a detailed RG
analysis, and revealed unusual NFL behaviors of massless fermions
and other interesting quantum critical phenomena.

In our present DS equation studies, we have neglected a number of
physical effects to simplify numerical computation, including the
fermion velocity renormalization \cite{Sabio10, WangLiu12,
Popovici13}, strong fermion damping \cite{WangLiu12, Gonzalez15,
Carrington16}, and vertex corrections \cite{WangLiu12,
Carrington16}. These effects formally embodied in the functions
$A_{0,1,2}(\omega,k_x,k_y)$ appearing in
Eq.~(\ref{Eq:FermionPropagatorFull}) and in the function
$\Gamma(\varepsilon,\mathbf{p};\omega,\mathbf{k})$ appearing in
Eq.~(\ref{Eq:GapEqOriginal}). To the leading order of $1/N$
expansion, it is safe to set $A_{0,1,2}(\omega,k_x,k_y) =
\Gamma(\varepsilon,\mathbf{p};\omega,\mathbf{k}) = 1$. However,
because the physical flavor $N$ is usually not large, the higher
order corrections may be important. The impact of these corrections
can be examined by solving the self-consistent equations of
$A_{0,1,2}(\omega,k_x,k_y)$ and $m(\omega,k_x,k_y)$. In order not to
spoil the Ward identity, the vertex corrections need to be properly
included by introducing an appropriate \emph{ansatz}
\cite{WangLiu12, Carrington16, Fischer04}. The polarization function
may also be self-consistently computed by using the full fermion
propagator that contains $A_{0,1,2}$ and $m$ \cite{Gonzalez15,
Fischer04}. While it is formally straightforward to write down the
full set of self-consistent equations, it is highly nontrivial to
solve them with high precision. The renormalized functions
$A_{0,1,2}(\omega,k_x,k_y)$ and the dynamically generated gap
$m(\omega,k_x,k_y)$ depend on energy and two components of momenta
separately, as a consequence of explicit Lorentz symmetry breaking
and strong anisotropy in fermion dispersion. This makes it
technically much harder to solve the coupled integral equations of
$A_{0,1,2}$ and $m$.

To estimate the importance of higher order corrections, we now
compare with the cases of QED$_3$ and 2D Dirac semimetal. In the
case QED$_3$, the critical flavor for dynamical fermion mass
generation is $N_c\approx 3.24$ at the leading order of $1/N$
expansion \cite{Appelquist88} and becomes $N_c\approx 4.0$ after
including higher order corrections \cite{Fischer04}. In 2D Dirac
semimetal, we present in Table~\ref{Table:AlphaC} the values of
$\alpha_c$ computed previously under three different approximations
at the leading order of $1/N$ expansion \cite{Khveshchenko01,
Gorbar02, Khveshchenko09, Gamayun10} and those obtained in the
presence of higher order corrections \cite{WangLiu12, Gonzalez15,
Carrington16}. For convenience, the values of $\alpha_{c}$ obtained
in our current work are also listed in Table~\ref{Table:AlphaC}.
According to the research experience accumulated in these studies,
we find that higher order corrections do lead to quantitative change
of the critical condition for dynamical gap generation. However, the
analysis performed with truncation to the lowest order is still
scientifically significant for two reasons. Firstly, it suffices to
capture many important qualitative properties of dynamical gap
generation, which are not changed by higher order corrections.
Moreover, even the quantitative result, such as the critical
parameter $\alpha_c$, obtained at the lowest order truncation is
usually only moderately altered by higher order corrections. In the
case of QED$_{3}$, $N_{c}$ increases by only about one fourth due to
non-leading order corrections. In a 2D Dirac semimetal, $\alpha_c$
obtained in the instantaneous approximation increases by about one
third comparing to the value obtained in the presence of higher
order corrections. The value $\alpha_c$ evaluated in GGG
approximation is much smaller than the ones evaluated by using other
approximations, but it still gives us a lower bound for the real
critical point $\alpha_c$. Therefore, although the critical value of
$\alpha_c$ obtained at the leading order of $1/N$ expansion is
quantitatively not accurate, they already gave us valuable
information about the possibility of dynamical gap generation and
laid the foundation for subsequent, more in-depth studies. In view
of the quantitative difference between the values of $\alpha_{c}$
obtained under different approximations in the case 2D Dirac
semimetal, as presented in Table~\ref{Table:AlphaC}, we would
naively expect the critical parameter $\alpha_{c}$ for 2D semi-Dirac
semimetal to fall in the rage $0.7 \sim 1.0$ when higher order
corrections are considered. Getting a precise value of $\alpha_c$
requires a more elaborative numerical computation of the coupled DS
equations, which is now in progress and will be reported later.
Moreover, in order to realize interaction-induced excitonic
insulator, one could design certain new 2D semi-Dirac semimetal
materials and endeavor to make the physical value of $\alpha$ as
large as possible. Our theoretic analysis presented in this paper
provides a helpful guide for such an interesting exploration.

Apart from the DS equation approach, one can study dynamical
excitonic gap generation by employing other powerful tools, such as
RG approach \cite{Herbut14, Janssen16B, Vafek08, Kubota01, Kaveh05,
Herbut16, Janssen16C} and Monte Carlo simulation \cite{Drut09A,
Drut09B, Drut09C, Armour10, Armour11, Buividovich12, Ulybyshev13,
Smith14}. To examine whether a dynamical gap is opened, one could
consider all the possible four-fermion couplings, allowed by the
lattice symmetry, and study their interplay with the Coulomb
interaction. In the absence of Coulomb interaction, weak
four-fermion couplings are usually irrelevant perturbations to the
system. However, some of these couplings might be driven by the
Coulomb interaction to become relevant under certain conditions. In
that case, even an infinitely weak coupling flows to the strong
coupling regime at low energies, leading to the excitonic pairing
instability of the system. This is an efficient way to study
dynamical gap generation and has wide applications in QED$_{3}$
\cite{Kubota01, Kaveh05, Herbut16, Janssen16C} and 3D quadratic
semimetal \cite{Herbut14, Janssen16B}. Moreover, the RG approach
also proves to be very powerful in the studies of the impact of
short-range interaction on various phase-transition instabilities in
a number of semimetal materials, including 2D Dirac semimetal
\cite{Herbut06, Herbut09}, 3D Dirac/Weyl semimetal
\cite{Maciejko14}, 3D nodal line semimetal \cite{Roy16A, Sur16}, and
3D double/triple-Weyl semimetal \cite{Roy16B}. A comprehensive RG
analysis of the role played by short-range interactions in 2D
semi-Dirac semiemtal is still lacking. We are now working on this
problem and would present the work in a separate paper.

\section*{ACKNOWLEDGEMENTS}

We would like to acknowledge the financial support by National key
Research and Development Program of China under Grant No.
2016YFA0300404, and the support by the National Natural Science
Foundation of China under Grants 11574285, 11274286, 11504379,
11674327, and U1532267. J.R.W. is also supported by the Natural
Science Foundation of Anhui Province under Grant 1608085MA19.

\appendix

\begin{widetext}

\section{Calculation of polarization $\Pi_{33}$\label{App:Polarization33}}

We now compute the polarization for the coupling between semi-Dirac
fermions $\psi$ and quantum fluctuation of insulating order
parameter $\phi$, which is defined as
\begin{eqnarray}
\Pi_{33}(\Omega,\mathbf{q}) &=& N\int\frac{d\omega}{2\pi}\int
\frac{d^2\mathbf{k}}{(2\pi)^{2}}
\mathrm{Tr}\left[\tau_{3}G_{0}(\omega,\mathbf{k})
\tau_{3}G_{0}\left(\omega + \Omega,\mathbf{k} +
\mathbf{q}\right)\right].\label{Eq:Pi33DefAppen}
\end{eqnarray}
Substituting the fermion propagator
Eq.~(\ref{Eq:FreeFermonPropagatorDef}) into
Eq.~(\ref{Eq:Pi33DefAppen}), we obtain
\begin{eqnarray}
\Pi_{33}(\Omega,q_{x},q_{y}) &=& -2N\int\frac{d\omega}{2\pi}\int
\frac{d^2\mathbf{k}}{(2\pi)^{2}} \frac{\omega\left(\omega +
\Omega\right) + a^2 k_{x}^{2}\left(k_{x}+q_{x}\right)^2 +
v^2k_{y}\left(k_{y}+q_{y}\right)} {\left[\omega^2 + a^2k_{x}^{4} +
v^2k_{y}^{2}\right] \left[\left(\omega+\Omega\right)^2 +
a^2\left(k_{x} + q_{x}\right)^{4} + v\left(k_{y} +
q_{y}\right)^2\right]}.
\end{eqnarray}
Using the Feynman parameterization
\begin{eqnarray}
\frac{1}{AB} = \int_{0}^{1}dx\frac{1}{\left[xA+(1-x)B\right]^2},
\end{eqnarray}
we further write the polarization as
\begin{eqnarray}
\Pi_{33}(\Omega,\frac{q_{x}}{\sqrt{a}},\frac{q_{y}}{v}) &=&
-\frac{2N}{\sqrt{a}v}\int_{0}^{1}dx\int\frac{dk_{x}}{2\pi}
\int\frac{d\omega}{2\pi}\int\frac{dk_{y}}{2\pi} \nonumber \\
&& \times \left\{\frac{\omega^2 + k_{y}^2 - x(1-x)\left(\Omega^2 +
q_{y}^{2}\right) + k_{x}^{2}\left(k_{x}+q_{x}\right)^2}
{\left[\omega^{2} + k_{y}^{2} +
x(1-x)\left(\Omega^2+q_{y}^{2}\right) + x\left(k_{x}+q_{x}\right)^4
+ (1-x)k_{x}^{4}\right]^2} - \frac{1}{\omega^{2} +
k_{y}^{2}+k_{x}^{4}}\right\},
\end{eqnarray}
where we have made the re-scaling manipulations:
\begin{eqnarray}
q_{x}\rightarrow\frac{q_{x}}{\sqrt{a}},\,\,\,
k_{x}\rightarrow\frac{k_{x}}{\sqrt{a}},\,\,\,
q_{y}\rightarrow\frac{q_{y}}{v},\,\,\,
k_{y}\rightarrow\frac{k_{y}}{v}.
\end{eqnarray}
Moreover, we have used
$\Pi_{33}(\Omega,\frac{q_{x}}{\sqrt{a}},\frac{q_{y}}{v})-\Pi_{33}(0,0,0)$
to replace $\Pi_{33}(\Omega,\frac{q_{x}}{\sqrt{a}},\frac{q_{y}}{v})$
to regularize the polarization function. We then define a new
variable $\mathbf{K} = (\omega, k_{y})$ and carry out the
integration over $\mathbf{K}$, which yields
\begin{eqnarray}
&&\Pi_{33}(\Omega,\frac{q_{x}}{\sqrt{a}},\frac{q_{y}}{v})\nonumber
\\
&=&-\frac{N}{\pi\sqrt{a}v}\int\frac{dk_{x}}{2\pi}\left\{\frac{1}{2}
\ln\left[\frac{\Lambda^{2} + \left(k_{x}+q_{x}\right)^4}
{\left(k_{x}+q_{x}\right)^4}\right]+\frac{1}{2}\int_{0}^{1}dx
\left[\frac{x\left(\Omega^2+q_{y}^{2}\right) + k_{x}^{4} -
k_{x}^{2}\left(k_{x}+q_{x}\right)^2}{\Lambda^{2} +
x(1-x)\left(\Omega^2+q_{y}^{2}\right) + x\left(k_{x} +
q_{x}\right)^4+(1-x)k_{x}^{4}}\right.\right.\nonumber
\\
&&\left.\left.-\frac{x\left(\Omega^2+q_{y}^{2}\right)
+k_{x}^{4}-k_{x}^{2}\left(k_{x}+q_{x}\right)^2}
{x(1-x)\left(\Omega^2+q_{y}^{2}\right) +
x\left(k_{x}+q_{x}\right)^4+(1-x)k_{x}^{4}}\right] -
\frac{1}{2}\ln\left(\frac{\Lambda^{2} +
k_{x}^{4}}{k_{x}^{4}}\right)\right\},
\end{eqnarray}
where $\Lambda$ is an UV cutoff. In the following, we calculate
$\Pi_{33}$ in several different limits.

\subsection{$q_{x}=0$}

In the limit $q_{x}=0$, we have
\begin{eqnarray}
\Pi_{33}(\Omega,0,\frac{q_{y}}{v}) &=&
-\frac{N}{2\pi\sqrt{a}v}\int\frac{dk_{x}}{2\pi}
\int_{0}^{1}dx\left[\frac{x\left(\Omega^2 +
q_{y}^{2}\right)}{\Lambda^{2} +
x(1-x)\left(\Omega^2+q_{y}^{2}\right) + k_{x}^{4}} -
\frac{x\left(\Omega^2+q_{y}^{2}\right)}{x(1-x)
\left(\Omega^2+q_{y}^{2}\right) + k_{x}^{4}}\right].
\end{eqnarray}
After integrating over $k_{x}$ and retaining the leading term, we
get
\begin{eqnarray}
\Pi_{33}(\Omega,0,\frac{q_{y}}{v}) &=& \frac{3N\left(\Omega^2 +
q_{y}^{2}\right)^{\frac{1}{4}}}{2\sqrt{2}\pi\sqrt{a}v}
\int_{0}^{1}dx\left[x(1-x)\right]^{\frac{1}{4}} \nonumber \\
&=& c_{1} \frac{N}{\sqrt{a}v}\left(\Omega^2 +
q_{y}^{2}\right)^{\frac{1}{4}},\label{Eq:PolaLimit1}
\end{eqnarray}
where $c_{1} = \frac{\Gamma(\frac{1}{4})}{8\sqrt{\pi}
\Gamma(\frac{3}{4})} \approx 0.208657$.

\subsection{$\Omega=0$ and $q_{y}=0$}

In the case $\Omega=0$ and $q_{y}=0$, the polarization is given by
\begin{eqnarray}
\Pi_{33}(0,\frac{q_{x}}{\sqrt{a}},0) &=&
-\frac{N}{\pi\sqrt{a}v}\int\frac{dk_{x}}{2\pi}\left\{\frac{1}{2}
\ln\left[\frac{\Lambda^{2} +\left(k_{x}+q_{x}\right)^4}
{\left(k_{x}+q_{x}\right)^4}\right]+\frac{1}{2}\int_{0}^{1}dx
\left[\frac{k_{x}^{4}-k_{x}^{2}\left(k_{x}+q_{x}\right)^2}{\Lambda^{2}
+x\left(k_{x}+q_{x}\right)^4+(1-x)k_{x}^{4}}\right.\right.\nonumber
\\
&&\left.\left.-\frac{k_{x}^{4} - k_{x}^{2} \left(k_{x} +
q_{x}\right)^2} {x\left(k_{x} + q_{x}\right)^4 +
(1-x)k_{x}^{4}}\right] - \frac{1}{2}\ln\left(\frac{\Lambda^{2} +
k_{x}^{4}}{k_{x}^{4}}\right)\right\},
\end{eqnarray}
Integrating over $x$ and retaining the leading term gives rise to
\begin{eqnarray}
\Pi_{33}(0,\frac{q_{x}}{\sqrt{a}},0) &=& \frac{N}{4\pi^{2}\sqrt{a}v}
\int_{-\infty}^{+\infty}dk_{x}\frac{\left(k_{x}+q_{x}\right)^4 -
k_{x}^{2}\left(k_{x}+q_{x}\right)^2}{\left(k_{x}+q_{x}\right)^4 -
k_{x}^{4}} \ln\left[\frac{\left(k_{x} +
q_{x}\right)^4}{k_{x}^{4}}\right].
\end{eqnarray}
We then assume $k_{x} = \left|q_{x}\right|y$, valid for both
positive and negative $q_{x}$, and convert $\Pi_{33}$ to
\begin{eqnarray}
\Pi_{33}(0,\frac{q_{x}}{\sqrt{a}},0) &=&
\frac{N}{4\pi^{2}}\left\{\int_{0}^{1}dy \frac{y^2} {y^2+(y-1)^{2}}
\ln\left[\frac{y^{4}}{(y-1)^{4}}\right] +
\int_{0}^{\infty}dy\frac{2y+1} {(y+1)^2 + y^{2}}
\ln\left[\frac{(y+1)^{4}}{y^{4}}\right]\right\}
\frac{|q_{x}|}{\sqrt{a}v},
\end{eqnarray}
which leads
\begin{eqnarray}
\Pi_{33}(0,\frac{q_{x}}{\sqrt{a}},0) =
c_{2}\frac{N}{\sqrt{a}v}|q_{x}|,\label{Eq:PolaLimit2}
\end{eqnarray}
where $c_{2}=\frac{1}{4}$.

\subsection{$q_{y}=0$ and $\Omega \gg aq_{x}^{2}$}

In the limit $q_{y} = 0$, we can expand the polarization in powers
of $\frac{aq_{x}^{2}}{\Omega}$ and retain the leading and
sub-leading terms, which simplifies $\Pi_{33}$ to the following form
\begin{eqnarray}
\Pi_{33}(\Omega,\frac{q_{x}}{\sqrt{a}},0) &=&
c_{1}\frac{N}{\sqrt{a}v}\Omega^{\frac{1}{2}} + \frac{1}{4\pi^{2}}
\frac{N}{\sqrt{a}v}\frac{q_{x}^{2}}{\Omega^{\frac{1}{2}}}I_{1},
\end{eqnarray}
where
\begin{eqnarray}
I_{1} &=& \int_{-\infty}^{+\infty}dy\int_{0}^{1}dx
\left[\frac{-y^{2}}{\left[x(1-x)+y^{4}\right]} + \frac{-6x^{2}y^{2}
+ 8xy^{6}}{\left[x(1-x) + y^{4}\right]^{2}} + \frac{16x^{3}y^{6}}
{\left[x(1-x)+y^{4}\right]^{3}}\right] = \frac{5\pi^{\frac{3}{2}}
\Gamma(\frac{3}{4})}{8\Gamma(\frac{5}{4})}.
\end{eqnarray}
Now $\Pi_{33}$ can be approximately written as
\begin{eqnarray}
\Pi_{33}(\Omega,\frac{q_{x}}{\sqrt{a}},0) &\approx&
c_{1}\frac{N}{\sqrt{a}v}\Omega^{\frac{1}{2}} +
c_{3}\frac{N}{\sqrt{a}v}\frac{q_{x}^{2}}{\Omega^{\frac{1}{2}}}
\approx c_{1}\frac{N}{\sqrt{a}v}\Omega^{\frac{1}{2}},
\label{Eq:PolaLimit3}
\end{eqnarray}
where $c_{3} = \frac{5\Gamma(\frac{3}{4})}{32\sqrt{\pi}
\Gamma(\frac{5}{4})}$.

\subsection{Ansatz for $\Pi_{33}$}

According to the polarization calculated in different limits, as
shown in Eqs.~(\ref{Eq:PolaLimit1}), (\ref{Eq:PolaLimit2}), and
(\ref{Eq:PolaLimit3}), we find it appropriate to express $\Pi_{33}$
by the following ansatz:
\begin{eqnarray}
\Pi_{33}(\Omega,\frac{q_{x}}{\sqrt{a}},\frac{q_{y}}{v}) =
\frac{N}{\sqrt{a}v}\left[b_{1}\left(\Omega^2 + q_{y}^{2}\right) +
b_{2}q_{x}^{4}\right]^{\frac{1}{4}},
\end{eqnarray}
where $b_{1} = c_{1}^{4}$ and $b_{2} = c_{2}^{4}$. Using the
re-scaling relations $\frac{q_{x}}{\sqrt{a}}\rightarrow q_{x}$ and
$\frac{q_{y}}{v}\rightarrow q_{y}$, we eventually write $\Pi_{33}$
in the form
\begin{eqnarray}
\Pi_{33}(\Omega,q_{x},q_{y}) = \frac{N}{\sqrt{a}v}\left[b_{1}
\left(\Omega^2+v^{2}q_{y}^{2}\right) +
b_{2}a^{2}q_{x}^{4}\right]^{\frac{1}{4}}.
\end{eqnarray}

\section{Self-Energy of Semi-Dirac fermion\label{App:FermionSelfEnergy}}

The self-energy of semi-Dirac fermion induced by the quantum
fluctuation of excitonic insulating order parameter is defined as
\begin{eqnarray}
\Sigma_{fb}(\omega,\mathbf{k}) =
\int'\frac{d\Omega}{2\pi}\frac{d^2\mathbf{q}}{(2\pi)^{2}} \tau_{3}
G_{0}\left(\omega+\Omega,\mathbf{k}+\mathbf{q}\right)
\tau_{3}D(\Omega,\mathbf{q})\label{Eq:FermionSelfEnergyDefAppen}
\end{eqnarray}
to the leading order of perturbative expansion, where
\begin{eqnarray}
D(\Omega,\mathbf{q}) = \frac{1}{\Pi_{33}(\Omega,\mathbf{q})}.
\end{eqnarray}
Substituting Eq.~(\ref{Eq:FreeFermonPropagatorDef}) into
Eq.~(\ref{Eq:FermionSelfEnergyDefAppen}), we obtain
\begin{eqnarray}
\Sigma_{fb}(\omega,\mathbf{k}) = \int'\frac{d\Omega}{2\pi}
\frac{d^2\mathbf{q}}{(2\pi)^{2}} \frac{i\left(\omega+\Omega\right) -
a\left(k_{x}+q_{x}\right)^{2}\tau_{1} -
v\left(k_{y}+q_{y}\right)\tau_{2}} {\left(\omega+\Omega\right)^2 +
a^2\left(k_{x}+q_{x}\right)^{4} +
v^2\left(k_{y}+q_{y}\right)^2}D(\Omega,\mathbf{q}).
\end{eqnarray}
It is easy to verify that
\begin{eqnarray}
\Sigma_{fb}(0,0) = -\int'\frac{d\Omega}{2\pi}
\frac{d^2\mathbf{q}}{(2\pi)^{2}}\frac{aq_{x}^{2}\tau_{1}}
{\Omega^2+a^2q_{x}^{4}+v^2q_{y}^2}D(\Omega,\mathbf{q}),
\end{eqnarray}
which is not divergent in the lowest energy limit. This constant
contribution plays no role in the low-energy region, and thus can be
safely dropped in the RG analysis. Expanding
$\Sigma_{fb}(\omega,\mathbf{k})$ in powers of small values of
$i\omega$, $k_{x}$, and $k_{y}$, and retaining the leading term, we
find that
\begin{eqnarray}
\Sigma_{fb}(\omega,\mathbf{k}) \approx
(-i\omega)\Sigma_{1} + ak_{x}^{2}\Sigma_{2}\tau_{1} + vk_{y}\Sigma_{3}\tau_{2},
\end{eqnarray}
where
\begin{eqnarray}
\Sigma_{1} &=& \int'\frac{d\Omega}{2\pi}
\frac{d^2\mathbf{q}}{(2\pi)^{2}}\frac{\Omega^2 - a^2q_{x}^{4} -
v^2q_{y}^2}{\left(\Omega^2 + a^2q_{x}^{4} + v^2q_{y}^2\right)^2}
D(\Omega,\mathbf{q}),\label{Eq:E1Appen} \\
\Sigma_{2}&=&\int'\frac{d\Omega}{2\pi}\frac{d^2\mathbf{q}}{(2\pi)^{2}}
\frac{-\Omega^{4}-3a^4q_{x}^{8}+12\Omega^2a^2q_{x}^{4} +
12a^2q_{x}^{4}v^2q_{y}^{2}-2\Omega^2v^2q_{y}^{2} -
v^4q_{y}^4}{\left(\Omega^2+a^2q_{x}^{4} +
v^2q_{y}^2\right)^3}D(\Omega,\mathbf{q}),\label{Eq:E2Appen} \\
\Sigma_{3} &=& \int'\frac{d\Omega}{2\pi}
\frac{d^2\mathbf{q}}{(2\pi)^{2}}\frac{-\Omega^2 -
a^{2}q_{x}^{4}+v^{2}q_{y}^{2}}{\left(\Omega^2 + a^2q_{x}^{4} +
v^2q_{y}^{2}\right)^{2}}D(\Omega,\mathbf{q}).\label{Eq:E3Appen}
\end{eqnarray}

We now need to perform RG re-scaling manipulations. Firstly, we
adopt a RG scheme which integrate over $\Omega$ and momenta in the
following way:
\begin{eqnarray}
-\infty < \Omega < \infty, \qquad b\Lambda < E_{q} < \Lambda,\quad
\mathrm{with} \quad E_{q} = \sqrt{a^2q_{x}^{4} + v^2q_{y}^{2}},
\label{Eq:RGSchemeAAppen}
\end{eqnarray}
where $b=e^{-l}$. If we define
\begin{eqnarray}
E_{q} = \sqrt{a^2q_{x}^{4} + v^{2}q_{y}^{2}}, \qquad \delta =
\frac{aq_{x}^{2}}{v\left|q_{y}\right|},\label{Eq:IntegalTransformA}
\end{eqnarray}
the two components of momenta are given by
\begin{eqnarray}
\left|q_{x}\right| = \frac{\sqrt{\delta}\sqrt{E_{q}}}{\sqrt{a}
\left(1 + \delta^2\right)^{\frac{1}{4}}}, \qquad \left|q_{y}\right|
= \frac{E_{q}}{v\sqrt{1+\delta^2}}.\label{Eq:IntegalTransformB}
\end{eqnarray}
Therefore, the integration over $q_{x}$ and $q_{y}$ can be converted
to the integration of over $E_{q}$ and $\delta$, through the relation:
\begin{eqnarray}
d|q_{x}|d|q_{y}|&=&
\left|\left|
\begin{array}{cc}
\frac{\partial |q_{x}|}{\partial E_{q}} & \frac{\partial |q_{x}|}{\partial \delta}
\\
\frac{\partial |q_{y}|}{\partial E_{q}} & \frac{\partial
|q_{y}|}{\partial \delta}
\end{array}
\right|\right|dE_{q}d\delta \nonumber \\
&=& \left|\frac{\partial|q_{x}|}{\partial E_{q}}\frac{\partial
|q_{y}|}{\partial\delta} - \frac{\partial |q_{x}|}{\partial
\delta}\frac{\partial|q_{y}|}{\partial
E_{q}}\right|dE_{q}d\delta \nonumber \\
&=& \frac{\sqrt{E_{q}}}{2v\sqrt{a} \sqrt{\delta}\left(1 +
\delta^2\right)^{\frac{3}{4}}}dE_{q}d\delta.
\label{Eq:IntegralTranformMeasure}
\end{eqnarray}
We now can calculate Eqs.~(\ref{Eq:E1Appen})-(\ref{Eq:E3Appen}) by
using the transformations given by Eq.~(\ref{Eq:IntegalTransformB})
and Eq.~(\ref{Eq:IntegralTranformMeasure}) along with the RG scheme
(\ref{Eq:RGSchemeAAppen}), and obtain
\begin{eqnarray}
\Sigma_{1} = C_{1}\ln(b^{-1}),\qquad \Sigma_{2} = C_{2}\ln(b^{-1}),
\qquad \Sigma_{3} = C_{3}\ln(b^{-1}),
\end{eqnarray}
where
\begin{eqnarray}
C_{1} &=& \frac{1}{4 N\pi^{3}}\int_{-\infty}^{+\infty}dx
\int_{0}^{+\infty}d\delta \frac{1}{\delta^{\frac{1}{2}}
\left(1+\delta^2\right)^{\frac{1}{2}}}\frac{x^{2} -
1}{\left(x^{2}+1\right)^2}\mathcal{G}(x,\delta),\label{Eq:C1RGSchemeAAppen}
\\
C_{2} &=&
\frac{1}{4N\pi^3}\int_{-\infty}^{+\infty}dx\int_{0}^{+\infty}
d\delta \frac{1}{\delta^{\frac{1}{2}}
\left(1+\delta^2\right)^{\frac{5}{2}}}\nonumber \\
&&\times\frac{-x^{4}\left(1+\delta^{2}\right)^{2} - 3\delta^{4} +
12x^{2}\delta^{2}\left(1+\delta^2\right)+12\delta^{2} -
2x^{2}\left(1+\delta^2\right) - 1}{\left(x^{2} + 1\right)^3}
\mathcal{G}(x,\delta),\label{Eq:C2RGSchemeAAppen} \\
C_{3}&=&\frac{1}{4N\pi^3}\int_{-\infty}^{+\infty}dx\int_{0}^{+\infty}
d\delta \frac{1}{\delta^{\frac{1}{2}}
\left(1+\delta^2\right)^{\frac{3}{2}}} \frac{-x^{2}(1+\delta^2) -
\delta^{2}+1}{\left(x^{2}+1\right)^{2}}
\mathcal{G}(x,\delta),\label{Eq:C3RGSchemeAAppen}\\
\mathcal{G}^{-1} &=& \left[b_{1}x^{2}(1+\delta^2)+b_{2} \delta^{2} +
b_{1}\right]^{\frac{1}{4}}.\label{Eq:GRGSchemeAAppen}
\end{eqnarray}

\section{Calculation of $C_{1, 2, 3}$ using different RG schemes\label{App:C1C2C3Expressions}}

As discussed in Sec.~\ref{Sec:SubSecNFLSoleQuantumFL}, there are a
number of different RG schemes, which are distinguished by the
different manners to integrate over energy and momenta. Here, we
provide the expressions for $C_{1,2,3}$ obtained separately by
employing these RG schemes.

For the RG scheme
\begin{eqnarray}
\int' d\Omega d^2\mathbf{q} = \int_{-\infty}^{+\infty}d\Omega
\left(\int_{-\Lambda}^{-\sqrt{b}\Lambda}+\int_{\sqrt{b}\Lambda}^{\Lambda}\right)dq_{x}
\int_{-\infty}^{+\infty}dq_{y},\label{Eq:RGSchemeBAppen}
\end{eqnarray}
the expressions of $C_{i}$ with ($i=1, 2, 3 $) are given by
\begin{eqnarray}
C_{1}&=&\frac{1}{8N\pi^3}\int_{-\infty}^{+\infty} dx
\int_{-\infty}^{+\infty}dy\frac{x^{2} - 1 - y^{2}}
{\left(x^{2}+1+y^2\right)^2}\mathcal{G}(x,y),\label{Eq:C1RGSchemeBAppen}\\
C_{2}&=&\frac{1}{8N\pi^3}\int_{-\infty}^{+\infty}dx
\int_{-\infty}^{+\infty}dy \frac{-x^{4}-3 + 12x^{2} + 12y^{2} -
2x^{2}y^{2} - y^{4}} {\left(x^{2} + 1 + y^{2}\right)^3}
\mathcal{G}(x,y),\label{Eq:C2RGSchemeBAppen} \\
C_{3}&=&\frac{1}{8N\pi^{3}}\int_{-\infty}^{+\infty}dx
\int_{-\infty}^{+\infty}
\frac{-x^{2}-1+y^{2}}{\left(x^{2}+1+y^{2}\right)^{2}}
\mathcal{G}(x,y),\label{Eq:C3RGSchemeBAppen} \\
\mathcal{G}^{-1} &=& \left[b_{1}\left(x^{2}+y^{2}\right)
+b_{2}\right]^{\frac{1}{4}}.\label{Eq:GRGSchemeBAppen}
\end{eqnarray}
By virtue of the exchange symmetry between variables $x$ and $y$, it
is easy to find that
\begin{eqnarray}
C_{1} = C_{3} = \frac{1}{4N\pi^3}\int_{-\infty}^{+\infty} dx
\int_{-\infty}^{+\infty}dy\frac{ -1} {\left(x^{2}+1+y^2\right)^2}
\frac{1}{\left[b_{1}\left(x^{2} +
y^{2}\right)+b_{2}\right]^{\frac{1}{4}}}.
\end{eqnarray}

For the RG scheme given by
\begin{eqnarray}
\int' d\Omega d^2 \mathbf{q} = \int_{-\infty}^{+\infty}d\Omega
\int_{-\infty}^{+\infty}dq_{x}\left(\int_{-\Lambda}^{-b\Lambda} +
\int_{b\Lambda}^{\Lambda}\right)dq_{y},\label{Eq:RGSchemeCAppen}
\end{eqnarray}
we accordingly find that
\begin{eqnarray}
C_{1} &=& \frac{1}{4N\pi^3}\int_{-\infty}^{+\infty}dx
\int_{-\infty}^{+\infty}dy\frac{x^{2} - y^{4} - 1}{\left(x^{2} +
y^{4}+1\right)^2} \mathcal{G}(x,y), \label{Eq:C1RGSchemeCAppen} \\
C_{2}&=&\frac{1}{4N\pi^{3}} \int_{-\infty}^{+\infty}dx
\int_{-\infty}^{+\infty}dy\frac{-x^{4}-3y^{8}+12x^{2}y^{4}
+12y^{4}-2x^{2}-1} {\left(x^{2}+y^{4}+1\right)^3}\frac{1}{
\left[b_{1}\left(x^{2}+1\right) + b_{2}y^{4}\right]^{\frac{1}{4}}}
\mathcal{G}(x,y), \label{Eq:C2RGSchemeCAppen} \\
C_{3} &=& \frac{1}{4N\pi^3} \int_{-\infty}^{+\infty}dx
\int_{-\infty}^{+\infty}dy\frac{-x^2-y^{4}+1}
{\left(x^{2}+y^{4}+1\right)^{2} }\frac{1}{\left[b_{1}\left(x^{2} +
1\right) + b_{2}y^{4}\right]^{\frac{1}{4}}}\mathcal{G}(x,y),
\label{Eq:C3RGSchemeCAppen} \\
\mathcal{G}^{-1} &=& \left[b_{1}\left(x^{2}+1\right) +
b_{2}y^{4}\right]^{\frac{1}{4}}.\label{Eq:GRGSchemeCAppen}
\end{eqnarray}

For the RG scheme
\begin{eqnarray}
\int' d\Omega d^2\mathbf{q} = \left(\int_{-\Lambda}^{-b\Lambda} +
\int_{b\Lambda}^{\Lambda}\right)d\Omega
\int_{-\infty}^{+\infty}dq_{x}\int_{-\infty}^{+\infty}dq_{y},
\label{Eq:RGSchemeDAppen}
\end{eqnarray}
we obtain
\begin{eqnarray}
C_{1}&=&\frac{1}{4N\pi^3} \int_{-\infty}^{+\infty}
dx\int_{-\infty}^{+\infty}dy\frac{1 -x^4-y^2}
{\left(1+x^4+y^2\right)^2}\mathcal{G}(x,y),\label{Eq:C1RGSchemeDAppen}
\\
C_{2}&=&\frac{1}{4N\pi^3} \int_{-\infty}^{+\infty}
dx\int_{-\infty}^{+\infty}dy \frac{-1-3x^8 + 12x^4 + 12x^4 y^2 -
2y^2 - y^4}{\left(1+x^4+y^2\right)^3}
\mathcal{G}(x,y),\label{Eq:C2RGSchemeDAppen} \\
C_{3}&=&\frac{1}{4N\pi^3}\int_{-\infty}^{+\infty}
dx\int_{-\infty}^{+\infty}dy\frac{-1 -x^4+y^2}{\left(1 + x^4 +
y^2\right)^2}\mathcal{G}(x,y),\label{Eq:C3RGSchemeDAppen}
\\
\mathcal{G}^{-1} &=& \left[b_{1}\left(1 + y^2\right) +
b_{2}x^4\right]^{\frac{1}{4}}.\label{Eq:GRGSchemeDAppen}
\end{eqnarray}

All the above expressions for $C_{1,2,3}$ are used in
Sec.~\ref{Sec:SubSecNFLSoleQuantumFL} to investigate the
Yukawa-coupling between the semi-Dirac fermions and the quantum
fluctuation of excitonic order parameter at the semimetal-insulator
QCP. We find that these different RG schemes lead to essentially the
same conclusion.

\section{Expressions of $C'_{1, 2, 3}$ in four different RG schemes
\label{App:C1C2C3PrimeExpressions}}

In RG scheme $1$ shown in Eq.~(\ref{Eq:RGSchemeAAppen}), $C_{1}'$
can be computed through $C_{1}$ shown in
Eq.~(\ref{Eq:C1RGSchemeAAppen}) by replacing $\mathcal{G}(x,\delta)$
with $-\mathcal{G}'(x,\delta)$, and $C_{2,3}'$ can be computed
through $C_{2,3}$ shown in Eqs.~(\ref{Eq:C2RGSchemeAAppen}) and
(\ref{Eq:C3RGSchemeAAppen}) by replacing $\mathcal{G}(x,\delta)$
with $\mathcal{G}'(x,\delta)$, where $\mathcal{G}'(x,\delta)$ is
given by
\begin{eqnarray}
\mathcal{G}'^{-1}(x,\delta) &=& \frac{\sqrt{\delta +
\frac{\beta}{\left(1+\delta^2\right)^{\frac{1}{2}}}}}{2\pi N\alpha}
+ \left[\frac{d_{x}\delta}{\left(x^2\left(1+\delta^2\right) +
c_{0}\delta^2 + 1\right)^{\frac{1}{4}}}+\frac{d_{y}}{\left(x^2
\left(1+\delta^2\right) + c_{0}\delta^2 +
1\right)^{\frac{3}{4}}}\right],
\end{eqnarray}
in which $\beta = \frac{a\Lambda}{v^2}$.

In RG scheme $2$ shown in Eq.~(\ref{Eq:RGSchemeBAppen}), $C_{1}'$
can be computed through $C_{1}$ given by
Eq.~(\ref{Eq:C1RGSchemeBAppen}) by replacing $\mathcal{G}(x, y)$
with $-\mathcal{G}'(x, y)$, and $C_{2,3}'$ can be computed through
$C_{2,3}$ given by Eqs.~(\ref{Eq:C2RGSchemeBAppen}) and
(\ref{Eq:C3RGSchemeBAppen}) by replacing $\mathcal{G}(x, y)$ with
$\mathcal{G}'(x, y)$, where $\mathcal{G}'(x, y)$ can be written as
\begin{eqnarray}
\mathcal{G}'^{-1}(x,y) = \frac{\sqrt{1+\beta^{2}y^2}}{2\pi
N\alpha}+\left[\frac{d_{x}} {\left(x^2 + c_{0} +
y^2\right)^{\frac{1}{4}}} + \frac{d_{y}y^2}{\left(x^2 + c_{0} +
y^2\right)^{\frac{3}{4}}}\right],
\end{eqnarray}
in which $\beta=\frac{av}{\Lambda}$.

In RG scheme $3$ shown in Eq.~(\ref{Eq:RGSchemeCAppen}), $C_{1}'$
can be computed through $C_{1}$ given by
Eq.~(\ref{Eq:C1RGSchemeCAppen}) by replacing $\mathcal{G}(x, y)$
with $-\mathcal{G}'(x, y)$, and $C_{2,3}'$ can be computed through
$C_{2,3}$ given by Eqs.~(\ref{Eq:C2RGSchemeCAppen}) and
(\ref{Eq:C3RGSchemeCAppen}) by replacing $\mathcal{G}(x, y)$ with
$\mathcal{G}'(x, y)$, where $\mathcal{G}'(x, y)$ takes the form
\begin{eqnarray}
\mathcal{G}'^{-1}(x,y)=\frac{\sqrt{y^2+\beta}}{2\pi N\alpha} +
\left[\frac{d_{x}y^2}
{\left(x^{2}+c_{0}y^{4}+1\right)^{\frac{1}{4}}} +
\frac{d_{y}}{\left(x^{2}+c_{0}y^{4}+1\right)^{\frac{3}{4}}}\right],
\end{eqnarray}
in which $\beta=\frac{a\Lambda}{v}$.

In RG scheme $4$ shown in Eq.~(\ref{Eq:RGSchemeDAppen}), $C_{1}'$
can be computed through $C_{1}$ shown in
Eq.~(\ref{Eq:C1RGSchemeDAppen}) by replacing $\mathcal{G}(x,y)$ with
$-\mathcal{G}'(x,y)$, and $C_{2,3}$ can be computed through $C_{2,3}$
given by Eqs.~(\ref{Eq:C2RGSchemeDAppen}) and
(\ref{Eq:C3RGSchemeDAppen}) by replacing $\mathcal{G}(x,y)$ with
$\mathcal{G}'(x,y)$, where $\mathcal{G}'(x,y)$ has the expression
\begin{eqnarray}
\mathcal{G}'^{-1}(x,y) &=& \frac{\sqrt{x^2+\beta
y^2}}{2N\pi\alpha}+\left[\frac{d_{x}x^2} {\left(1+c_{0}x^4 +
y^2\right)^{\frac{1}{4}}} + \frac{d_{y}y^2}{\left(1 + c_{0}x^4 +
y^2\right)^{3/4}}\right],
\end{eqnarray}
in which $\beta = \frac{a\Lambda}{v^2}$.
\end{widetext}

\end{document}